\documentclass[twoside]{article}
\usepackage{pslatex}
\usepackage[a4paper]{anysize}
\marginsize{25mm}{20mm}{15mm}{16mm}
\usepackage{rotating}
\usepackage{natbib}
\usepackage[bf]{caption}
\usepackage{graphicx}
\usepackage{color}
\usepackage{epsfig}
\usepackage{amsfonts}
\usepackage{amsmath}
\usepackage{hyperref}
\usepackage{booktabs}
\usepackage{titling}
\usepackage{multicol}
\raggedcolumns
\usepackage{fancyhdr}
\newcommand{\mytitle}{Calibration of the strain amplitude recorded with DAS
using a strainmeter array}
\newcommand{\myauthors}{T. Forbriger, N. Karamzadeh, J. Azzola, E. Gaucher, R.
Widmer‐Schnidrig, A. Rietbrock, 2025}
\newenvironment{datres}{\section*{Data and Resources}}{}
\newenvironment{ack}{\section*{Acknowledgments}}{}
\newcommand{\tbl}[1]{\caption{#1}}
\newcommand{\colrule}{\toprule}
\newcommand{\botrule}{\bottomrule}
\newcommand{\trowsep}{\\ \bottomrule\rule{0pt}{12pt}\\}
\renewcommand{\eqref}[1]{Eq.~(\ref{#1})}
\newcommand{\doiurl}[1]{\url{https://dx.doi.org/#1}}
\newcommand{\qmarks}[1]{`#1'}
\newcommand{\tunnellocation}[1]{\qmarks{#1}}
\newcommand{\Lanton}{\tunnellocation{Anton Gang}}
\newcommand{\Lvorstollen}{\tunnellocation{Vorstollen}}
\newcommand{\Tfiberstrain}{\qmarks{fiber strain}}
\newcommand{\Trockstrain}{\qmarks{rock strain}}
\newcommand{\Tstraintransfer}{\qmarks{strain transfer rate}}
\newcommand{\SMRcablespec}{\ref{sec:cablespec}}
\newcommand{\SMRfigmapzoomA}{\ref{fig:mapzoom1}}
\newcommand{\SMRfigmapzoomB}{\ref{fig:mapzoom2}}
\newcommand{\SMRtablocations}{\ref{tab:locations}}
\newcommand{\SMRtabUsedEvents}{\ref{tab:UsedEvents}}
\newcommand{\SMRsubsecnoisereduction}{\ref{subsec:noise:reduction}}
\newcommand{\SMRsubsecstrainmetercalibration}{\ref{subsec:strainmeter:calibration}}
\newcommand{\SMRsubseccavityeffects}{\ref{subsec:cavity:effects}}
\newcommand{\SMRfignoiseredP}{\ref{fig:noisered:P}}
\newcommand{\SMRfignoiseredall}{\ref{fig:noisered:all}}
\newcommand{\SMRfigstrvsbaz}{\ref{fig:str:vs:baz}}
\newcommand{\SMRfigSTRfNCC}{\ref{fig:STR_f_NCC}}
\newcommand{\SMRfigNCCvsmaxAmp}{\ref{fig:NCC_vs_maxAmp}}
\newcommand{\SMRfigturkeyNCC}{\ref{fig:turkey_NCC}}
\newcommand{\SMRfiggfzAvpri}{\ref{fig:gfzAvpri}}
\newcommand{\SMRsubsecscalingparticlevelocity}{\ref{subsec:scaling:particle:velocity}}
\newcommand{\SMRfigRCDASvsseismometerandstrainmeter}{\ref{fig:RC:DAS:vs:seismometer:and:strainmeter}}
\newcommand{\SMRfigNCCDASseismometerandstrainmeter}{\ref{fig:NCC:DAS:seismometer:and:strainmeter}}
\newcommand{\Unstrain}{nstrain}
\newcommand{\degree}[1]{\ensuremath{#1^\circ}}
\newcommand{\azimuth}[1]{\text{N}\degree{#1}\text{E}}
\newcommand{\eventlink}[2]{\href{http://geofon.gfz-potsdam.de/eqinfo/event.php?id=#2}{#1}}
\newcommand{\vect}[1]{\mathbf{#1}}
\DeclareMathOperator{\Sd}{d}
\newcommand{\AZI}{\psi}
\newcommand{\ett}{\epsilon_{\theta\theta}}
\newcommand{\epp}{\epsilon_{\phi\phi}}
\newcommand{\etp}{\epsilon_{\theta\phi}}
\newcommand{\epsi}{\epsilon(\AZI)}
\newcommand{\eA}{\epsilon_A}
\newcommand{\eB}{\epsilon_B}
\newcommand{\eC}{\epsilon_C}
\newcommand{\Sstr}{r}

\newcommand{\Mrot}{\mathbf{M}}

\begin{document}
\title{Calibration of the strain amplitude recorded with DAS\\
using a strainmeter array}
\author{Thomas Forbriger$^{*1}$,
Nasim Karamzadeh$^{1,2}$,
Jérôme Azzola$^{3}$,\\
Emmanuel Gaucher$^{3}$,
Rudolf Widmer-Schnidrig$^{4}$,
Andreas Rietbrock$^{1}$}
\predate{\begin{flushleft}}
\postdate{\end{flushleft}}
\date{$^{1}$Karlsruhe Institute of Technology (KIT), Geophysical Institute,
Karlsruhe, Germany\\
$^{2}$now at University of Münster Institut für Geophysik,
Münster, Germany\\
$^{3}$Karlsruhe Institute of Technology (KIT), Institute for Applied
Geosciences,
Karlsruhe, Germany\\
$^{4}$Institute of Geodesy, University of Stuttgart,
Stuttgart, Germany\\[4pt]
$^{*}$Corresponding author: 
Thomas.Forbriger@kit.edu,
Karlsruhe Institute of Technology (KIT),
Geophysical Institute (GPI),
Black Forest Observatory (BFO),
Heubach 206,
77709 Wolfach,
Germany}
\maketitle
%
\noindent
Preprint of:\\
T. Forbriger, N. Karamzadeh, J. Azzola, E. Gaucher, R. Widmer‐Schnidrig, A.
Rietbrock, 2025. Calibration of the Strain Amplitude Recorded with DAS Using
a Strainmeter Array. Seism. Res. Lett., 96(4), 2356--2367. 
doi: \href{http://dx.doi.org/10.1785/0220240308}{10.1785/0220240308}
  \subsection*{Abstract}
  The power of distributed acoustic sensing (DAS) lies in its ability to
  sample deformation signals along an optical fiber at hundreds of locations
  with only one interrogation unit (IU).
  While the IU is calibrated to record \Tfiberstrain, the properties
  of the cable and its coupling to the rock control the \Tstraintransfer\ and
  hence how much of \Trockstrain\ is represented in the recorded signal.
  We use DAS recordings in an underground installation near an array
  of strainmeters in order to calibrate the \Tstraintransfer\ in situ, using
  earthquake signals between 0.05\,Hz and 0.1\,Hz.
  A tight-buffered cable and a standard loose-tube telecommunication cable
  (running in parallel) are used, where a section of both cables 
  loaded down by loose sand and sand bags
  is compared to a section, where cables are just unreeled
  on the floor.
  The \Tstraintransfer\ varies between 0.13 and 0.53 depending on cable and
  installation type.
  The sandbags show no obvious effect and the tight-buffered cable generally
  provides a larger \Tstraintransfer. 
  Calibration of the \Tstraintransfer\ with respect to the strainmeter does not
  depend on wave propagation parameters.
  Hence it is applicable to the large
  amplitude surface wave signal in a strain component almost perpendicular to
  the great-circle direction for which a waveform comparison with
  seismometer data does not work.
  The noise background for \Trockstrain\ in the investigated band is found at
  about an rms-amplitude of 0.1\,\Unstrain{} 
  in 1/6 decade for the tight-buffered
  cable.
  This allows a detection of marine microseisms at times of high
  microseism amplitude.

\begin{multicols}{2}
\section{Introduction}
Distributed Acoustic Sensing (DAS) measures dynamic strain in an optical
fiber. The DAS interrogation unit (IU) sends laser-generated coherent light
into an optical fiber possibly extending for tens of kilometers.
The light gets partly back-scattered by Rayleigh scattering due to manufacturing
imperfections along the optical fiber and is sensitive to rapid variations in strain, 
commonly referred to as dynamic strain. 
Pioneering work by \cite{dakin1990summary}, followed by
\cite{taylor1993apparatus}, recognized the potential of coherent Rayleigh
back-scatter in assessing spatial disturbances in optical fibers. 
Following this work a range of techniques has been developed and implemented in
IUs to measure the phase of the back-scattered signal.
\cite{hartog2017introduction}
summarizes these progressions and outlines applications of 
distributed optical fiber sensors across various domains.

In seismology, the power of DAS lies in its ability to sample deformation
signals along an optical fiber at hundreds of locations over distances of many
kilometers with a single IU.
In some applications, unused telecommunication infrastructures (so-called
dark-fibers) can be leveraged, which significantly reduces the necessary
effort for field deployment.
\cite{lindsey_review_2021} and \cite{li2021literaturereview} present an
overview of fields of application in geosciences.
Most of these applications rely only on the phase information in the recorded
data.

Use cases, which rely on amplitude are less frequent.
Such a use case would be the measurement of volume strain for the reduction of
Newtonian Noise \citep{harms2015,harms2022}, 
as being planned for the Einstein Telescope \citep{et2020}.
With well calibrated DAS strain recordings on cables running in six different directions
from a single location, the full strain tensor could be composed.
This requires a proper calibration of the \Tstraintransfer{} 
(the fraction of \Trockstrain{} picked up as \Tfiberstrain{}) 
and a sufficiently low detection level, such that the strain background signal
is resolved.
\citet{rademacher2024} recently pointed at the limitations which still exist
for DAS in both respects.
We investigate these in the current study.

Signals recorded by the DAS IU represent the deformation
of the fiber (denoted here as \Tfiberstrain).
However, observations showing that the signal from two colocated cables might
differ in
amplitude \citep[and the current study]{azzola_comparison_2022} suggest that 
\Tfiberstrain\ does not always represent the actual \Trockstrain. 
This discrepancy can arise from the coupling of the fiber to the ground
through the various layers of the cable.

\cite{reinsch2017}, for example, focus on the internal structure of an optic 
fiber cable and discuss this discrepancy based on a physical
model \citep{li2006} of a DAS cable, based on actual material properties.
They derive the \Tstraintransfer\ as the ratio between \Tfiberstrain\ and
\Trockstrain.
In particular gel layers in loose-tube cables let the \Tstraintransfer\ 
be less than 1 due to their small value of Young's modulus.  
These thixotropic fluids are used to protect the fibers from damage.
\cite{reinsch2017} estimate that during measurement their yield point is not
exceeded, such that they behave in a linear elastic way.
In practice the parameters of the different layers in the cable are not
available from manufacturers and the coupling to the rock is controlled by the
actual installation conditions.
For this reason the \Tstraintransfer{} can be determined empirically only.

The studies by
\citet{lindsey2020} and \citet{paitz2020} are two examples for in-situ
calibration experiments. 
Both use a Silixa iDAS IUs and both primarily use surface wave signals.

\citet{lindsey2020} focus on the low-frequency band from about 0.08~Hz to
1~Hz.
They not only describe the calibration experiment, but also provide a well-written
introduction to DAS including 
the conversion from optical phase to strain,
the coupling of rock to fiber, and
typical optical sources of noise.
The DAS signals in their study are recorded on a gel-filled loose-tube dark
fiber installed in a conduit.
They convert the signals to particle velocity based on the 
assumption of plane waves
and compare with the recordings of a broad-band
seismometer at 66~m distance. 
The needed scaling factor, which is phase velocity along the fiber, is
obtained by fk-analysis of 1~km of DAS data, which allows to capture wave
dispersion and does not rely on the assumption of great-circle propagation.
Though the superposition of Rayleigh- and Love-waves 
(both propagating at difference phase velocity)
might still present a problem.
The authors use surface waves from four teleseismic earthquakes to demonstrate
that the amplitude response of the DAS is like nominally expected for
frequencies below 0.1~Hz, although there is a significant amplitude
fluctuation of a factor of 10 along the fiber \cite[][their
figure~6b]{lindsey2020}.
For frequencies higher than 0.1~Hz, they find DAS amplitudes larger than
nominal by up to about 10~dB (about a factor of 3).
The analysis of earthquakes and marine microseisms in this band yields
inconsistent results in that the amplification seen for marine microseisms is
larger.
As a potential cause, the authors mention possible problems with the
fk-analysis at near-perpendicular incidence of marine microseisms.
While they prefer fiber coupling issues as the explanation for the
amplification, they do not clarify how a passive mechanism could account for the signal amplification.

\citet{paitz2020} derive a frequency dependent transfer function of DAS
signals with respect to reference recordings for seven frequency bands,
spanning a total range from 0.34\,mHz to 60\,Hz.
As test signals, they use hydraulic stimulation and surface waves from four
earthquakes (one per analyzed frequency band) and one icequake.
The installation types of DAS cables in the various experiments differ
significantly, without the possibility to track down systematic variations to
the type of installation.
The hydraulic stimulation experiment carried out for the lowest frequency
bands is exceptional in that the DAS fiber is cemented in a borehole and in
that a fiber-Bragg-grating (FBG) strainmeter could be used as a reference,
while for the others strain had to be estimated from particle velocity.
Though the results for this hydraulic stimulation does not appear exceptional
in terms of deviation from the expected nominal response.
Overall the authors report a variation of the amplitude response by
$\pm$10~dB (about a factor of 0.3 to 3) around the nominal value.

When comparing DAS recordings with seismometer recordings a translating of
strain or strain rate into particle displacement or velocity (or vice versa)
is necessary.
This is only possible for non-dispersive plane waves with known ray-parameter. 
The limitations of this approach and the resulting waveform dissimilarity are
shown by \citet[][their figure~4a]{lindsey2020} for example.
\citet{paitz2020} mention incorrect phase-velocity estimates as a
possible source of inaccuracy. 

In the current study, we directly compare the recordings of a DAS IU 
with those from the long-running array of Invar-wire strainmeters 
at the Black Forest Observatory (BFO). 
Both, the DAS recorded strain and the signal from the strainmeter can be
expected to represent the same rock deformation, independent of its cause and
independent of wave parameters in particular.
Hence, this comparison allows for direct calibration of the DAS system’s
amplitude response, not limited to a specific wave type and without additional
uncertainties due to limited accuracy of ray-parameters.
Based on this calibration we estimate the detection threshold for strain
recordings.
The signal-to-noise ratio of both systems is appropriate for this purpose
in the frequency band between 0.05\,Hz and 0.1\,Hz.

Our study further extends beyond the work of \citet{lindsey2020} and
\citet{paitz2020} by interrogating two fibers in each cable, allowing us to distinguish coupling issues that effect all fibers in a cable, from
photonic effects (like optical fading), which are specific to each fiber.
We use a significantly larger dataset, comprising 19 earthquakes, covering
almost all backazimuths and include body waves in the analysis.
The simultaneous interrogation of fibers in two cables at two locations enables
us to investigate the direct differences observed under four different coupling
conditions.

\section{DAS installation\\ and instruments}
\begin{figure*}
\centering
  {\includegraphics[width=\textwidth,clip,trim=20 140 20 140]{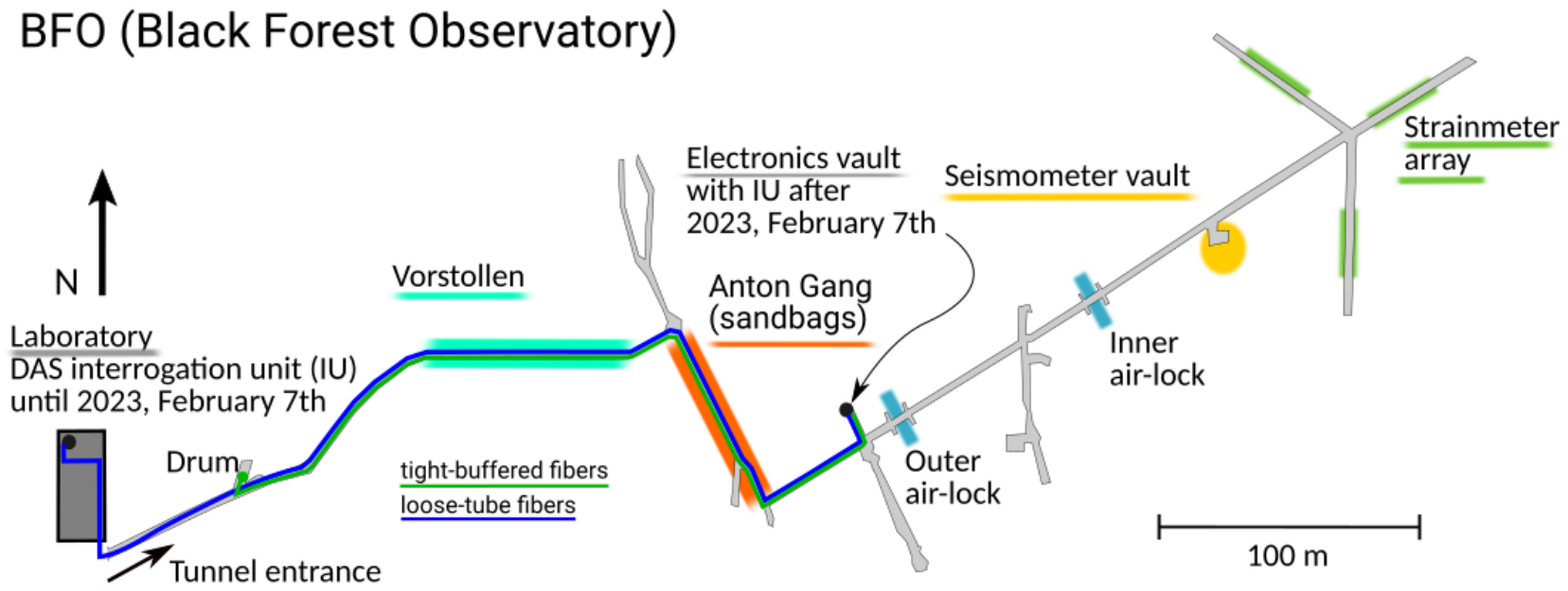}}  %
\caption{Floor map of the gallery of the former silver mine in which the
  instruments are installed.
  Two fiber optic cables are installed in the part in front of the air-locks.
  The cables are loaded down with tightly packed sandbags in the section called
  \Lanton{} (azimuth \azimuth{330}).
  We compare signals recorded in the \Lanton\ (69\,m long)
  with signals recorded in the
  straight section called \Lvorstollen\ (80\,m long,
  azimuth \azimuth{90}),
  strain derived from the \tunnellocation{Strainmeter array}, 
  and strain simulated from the {STS-2} broad-band seismometer in the
  \tunnellocation{Seismometer vault}.
  The overburden increases from the \tunnellocation{Tunnel entrance} (0\,m)
  to the \Lanton\ (100\,m) to the \tunnellocation{Strainmeter array} (170\,m).
  585\,m of the tight-buffered cable are rolled up on a 
  \tunnellocation{Drum} that is used to
  remove coherent laser noise.}
\label{fig:BFOfloorplan}
\end{figure*}
\begin{table*}
  \tbl{Recording parameters for the Febus A1-R.
  \label{tab:DASrecording}}
{ \begin{tabular}{ll}
    \colrule
    \multicolumn{2}{l}{\textbf{Interrogation unit (IU) Febus A1-R}} \\
    Software: & version 2.2.2\\
    optical wavelength: & 1550.12~nm\\
    Fiber length: & 2200~m\\
    Pulse width: & 5~m\\
    Block rate: & 1~Hz\\
    Pulse rate frequency: & 5~kHz\\
    Ampli power: & 28~dBm\\
    Sampling resolution: & 80~cm\\
    Gauge length (GL): & 50~m\\
    Derivation time (DT): & 20~ms \\
    Channel spacing: & 24.8~m\\
    \botrule
\end{tabular}
  }{}
\end{table*}
Black Forest Observatory (BFO) is situated in a former silver mine in the
central Black Forest, Germany \citep{emter1994}.
The gallery that hosts the instruments (Fig.~\ref{fig:BFOfloorplan})
is mostly horizontal and is excavated in granite.
The granite is covered with Triassic sedimentary rocks 
\citep{emter1994}.
The study focuses on two straight sections of the tunnel, \Lanton{} and \Lvorstollen{}.  
Overburden increases with distance to the tunnel entrance, is about 100\,m at
the \Lanton\ and reaches 170\,m at the \tunnellocation{Strainmeter array}.

\subsection{Distributed acoustic sensing (DAS)}
Two fiber optic cables are installed in the tunnel.
They are unreeled on the floor of the gallery and are in direct contact with the formation.
In the section called \Lanton\ the cables are loaded down by sand and tightly
spaced sandbags in order to improve the mechanical coupling to the rock,
provide thermal shielding and cover the cable against water dripping from the gallery ceiling.
The blue cable in Fig.~\ref{fig:BFOfloorplan} is a standard flexible
telecommunication cable with loose-tube fibers embedded in gel.
The green cable in Fig.~\ref{fig:BFOfloorplan} has a stiff jacket containing
tight-buffered fibers (see section~\SMRcablespec{} in the 
supplemental material for additional details). 
We use two single mode fibers in each of the cables 
and splice them in series, such that the \Lanton\ and the \Lvorstollen\  
are both sampled four times.
The signals from the fibers are sampled by a {Febus~A1-R} interrogation unit
(IU) installed in the laboratory (Fig.~\ref{fig:BFOfloorplan}).
The DAS recording parameters are listed in Table~\ref{tab:DASrecording}. 
The {Febus~A1-R} recording time is synchronized by GPS, like all other
digitizers involved in the current study.
When the IU was installed in the mine, the GPS timing signal was provided
through an optical link (Meinberg GOAL: GPS Optical Antenna Link).

The Febus A1-R IU operates on the principle of differential
phase-measuring distributed acoustic sensing. 
The light that is back-scattered in the optical fiber is mixed with a reference
signal to measure the differential phase over a gauge length, subsequently
leading to distributed strain-rate measurements, 
which is converted to strain by integration over time. 
The A1-R falls within the category of heterodyne Distributed Vibration Sensing
(hDVS) systems, using a single-pulse heterodyne approach for phase detection,
as described by \cite{pan2011hDVS}. 

\subsection{Reference instruments}
BFO operates an array of three well calibrated,
10\,m long, horizontal, Invar-wire strainmeters
(Fig.~\ref{fig:BFOfloorplan}).
The SEED codes for the instruments are 
II.BFO.00.BSA,
II.BFO.00.BSB, and
II.BFO.00.BSC.
Their design is based on the instruments by \cite{king1976} and are discussed
in more detail by \cite{zuern2015}.
\cite{agnew1986} and \cite{zuern2012} discuss instruments of this type and
their properties.
These instruments are primarily designed to record very-long period signals,
such as tidal strain or Earth's free oscillations.
For this reason they are equipped with an in-situ calibration device, which
makes use of interferometrically calibrated 'Crapaudines'
\citep{verbaandert1959etalonnage}. 
The accuracy of this calibration for the strainmeters is about 2~per cent.
Comparison against theoretical tidal strain ensures a stability of the
calibration of about 5~per cent in the long run.
Section~\SMRsubsecstrainmetercalibration{} in the supplemental material gives
further details.

At short signal period the Invar-wire strainmeters show a linear parasitic
sensitivity to vertical ground acceleration because of the inertia of the
pick-up system.
They provide useful strain signals at frequencies below 1\,Hz.
We use the instruments at frequencies below 0.1\,Hz
(which provides a safety-margin to the parasitic response)
as a reference to present \Trockstrain. 

We use signals recorded by the {STS-2}
broad-band seismometer (Fig.~\ref{fig:BFOfloorplan}) to demonstrate the
limitations of waveform comparisons with particle velocity derived strain.
The SEED codes for its three components are
GR.BFO..BHZ,
GR.BFO..BHN, and
GR.BFO..BHE.

\section{Data and data processing }
\phantomsection
\subsection{Available data}
The installation of DAS cables as shown in Fig.~\ref{fig:BFOfloorplan}
was used for recording from May 22, 2022 until March 13, 2023.
The IU was moved on February 7, 2023 from the laboratory
to the electronics vault in the mine (see Fig.~\ref{fig:BFOfloorplan} and 
Figs.~\SMRfigmapzoomA{} and \SMRfigmapzoomB{} in the
supplemental material for additional details).
In both, \Lanton{} and \Lvorstollen{}, we focus on four almost colocated 
read-out locations
that are separated by 5~m to 15~m. 
In all cases, the section of gauge length completely falls into the \Lanton\ or
\Lvorstollen{}, such that the strain recordings are representative 
of the azimuth of these gallery sections.

From the GEOFON catalog \citep{quinteros2020} we find 84 earthquakes with
moment magnitude larger than 6 in the recording time period.
21 of them show a maximum strain amplitude of larger than 1\,\Unstrain{} in a
visual inspection, which is considered large enough to provide a sufficient
signal-to-noise ratio in the DAS data.
For 19 of the events DAS data is available and allows an analysis. 
Their backazimuths (BAZ) cover all directions with an azimuthal 
gap of \degree{107} (BAZ between \azimuth{110} and \azimuth{217}).
Event characteristics are detailed in the supplementary material 
(see Table~\SMRtabUsedEvents{}).
The largest amplitudes are found in the surface wave train.
For body-waves, the horizontal strain amplitude is the smaller the steeper the
ray incidence.

The strongest signals are recorded from the
main shocks (Mw~7.7 and Mw~7.6) of the
Kahramanmaraş earthquake sequence
\citep{melgar2023} on February 6th 2023.
Their body-wave signals have large enough amplitude of a few \Unstrain{} to be included
in the analysis.
This is due to the larger earthquake magnitude and the smaller epicentral
distance (\degree{23}), 
compared to the other analyzed events in this study,  
which both result in rather large amplitude in
general.
In the set of analyzed events 
their surface waves provide by far the largest strain amplitudes
of about
130\,\Unstrain{} and about 250\,\Unstrain{}, respectively, 
as measured by the Invar-wire strainmeters in the investigated
frequency band of 0.05\,Hz to 0.1\,Hz.

For the comparison of strain waveforms, we can use 46~recordings for each
combination of cable (tight-buffered and loose-tube) and location (\Lanton{} and \Lvorstollen).
For the loose-tube cable in \Lanton{} we use only 44~recordings, 
as the configuration was modified on 2023, February 7th.

\subsection{Pre-processing of DAS data}
Waveforms sampled simultaneously on a section of the tight-buffered
cable rolled up on a
\tunnellocation{Drum} (see Fig.~\ref{fig:BFOfloorplan}
and Table~\SMRtablocations{} in the supplemental material)
get averaged and subtracted from
the recordings in order to reduce common mode laser noise \citep[][their
section 2.6 Optical Noise]{lindsey2020}.
For a gauge-length of 50\,m 
this coherent noise component clearly dominates the background level of the
DAS signal at frequencies below 0.5\,Hz and peaks at about 0.07\,Hz with
\Tfiberstrain-amplitudes of a few \Unstrain.
By subtracting this coherent signal component, the background level is lowered
by up to 20\,dB near 0.1\,Hz.
In the current application, it is essential that the drum is sufficiently
decoupled from the rock and that the fiber coiled on the drum does not pick up
the earthquake signal. 
Otherwise the correction procedure would affect the earthquake signal
amplitude.
In the supplemental material (section~ \SMRsubsecnoisereduction{} and Figures
\SMRfignoiseredP{} and \SMRfignoiseredall{}) we demonstrate that no signature
of the earthquake is apparent in the correction signal.

Data is converted from strain rate to strain for which the noise floor rapidly
increases at low frequencies, such that we focus our analysis on 
frequencies higher than 0.05\,Hz.

\subsection{Linear strain from the\\ strainmeter array}
In order to compare the DAS recordings with recordings of the strainmeters, we
derive linear strain in either the direction \Lanton{} or \Lvorstollen{} by a
linear combination of the signals recorded by the three strainmeters.
\citet[][their eq.~2]{zuern2015} specify linear strain 
\begin{equation}
  \epsi=\ett\cos^2(\AZI)+\epp\sin^2(\AZI)-\etp\sin(2\AZI)
\end{equation}
in azimuth $\AZI$ as a function of the components $\ett$, $\epp$, and $\etp$
of the 2D strain tensor. 
Here, \Lanton{} and  \Lvorstollen{} are in azimuth \azimuth{330} and \azimuth{90}, 
respectively.
Based on this, we compute linear strain in azimuth $\AZI$
\begin{equation}
  \epsi=
    \begin{pmatrix}
  \cos^2(\AZI) \\ \sin^2(\AZI) \\ -\sin(2\AZI)
  \end{pmatrix}
  \,
  \Mrot^{-1}\,
  \begin{pmatrix}
    \eA \\ \eB \\ \eC
  \end{pmatrix},
  \label{eq:linear:strain}
\end{equation}
from the strain recorded by the three array-instruments, where
$\eA$, $\eB$, and $\eC$ are in azimuth
N$2^\circ$E, N$60^\circ$E, and N$300^\circ$E, respectively. 
They correspond to the SEED channel names BSA, BSB, and BSC.
The rotation matrix
\begin{equation}
  \Mrot^{-1}
  = 
    \begin{pmatrix}
     1.002  & -0.041  &  0.040      \\
    -0.334  &  0.680  &  0.653      \\
     0.000  & -0.577  &  0.577
  \end{pmatrix}
\end{equation}
is the inverse of
\begin{equation}
  \Mrot=
  \begin{pmatrix}
    \cos^2(2^\circ) & \sin^2(2^\circ) & -\sin(2\times 2^\circ) \\
    \cos^2(60^\circ) & \sin^2(60^\circ) & -\sin(2\times 60^\circ) \\
    \cos^2(300^\circ) & \sin^2(300^\circ) & -\sin(2\times 300^\circ) \\
  \end{pmatrix}.
\end{equation}

\section{Comparison of\\ strain measurements}
\subsection{Strain transfer rate}
By fitting the strainmeter data to the DAS data with a linear regression,
we compute the \Tstraintransfer{} 
\begin{equation}
  \Sstr=\frac{\sum\limits_k \, x_k\,y_k}{\sum\limits_k y^2_k},
  \label{eq:strain:transfer}
\end{equation}
where $x_k$ is the DAS time series of \Tfiberstrain\
and $y_k$ is the strainmeter time series of \Trockstrain.
Data are filtered by a Butterworth high-pass (0.05~Hz, 4th order) and low-pass
(0.1~Hz, 4th order).
The average is thus removed from the signals prior to the computation.
Fig.~\ref{fig:str:ncc} (left) shows the \Tstraintransfer\ $\Sstr$ and
Table~\ref{tab:strain:transfer:rate} summarizes the ranges of values.
They are between 0.13 and 0.53 in all cases and
primarily depend on location and cable type. 
The variability in the values calculated for a single 
installation (i.e., one cable at one location) is
depicted using a violin plot, where the edges illustrate the density of dots
in Fig.~\ref{fig:str:ncc}. 

The variability within individual violin plots is smaller than 
the differences between the median values for different cables and locations. 
This observation remains consistent across all considered installations, 
regardless of cable type or location. 
It underlines the significance of the differences observed between the installations.
Differences between colocated fibers for the same cable, location, and event
are smaller than 0.05 in almost all cases, 
which lets us rule out that fiber related causes
(like optical fading) would dominate the observed differences.
We find no discernible correlation of \Tstraintransfer\ with signal amplitude,
back-azimuth or other earthquake specific parameters (see 
Figs.~\SMRfigstrvsbaz{} and \SMRfigSTRfNCC{}
in the supplemental material for example).

The largest values of \Tstraintransfer\ are found for the tight-buffered
cable, which was expected.
Although the loose-tube cable in the \Lanton\ is loaded down by sandbags as
well, its \Tstraintransfer\ is the lowest among all fibers. 
Hence, we do not observe a significant benefit from sand and sandbags to improve overall
coupling.

\subsection{Waveform similarity}
\begin{figure*}
  \begin{tabular}{cc}
  {\includegraphics[trim=0 0 0 30,clip,width=0.48\textwidth]{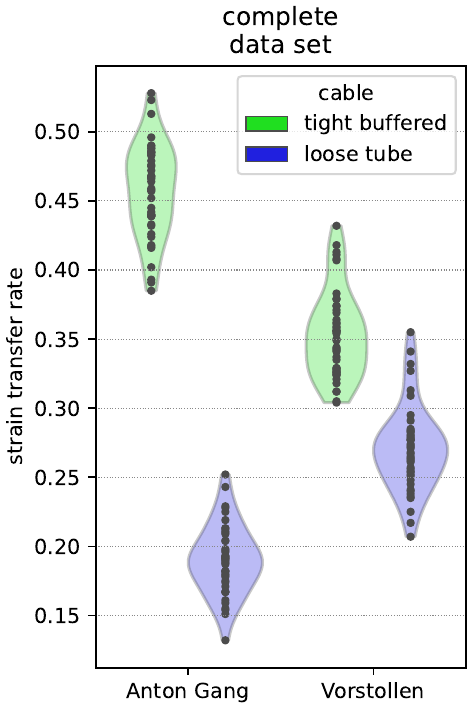}} &
  {\includegraphics[trim=0 0 0 30,clip,width=0.48\textwidth]{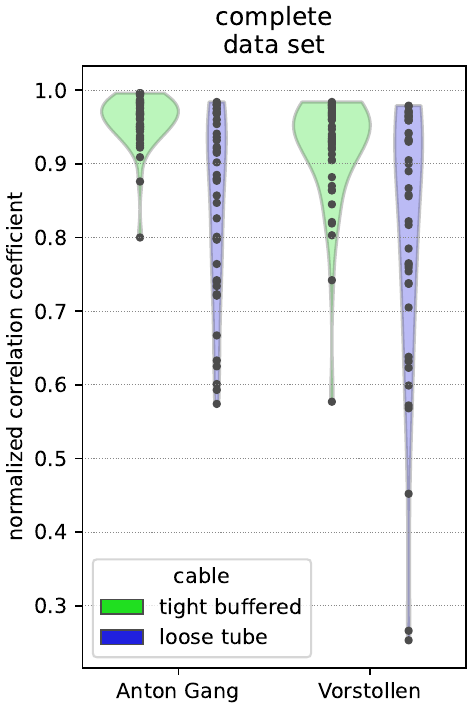}}
  \end{tabular}
  \caption{Values of \Tstraintransfer\ (left diagram, representing the ratio
  of \Tfiberstrain\ to \Trockstrain\ as recorded by the strainmeter) derived
  by \eqref{eq:strain:transfer} and the normalized correlation coefficients
  (right diagram) calculated using  \eqref{eq:normalized:correlation}.
  The results are shown separately for two locations: 
  \Lanton\ (left, \azimuth{330}) and \Lvorstollen\ (right, \azimuth{90}). 
  They are presented for two different cables: a tight-buffered (green, left)
  and a loose-tube cable (blue, right) at each location. 
  These values are derived in two fibers per cable, for surface wave signals
  of 19 teleseismic earthquakes and the P- and S-wave signals of the two
  Turkey-events on 2023-02-06.
  All signals are consistently filtered with 4th order Butterworth high-pass
  (0.05~Hz) and low-pass (0.1~Hz) filters. 
  For each set of cable and location, the black dots represent individual
  results from the 46 analyses and a summary of the values is provided in
  Table~\ref{tab:strain:transfer:rate}. 
  The kernel density estimates that define the edges of the violin plot
  illustrate the spread in the distribution of numerical values. 
  }
  \label{fig:str:ncc}
\end{figure*}

\begin{figure*}
  \begin{center}
    {\includegraphics[clip,trim=10 10 10 50,width=\textwidth]{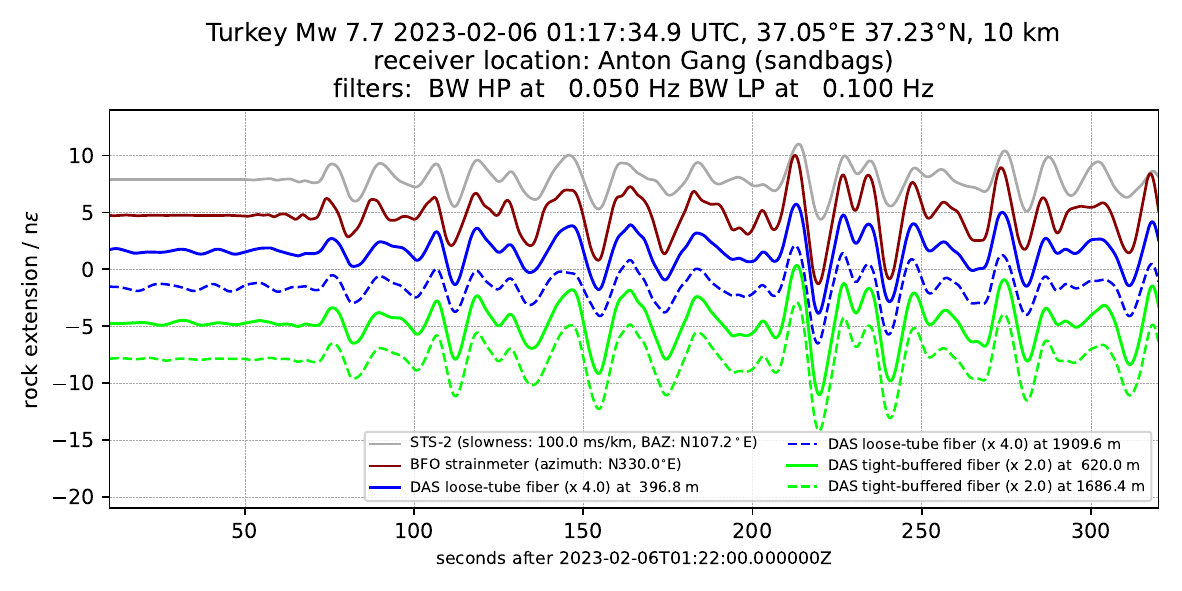}}
  \end{center}
  \caption{Waveforms of the P-waves radiated by the Mw~7.7 Pazarcık earthquake
  on February 6th, 2023.
  Traces are shifted vertically for better visibility.
  The legend (top to bottom, left to right) specifies the traces in order from
  top to bottom.
  All signals are consistently filtered with Butterworth high-pass
  (0.05~Hz, 4th order) and low-pass (0.1~Hz, 4th order)
  filters.
  The seismometer response is removed from the STS-2 data.
  The DAS signals of \Tfiberstrain\ from the tight-buffered cable (green,
  signal amplified by a factor of 2) and the loose-tube
  cable (blue, signal amplified by 4) 
  are taken with a gauge length of 50~m at a location
  in the center of the \Lanton\ (Fig.~\ref{fig:BFOfloorplan}).
  Linear strain in azimuth \azimuth{330} of \Lanton\ 
  is obtained from the BFO strainmeter array by \eqref{eq:linear:strain}.
  For comparison we show an estimate of linear strain simulated from ground
  velocity for a plane wave of slowness 100~ms~km$^{-1}$ and incoming from
  backazimuth \azimuth{107.2}.
  The basis for this simulation is the recording of the STS-2 seismometer
  converted by \eqref{eq:plane:wave:strain}.
  The slowness of 100~ms~km$^{-1}$, which acts as a scaling factor in
  \eqref{eq:plane:wave:strain} is not appropriate throughout the
  entire time window.
  This makes calibration with respect to strain simulated from ground velocity
  disputable.
  While the first P-wave onset is apparent near 01:22:50~UT (50~s on timescale)
  in the strainmeter 
  and STS-2 data, the noise level in the DAS data is too high to detect the
  small amplitude signals before 01:23:10~UT (70~s).
  }
  \label{fig:Turkey77:Anton:P}
\end{figure*}

The linear regression used to derive the \Tstraintransfer{} 
requires that the \Tfiberstrain{} waveform as recorded by DAS is
consistent with the \Trockstrain{} waveform 
obtained from the strainmeters.
We use the normalized correlation coefficient
\begin{equation}
  c=\frac{\sum\limits_k \, x_k\,y_k}{\sqrt{\sum\limits_k
  x^2_k}\,\sqrt{\sum\limits_k y^2_k}}
  \label{eq:normalized:correlation}
\end{equation}
as a measure of waveform similarity.
Fig.~\ref{fig:Turkey77:Anton:P} compares the P-waves 
radiated by the Mw~7.7 Pazarcık earthquake 
across all measurement types and illustrates differences in SNR.
Fig.~\ref{fig:str:ncc} (right) shows the distribution of the 
normalized correlation coefficients obtained for all tested configurations.
They primarily depend on signal amplitude and thus on signal-to-noise
ratio for the DAS data (see Fig.~\SMRfigNCCvsmaxAmp{}
in the supplemental material).
The values are largest and larger than 0.92 for all signals of the Mw~7.7 and
Mw~7.6 earthquakes of the Kahramanmaraş earthquake sequence 
(see Fig.~\SMRfigturkeyNCC{} in the supplemental material).
For the surface waves recorded by the tight-buffered cable in the \Lanton\
they are even larger than 0.99.
The distribution shown in Fig.~\ref{fig:str:ncc} (right) indicates a generally
better signal quality for data recorded with the tight-buffered cable.
The loose-tube cable, if not protected by sand-bags (\Lvorstollen), appears to
be prone to glitches caused by water drops (see Fig.~\SMRfiggfzAvpri{}
in the supplemental material for example).

The signal-to-noise ratio of the strainmeter generally is better than that of
the DAS cables. 
This can be seen in Fig.~\ref{fig:Turkey77:Anton:P}, where the first P-wave
arrival is clearly seen in the seismometer and strainmeter data at 50\,s on
the time scale. 
Only after 70\,s the signal amplitude is large enough to allow the DAS
recording to capture the P-waves.
The high signal-to-noise ratio of the \Trockstrain\ allows the strainmeter
signal to lock onto the \Trockstrain-waveform in the DAS data in the linear
regression.
The high signal-to-noise ratio time series is in the denominator of
\eqref{eq:normalized:correlation} and thus there is no correlation between
\Tstraintransfer\ and normalized correlation coefficient found in the results.

\subsection{Intercomparison of fibers\\ at higher frequency}
\begin{figure*}
  \begin{center}
    {\includegraphics[clip,trim=0 0 0 45,width=\textwidth]{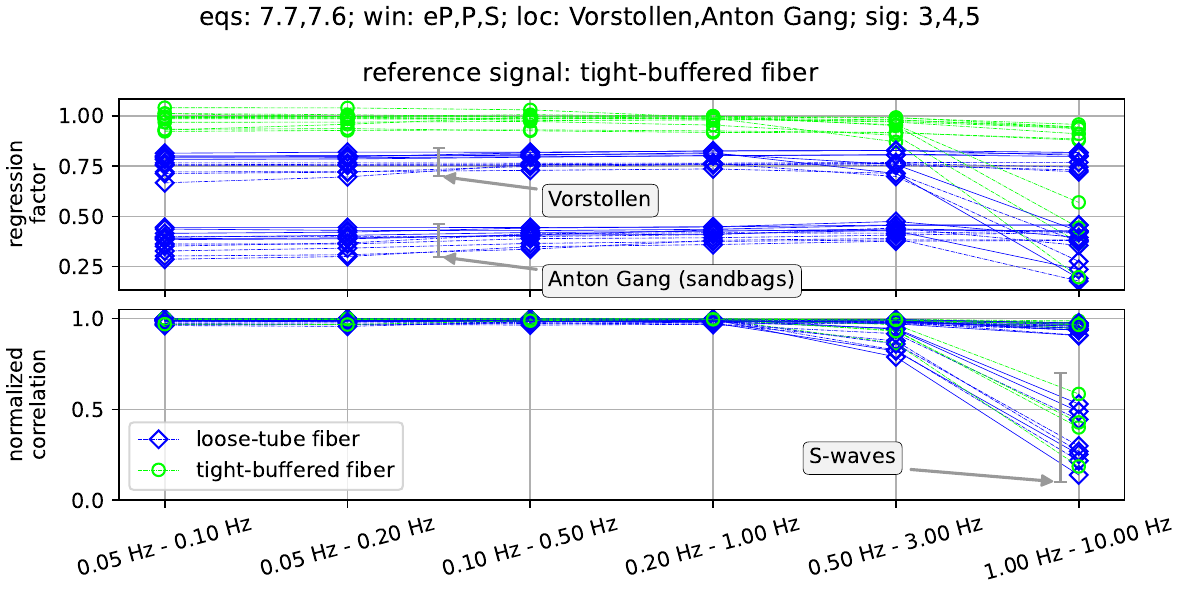}}
  \end{center}
  \caption{Regression factor derived by \eqref{eq:strain:transfer} and
  normalized correlation coefficient by \eqref{eq:normalized:correlation}
  with respect to one colocated
  fiber in the tight-buffered cable (location 620.0~m in the \Lanton\
  and 719.2~m in the \Lvorstollen, respectively,
  see Fig.~\ref{fig:BFOfloorplan}).
  We use body-wave signals of both main shocks (Mw 7.7 and Mw 7.6) of the
  Kahramanmaraş earthquake sequence on February 6th 2023 for both locations,
  fibers and cables.
  The S-waves contain less short-period energy.
  The signal-to-noise ratio for them is reduced above 1~Hz, which results in a
  reduced waveform similarity (normalized correlation coefficient).
  This also affects the regression factors derived for S-waves.
  A summary of the values up to 1~Hz is given in
  Table~\ref{tab:regression:factors}.
  }
  \label{fig:factors:wrt:green}
\end{figure*}

The comparison with strainmeter data is limited to frequencies below 0.1 Hz. 
Due to their short-period signal energy we can use the body-waves of the
Kahramanmaraş earthquakes to extend the study to a larger frequency range 
by comparing the DAS signals of the main shocks (Mw 7.7 and
Mw 7.6) with one another.
Hence, the regression is computed with respect to recordings 
of read-out locations on the tight-buffered fibers. 
Fig.~\ref{fig:factors:wrt:green} shows the analysis results for a comparison
between different fibers (all with respect to one of the tight-buffered
fibers).
The normalized correlation coefficients (waveform similarity) generally are
higher than 0.970 and reach 1.000 for some of the combinations. 
At frequencies above 1~Hz the signal-to-noise ratio for S-waves quickly drops
because of the decrease of S-wave energy with increasing signal frequency.
This results in reduced signal-to-noise ratio and thus 
waveform similarity and regression coefficients are reduced as well.
A frequency dependence of regression factors beyond the scatter is not
apparent up to 1\,Hz.

In the frequency range up to 1~Hz, colocated fibers in the same cable
practically pick up the same strain amplitude. 
Regression coefficients listed in Table~\ref{tab:regression:factors}
scatter about unity, where the scatter is smaller for the tight-buffered
cable (smallest in the \Lvorstollen{} with a range of 0.96 to 1.02) 
and largest for the loose-tube cable in
the \Lanton{} (0.80 to 1.23).
The ratios of \Tstraintransfer\ given in Table~\ref{tab:strain:transfer:rate}
are consistent with the regression factors for colocated cables in the
extended frequency range given in Table~\ref{tab:regression:factors}.

\begin{table*}
  \tbl{Ranges of \Tstraintransfer. 
  \label{tab:strain:transfer:rate}}
{ \begin{tabular}{llrrr}
  \colrule
    cable & location & minimum &median &maximum \\
    \cmidrule(r){1-1}
    \cmidrule(lr){2-2}
    \cmidrule(lr){3-3}
    \cmidrule(lr){4-4}
    \cmidrule(l){5-5}
    tight-buffered & \Lanton\   &  0.39 & 0.46 & 0.53  \\
    tight-buffered & \Lvorstollen\   &  0.30 & 0.35 & 0.43 \\
    loose-tube & \Lvorstollen\    &  0.21 & 0.27 & 0.36  \\
    loose-tube & \Lanton\    &  0.13 & 0.19 & 0.25  \\
    \bottomrule
\end{tabular}\par
\bigskip
  Values are as presented in Fig.~\ref{fig:str:ncc}
  (left).
  Strainmeter signals are fitted to DAS (distributed acoustic sensing)
  recorded waveforms by linear regression in the
  frequency band from 0.05~Hz to 0.1~Hz.
  }{}
\end{table*}
\begin{table*}
  \tbl{Ranges of regression coefficients for an intercomparison of fibers.\label{tab:regression:factors}}
{ \begin{tabular}{ll}
    \colrule
    \multicolumn{2}{l}{{DAS with respect to colocated DAS fibers (in
    the same cable)}}\\
    \colrule
    tight-buffered  \Lvorstollen\ & 0.96 -- 1.02\\
    tight-buffered  \Lanton\ & 0.92 -- 1.08\\
    loose-tube  \Lvorstollen\ &  0.89 -- 1.11\\
    loose-tube  \Lanton\ &  0.80 -- 1.23\trowsep
    \multicolumn{2}{l}{{DAS with respect to colocated DAS fibers (in
    the other cable)}}\\
    \colrule
    loose-tube vs. tight-buffered  \Lvorstollen\ & 0.72 -- 0.83\\
    loose-tube vs. tight-buffered  \Lanton\ & 0.32 -- 0.47\\
    \botrule
\end{tabular}\par\bigskip
  The presented values cover the frequency band from 0.05\,Hz to 1.0\,Hz.}{}
\end{table*}
\begin{figure*}
  \begin{center}
    \begin{tabular}{r}
    {\includegraphics[clip,trim=0 27 0 60,width=\textwidth]{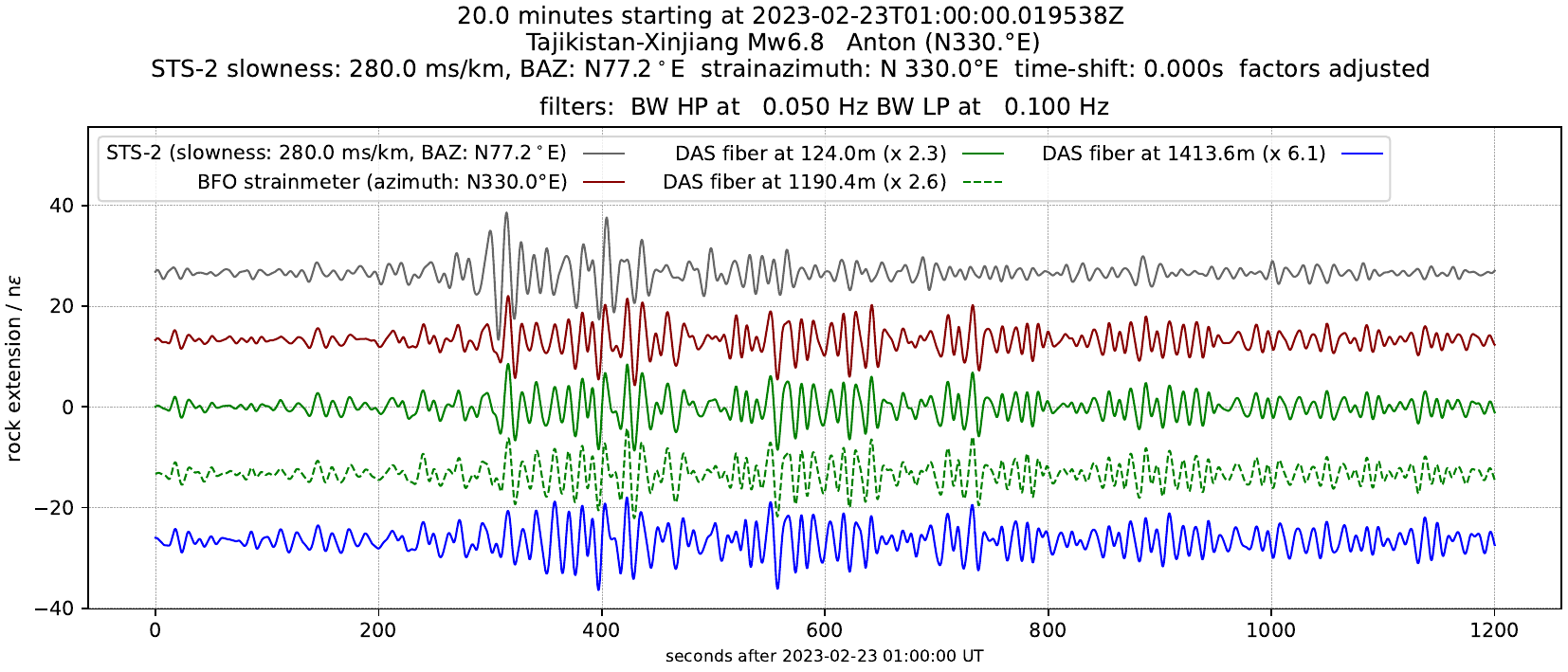}}\\
    {\includegraphics[clip,trim=0 0 0 75,width=\textwidth]{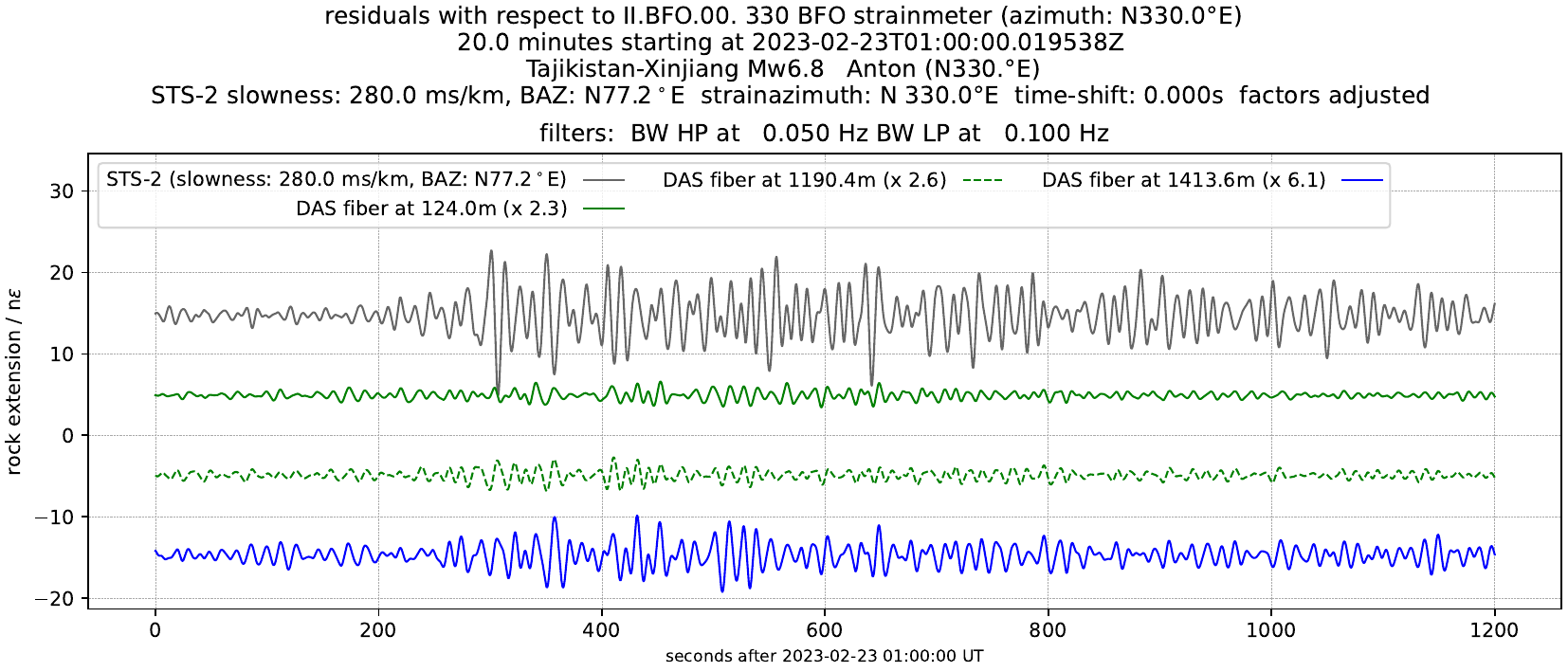}}
    \end{tabular}\par
  \end{center}
  \caption{Surface waves (top) and residuals with respect to the strainmeter
  data (bottom) for the
  \eventlink{Mw~6.79 Tajikistan-Xinjiang Border Region earthquake}{gfz2023dswx}
  (origin: 2023-02-23 00:37:39.01  UTC,
  \degree{ 38.06}N, \degree{ 73.29}E, 10.0~km depth,
  BAZ=\azimuth{ 77.24}, $\Delta$=\degree{ 47.06}).
  Traces are shifted vertically for better visibility.
  The legends (top to bottom, left to right) specify the traces in the order
  from top to bottom.
  Love-waves (300\,s) and Rayleigh-waves (400\,s) are inseparably superimposed.
  All direct body wave phases arrived prior to 1:00~UT (outside the displayed
  window).
  Displayed are strain signals for the azimuth \azimuth{330} (\Lanton).
  Top panel traces from top to bottom:
  gray: particle velocity scaled to apparent strain by
  \eqref{eq:plane:wave:strain} with $\AZI_{\text{BAZ}}=\azimuth{77.2}$ and
  slowness $s_h=280\,\text{ms}\,\text{km}^{-1}$.
  The backazimuth is only by $17^\circ$ off the direction perpendicular to the
  \Lanton\ azimuth of $\AZI_{x}=\azimuth{330}$.
  red: BFO strainmeters.
  green solid, green dashed: tight-buffered DAS cable (read-out locations
  124.0\,m and 1190.4\,m).
  blue: loose-tube DAS cable (1413.6\,m).
  The DAS signals are scaled to minimize the rms-residual with respect to the
  strainmeter (factors are given in the legend).
  Bottom panel: 
  Waveform difference of the traces in the top panel with respect to the
  strainmeter signal.
  }
  \label{fig:Tajikistan}
\end{figure*}
\subsection{Comparison with seismometer data}
Strainmeter installations recording at seismic frequencies
are rare and available only in a few observatories.
In the absence of strainmeters many studies use strain signals estimated from
particle velocity recorded by seismometers, which appears possible for plane
waves of known incidence and phase-velocity.
The scaling relation is given by 
\begin{equation}
  \epsilon_{xx}(\vect{r},t)=
  -s_h\,\cos\bigl(\AZI_{\text{BAZ}}-\AZI_{x}-180^{\circ}\bigr)
  \,v_{x}(\vect{r},t),
  \label{eq:plane:wave:strain}
\end{equation}
where $v_{x}$ and $\epsilon_{xx}$ are horizontal particle velocity and linear
strain in azimuth of $\AZI_{x}$, respectively, 
$\AZI_{\text{BAZ}}-180^{\circ}$ is the azimuth of wave
propagation, and $s_h$ is the horizontal component of slowness
(see section~\SMRsubsecscalingparticlevelocity{} in the
supplemental material for additional details).
The accuracy of the scaled seismometer signal in representing strain depends on the validity of the plane-wave assumption underlying the conversion.
While strainmeters and DAS can be expected to record the same quantity, namely
linear strain in a given azimuth, independent of the nature of the signal,
this is not the case for the seismometer derived strain signals.
This is not only the case because the recorded waveform is a superposition of
non-plane waves with different phase slowness, but also because of the local
free surface affecting the particle velocity and the strain in different ways
(see section~\SMRsubseccavityeffects{} in the supplemental material for
additional details).

The signal derived from the STS-2 recording in Fig.~\ref{fig:Turkey77:Anton:P}
matches the strain signals quite well near the first P-wave arrival (75\,s).
In later parts (220\,s) the P-wave signal of particle velocity obviously
provides only half the amplitude of the signals from the strainmeter and the
scaled DAS signals.

The surface waves with good signal-to-noise ratio used in the current study 
not only propagate at varying slowness (dispersion and superposition of Love-
and Rayleigh-waves), they also contain significant non-plane wavefield
components due to scattering in particular on continental paths.
\citet{wielandt1993} discusses the consequences of the non-plane nature of 
waves for the spatial derivatives of displacement.
The scaling in \eqref{eq:plane:wave:strain} lets $\epsilon_{xx}$ vanish for
$\AZI_{\text{BAZ}}$ being perpendicular to $\AZI_{x}$.
In cases where $\AZI_{\text{BAZ}}$ comes close to being perpendicular to
$\AZI_{x}$ the actual strain component is dominated by
off-great-circle propagation and non-plane wave contributions, 
which results in a large scatter of regression coefficients 
(see Fig.~\SMRfigRCDASvsseismometerandstrainmeter{}
in the supplemental material). 
In these cases,  
seismometers cannot reasonably be used to calibrate DAS signals.

Here, we illustrate the limitations of the comparison with seismometer data
using
the surface waves recorded on February 23rd 2023 after the
Mw~6.8 earthquake in the the Tajikistan-Xinjiang Border Region.
The great-circle backazimuth for this earthquake is
$\AZI_{\text{BAZ}}=\azimuth{77.2}$, which is
only by $17^\circ$ off the direction perpendicular to the \Lanton\ azimuth of
$\AZI_{x}=\azimuth{330}$.
Fig.~\ref{fig:Tajikistan} (top panel) shows the signal comparison.
Love-waves (300\,s) and Rayleigh-waves (400\,s) are inseparably superimposed.
Both are dispersive and arrive at different phase velocity with non-plane
wavefronts due to scattering on the continental path.
Slowness $s_h=280\,\text{ms}\,\text{km}^{-1}$ is used for scaling the
seismometer signal, which corresponds to the average phase velocity of about
3.6\,km\,s$^{-1}$, the value of Rayleigh-wave phase velocity at 0.05\,Hz in
Southern Germany \citep[their Table~1]{friederich1996}.
The bottom panel of Fig.~\ref{fig:Tajikistan} shows the waveform residuals
with respect to the strainmeter signal.
The residual amplitude for the tight-buffered cable is about three to five
times the noise level.
For the seismometer the largest residual amplitude is five times larger
and the waveform mismatch renders the seismometer signal unusable for a
calibration.
The adjustment of a scaling factor would not solve the problem.
The seismometer derived amplitude for the first wave group (300\,s -- 500\,s)
is larger than that of strainmeter and DAS, while it is too small after
500\,s.
Additionally, Fig.~\ref{fig:Tajikistan} shows a significant mismatch 
throughout the wavetrain, while
DAS and strainmeter waveforms are rather consistent.
In practically all cases examined the normalized correlation coefficient
between seismometer signal and DAS is smaller than between strainmeter
signal and DAS (see Fig.~\SMRfigNCCDASseismometerandstrainmeter{}
in the supplemental material).
In most cases it is significantly smaller and in some 
cases it is even negative
(anti-correlation).

\section{Detection threshold\\ for \Trockstrain}
\begin{figure*}
  \begin{center}
    \begin{tabular}{r}
    {\includegraphics[clip,trim=3 48 0 34,width=0.758\textwidth]{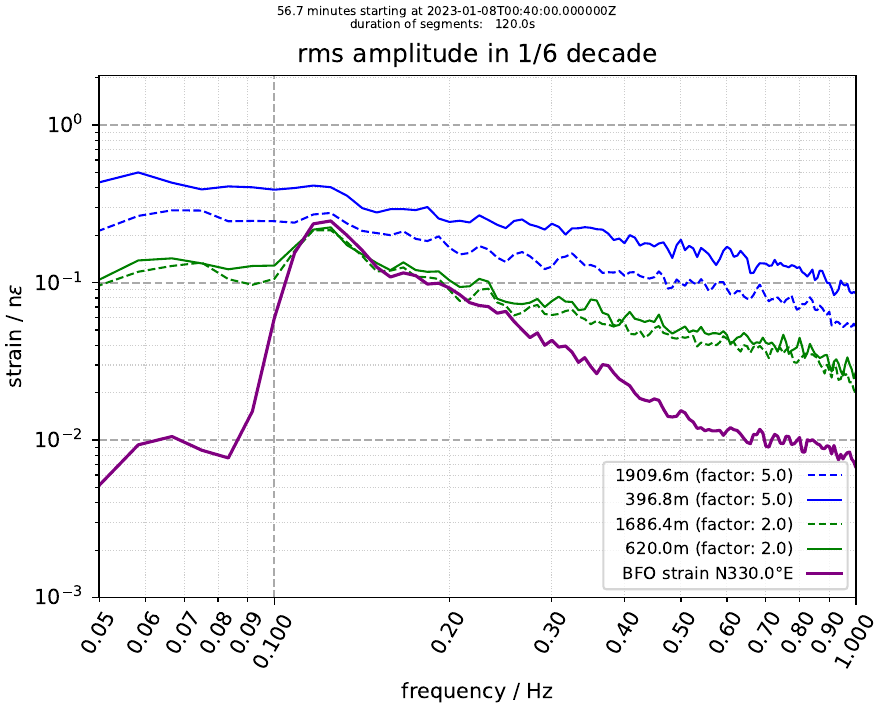}}
    \\
    {\rule{5pt}{0pt}\includegraphics[clip,trim=0 0 0 34,width=0.748\textwidth]{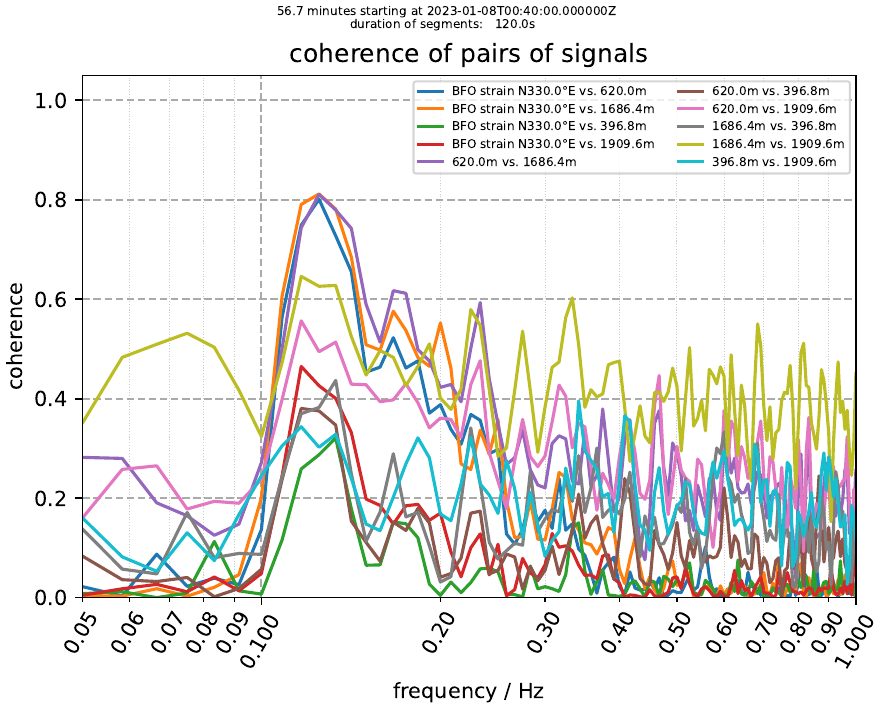}}
    \end{tabular}
  \end{center}
  \caption{Analyses of the strain background signal in the \Lanton\
  for 50~minutes after
  2023-01-08 0:40 UT at times of large marine microseism amplitude.
  Top: rms-amplitude in 1/6 decade.
  The peaks of the primary (0.07\,Hz) and secondary (0.14\,Hz) microseisms are
  observed in the BFO strainmeter data.
  DAS recorded signals are multiplied by the average reciprocal
  \Tstraintransfer\ as given in Table~\ref{tab:regression:factors}.
  The values from the loose-tube cable (1909.6\,m and 396.8\,m) are largest.
  The values for the BFO strainmeter are the smallest below 0.1\,Hz and above
  0.25\,Hz.
  At 0.14\,Hz the curves for the tight-buffered cable (1686.4\,m and 620.0\,m)
  coalesce with the curve for the BFO strainmeter.
  Bottom: magnitude-squared coherence of pairs of signals.
  The curves for the BFO strainmeter versus the tight-buffered cable (620.0\,m
  and 1686.4\,m) and the intercomparison of the tight-buffered fibers
  (620.0\,m versus 1686.4\,m) have a peak value of 0.8 at 0.14\,Hz.
  Values for all other pairs are smaller at this frequency.
  }
  \label{fig:background:spectra}
\end{figure*}
\begin{figure*}
  \begin{center}
    {\includegraphics[clip,trim=0 0 0 34,width=\textwidth]{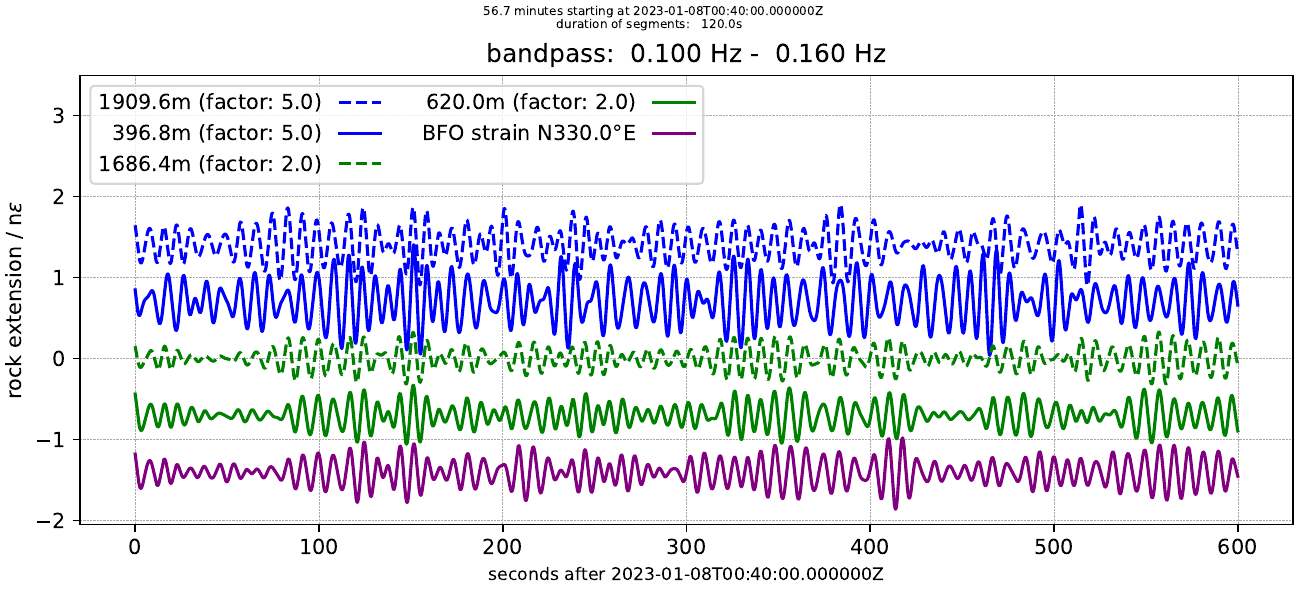}}
  \end{center}
  \caption{Waveforms of the recorded background signal at large amplitudes of
  marine microseisms.
  Traces are shifted vertically for better visibility.
  The legend (top to bottom, left to right) specifies the traces in order from
  top to bottom.
  Read-out locations 1909.6\,m and 396.8\,m are from the loose-tube cable 
  and locations 1686.4\,m and 620.0\,m are from the
  tight-buffered cable, both in the \Lanton.
  A narrow bandpass from 0.1\,Hz to 0.16\,Hz is applied in order to focus 
  on the signal of the secondary microseisms.
  DAS recorded signals are multiplied by the average reciprocal
  \Tstraintransfer{} as given in Table~\ref{tab:strain:transfer:rate}.
  }
  \label{fig:background:waveform}
\end{figure*}
  The signal of largest amplitude in the natural seismic background are the marine microseisms. 
  This can be a helpful test signal for the investigation of instrumental properties 
  at frequencies of 0.1\,Hz to 1\,Hz. 
  In applications for Newtonian Noise mitigation the strain background 
  at even smaller amplitude must be resolved.

  We analyze the signals at times of large marine microseism amplitude.
  The rms-amplitude levels are shown in Fig.~\ref{fig:background:spectra}
  (top) together with the pair-wise magnitude-squared coherence \citep[their
  eq.~2]{carter1973} of signals (bottom).
  The signal of the BFO strainmeters shows the signature of the primary
  microseism peak at about 0.07\,Hz as well as the secondary peak at 0.14\,Hz.
  The tight-buffered cable appears to detect the secondary microseism peak,
  while the scaled noise in the loose-tube fibers is above the background
  \Trockstrain\ level.
  The coherence between the BFO strainmeter signal and the tight-buffered fibers 
  (at 620.0 m and 1686.4 m), as well as among the tight-buffered fibers themselves, 
  peaks at 0.8 at the frequency of the secondary microseisms, i.e. 0.14\,Hz. 
  The coherence for all other signal combinations reaches a maximum of 0.6 or less.
  The lower envelope of the coherence curves peaks at 0.3 at 0.14 Hz, 
  which may indicate that all signals contain at least a portion of the marine microseisms.
  Nevertheless the similarity of the waveforms shown in
  Fig.~\ref{fig:background:waveform} is not as good as for the earthquake
  signals, even though a narrow bandpass is applied.
  This is consistent with the coherence not exceeding a value of 0.8.

  The amplitude levels of the marine microseisms at BFO strongly vary with
  weather conditions over the North Atlantic and are typically lower during
  northern hemisphere summer\footnote{The seasonal variation of marine
  microseism energy is seen in the analyses presented by the
  Waveform Quality Center at Lamont-Doherty Earth Observatory:
  \url{https://www.ldeo.columbia.edu/~ekstrom/Projects/WQC/MONTHLY_HTML/BFO-II.LHZ-00_time.html}.}.
  A similar analysis for signals recorded on 2022-07-10 shows the background
  level of the BFO strainmeter recordings at below 0.02\,\Unstrain\ with no
  indication of a microseism peak.
  At the same time, the amplitude levels of the DAS recordings range between
  0.05\,\Unstrain\ (0.3\,Hz) and 0.4\,\Unstrain\ (0.05\,Hz), with the lowest
  levels observed for the tight-buffered fibers and rms values decreasing with
  increasing frequency for all fibers.
  Coherence between any pair of fibers, or between DAS recording and BFO
  strainmeter, is below 0.25 at these times.
  Thus, at times of low marine microseism amplitude, the DAS fibers are not
  able to pick up the background \Trockstrain\ in the discussed frequency
  band.

\section{Conclusions}
  The waveform similarity between DAS- and strainmeter-recorded signals, as
  well as in the intercomparison of DAS signals, is high for signals with
  large amplitudes.
  For the body- and surface waves recorded after the main shocks (Mw 7.7 and
  Mw 7.6) of the Kahramanmaraş earthquake sequence on February 6th 2023
  the normalized correlation coefficient (NCC) with respect to the strainmeter
  is larger than 0.92 for all DAS fibers in the frequency band from 0.05~Hz to
  0.1~Hz.
  In an intercomparison of colocated DAS fibers for the body-wave signals up
  to 1~Hz, the NCC is larger than 0.97 and reaches 1.000 in some
  cases.
  For smaller signal amplitude the DAS signal-to-noise ratio worsens and the
  NCC becomes smaller.
  The signal quality in this respect is better for the tight-buffered cable
  and worst for the loose-tube cable in the \Lvorstollen, where it is not
  protected by a sand bed against water dripping from the ceiling.
  
  The DAS recorded \Tfiberstrain\ at large amplitude hence well represents the
  waveform and the signal phase of \Trockstrain.
  This is, however, not the case for the signal amplitude.
  The values of \Tstraintransfer\ (amplitude ratio of \Tfiberstrain\
  versus \Trockstrain), which we calibrate by linear regression,
  cover a range from 0.13 to 0.53
  (Table~\ref{tab:strain:transfer:rate}).
  None of the installed cables provides a \Tstraintransfer\ equal~1, which
  would make \Tfiberstrain\ equal \Trockstrain.
  The largest value is obtained for the tight-buffered cable in the \Lanton,
  where cables are loaded down by loose sand and sand bags.
  While the tight-buffered cable generally provides a larger
  \Tstraintransfer, the smallest values are obtained 
  for the loose-tube cable in the \Lanton.
  We hence find no consistent effect of the sand bags on 
  coupling the fiber to the rock.

  The scatter in \Tstraintransfer{} values 
  for a single cable at one location is smaller than 
  the differences observed between different cables and locations.
  By the high similarity of signals from two fibers in the same cable, we can
  rule out fiber related causes for the difference, like optical fading.
  We find no systematic dependency of \Tstraintransfer\ 
  on backazimuth of the event, nor do we find any other systematic
  correlation except with respect to cable type and location.
  The \Tstraintransfer\ of the cables cannot be specified
  independently of local installation conditions.

  An amplitude correction factor to be applied to DAS recorded strain not only
  adjusts the signal amplitude, but also amplifies the instrument generated
  background noise.
  Consequently the peak of the secondary marine microseisms is only detected
  by the installation with largest \Tstraintransfer\ in
  Table~\ref{tab:regression:factors}.
  We estimated the detection threshold for the tight-buffered cable in the
  \Lanton\ at 0.1~\Unstrain{} rms-amplitude in 1/6 decade near 0.1~Hz.

  Even if we attribute the \Tstraintransfer\ to the cable type and
  installation here, we cannot rule out the possibility that some other effect
  reduces the recorded \Tfiberstrain.
  In a next step, we plan to install a testbed of fibers (just core, cladding,
  coating, and tight-buffer) directly cemented into the floor of the mine.
  We expect that this would make the DAS recordings more sensitive and would
  make \Tfiberstrain\ equal \Trockstrain.

\begin{datres}
  All data were recorded locally at
  \cite[][\doiurl{10.5880/BFO}]{bfo1971}.
  Strainmeter and seismometer data are available through
  data centers at the \cite[][\doiurl{10.7914/SN/II}]{sio1986} 
  and the \cite[][\doiurl{10.25928/MBX6-HR74}]{bgr1976}, respectively.
  DAS data can be provided upon request.
  Event parameters were taken from the GEOFON catalog 
  \citep[][\url{https://geofon.gfz-potsdam.de/eqinfo/}]{quinteros2020}.
  Data analysis was carried out with ObsPy 
  \citep[][\doiurl{10.5281/zenodo.6327346}]{obspy2022}.
  We provide a supplemental text file with figures, tables, and math to
  further support the findings presented in this paper.
\end{datres}

\begin{ack}
We thank Peter Duffner for taking care of the BFO strainmeters and their
  calibration devices.
We are grateful to Walter Zürn for many fruitful discussions and his
  continuing support of the operation of the Invar-wire strainmeters.
Gaetan Calbris and Vincent Lanticq from Febus Optics 
provided helpful information on the measurement principle used in the
Febus~A1-R.
Peter Duffner, Felix Bögelspacher, and Leon Merkel supported the installation
of the experiment.
We are grateful for the well-considered comments provided by
two anonymous reviewers.
The cable with the tight-buffered fibers (green cable) has been made
available for the experiments through the INSIDE project.
INSIDE is supported by the German Federal Ministry for Economic Affairs and
Climate Action and the Project Management Jülich (PtJ) under the grant
agreement number 03EE4008C.
\end{ack}

\bibliography{DAS.bib}
\end{multicols}


\clearpage
\lfoot{\myauthors}
\rfoot{Supplement}
\newcommand{\gfzBdswxEQpara}{%
  \eventlink{Mw~6.79 Tajikistan-Xinjiang Border Region   2023-02-23 00:37:39.01  UTC}{gfz2023dswx}
  \degree{ 38.06}N \degree{ 73.29}E 10.0~km
  (BAZ=\azimuth{ 77.24}, $\Delta$=\degree{ 47.06})}
\newcommand{\gfzBdoqyEQpara}{%
  \eventlink{Mw~6.34 Turkey   2023-02-20 17:04:30.09  UTC}{gfz2023doqy}
  \degree{ 36.11}N \degree{ 35.93}E 10.0~km
  (BAZ=\azimuth{110.83}, $\Delta$=\degree{ 23.63})}
\newcommand{\gfzBcoosEQpara}{%
  \eventlink{Mw~7.59 Turkey   2023-02-06 10:24:49.96  UTC}{gfz2023coos}
  \degree{ 38.11}N \degree{ 37.22}E 10.0~km
  (BAZ=\azimuth{105.26}, $\Delta$=\degree{ 23.24})}
\newcommand{\gfzBcnwrEQpara}{%
  \eventlink{Mw~7.69 Turkey   2023-02-06 01:17:34.97  UTC}{gfz2023cnwr}
  \degree{ 37.23}N \degree{ 37.05}E 10.0~km
  (BAZ=\azimuth{107.24}, $\Delta$=\degree{ 23.64})}
\newcommand{\gfzBapzgEQpara}{%
  \eventlink{Mw~7.58 Banda Sea   2023-01-09 17:47:35.03  UTC}{gfz2023apzg}
  \degree{ -7.11}N \degree{129.98}E km~km
  (BAZ=\azimuth{ 69.87}, $\Delta$=\degree{115.92})}
\newcommand{\gfzAwxsgEQpara}{%
  \eventlink{Mw~6.06 Turkey   2022-11-23 01:08:16.89  UTC}{gfz2022wxsg}
  \degree{ 40.97}N \degree{ 31.05}E 10.0~km
  (BAZ=\azimuth{106.06}, $\Delta$=\degree{ 17.69})}
\newcommand{\gfzAwvyoEQpara}{%
  \eventlink{Mw~7.01 Solomon Islands   2022-11-22 02:03:07.65  UTC}{gfz2022wvyo}
  \degree{ -9.78}N \degree{159.57}E 9.7~km
  (BAZ=\azimuth{ 41.48}, $\Delta$=\degree{134.41})}
\newcommand{\gfzAwpnlEQpara}{%
  \eventlink{Mw~6.83 Southwest of Sumatra, Indonesia   2022-11-18 13:37:09.2  UTC}{gfz2022wpnl}
  \degree{ -4.77}N \degree{100.90}E 10.0~km
  (BAZ=\azimuth{ 91.10}, $\Delta$=\degree{ 95.16})}
\newcommand{\gfzAwcnjEQpara}{%
  \eventlink{Mw~7.28 Tonga Islands Region   2022-11-11 10:48:48.57  UTC}{gfz2022wcnj}
  \degree{-19.27}N \degree{-172.38}E 44.5~km
  (BAZ=\azimuth{  1.38}, $\Delta$=\degree{150.73})}
\newcommand{\gfzAvpriEQpara}{%
  \eventlink{Mw~6.08 Gulf of California   2022-11-04 10:02:51.02  UTC}{gfz2022vpri}
  \degree{ 28.13}N \degree{-112.15}E 10.0~km
  (BAZ=\azimuth{310.50}, $\Delta$=\degree{ 86.94})}
\newcommand{\gfzAuokoEQpara}{%
  \eventlink{Mw~6.78 South of Panama   2022-10-20 11:57:11.66  UTC}{gfz2022uoko}
  \degree{  7.66}N \degree{-82.32}E 10.0~km
  (BAZ=\azimuth{275.73}, $\Delta$=\degree{ 84.68})}
\newcommand{\gfzAsovfEQpara}{%
  \eventlink{Mw~6.79 Michoacan, Mexico   2022-09-22 06:16:08.89  UTC}{gfz2022sovf}
  \degree{ 18.33}N \degree{-102.94}E 15.0~km
  (BAZ=\azimuth{297.89}, $\Delta$=\degree{ 89.69})}
\newcommand{\gfzAskgcEQpara}{%
  \eventlink{Mw~7.59 Near Coast of Michoacan, Mexico   2022-09-19 18:05:08.76  UTC}{gfz2022skgc}
  \degree{ 18.44}N \degree{-103.01}E 20.0~km
  (BAZ=\azimuth{298.02}, $\Delta$=\degree{ 89.65})}
\newcommand{\gfzAshodEQpara}{%
  \eventlink{Mw~6.92 Taiwan   2022-09-18 06:44:16.55  UTC}{gfz2022shod}
  \degree{ 23.20}N \degree{121.35}E 10.0~km
  (BAZ=\azimuth{ 57.83}, $\Delta$=\degree{ 86.89})}
\newcommand{\gfzAsggkEQpara}{%
  \eventlink{Mw~6.46 Taiwan   2022-09-17 13:41:20.35  UTC}{gfz2022sggk}
  \degree{ 23.11}N \degree{121.29}E 10.0~km
  (BAZ=\azimuth{ 57.93}, $\Delta$=\degree{ 86.92})}
\newcommand{\gfzArufwEQpara}{%
  \eventlink{Mw~7.51 Papua New Guinea Region   2022-09-10 23:46:59.95  UTC}{gfz2022rufw}
  \degree{ -6.25}N \degree{146.48}E km~km
  (BAZ=\azimuth{ 53.88}, $\Delta$=\degree{124.91})}
\newcommand{\gfzArjqxEQpara}{%
  \eventlink{Mw~6.60 Sichuan, China   2022-09-05 04:52:19.87  UTC}{gfz2022rjqx}
  \degree{ 29.64}N \degree{102.16}E 10.0~km
  (BAZ=\azimuth{ 66.71}, $\Delta$=\degree{ 70.75})}
\newcommand{\gfzAriezEQpara}{%
  \eventlink{Mw~6.88 Central Mid-Atlantic Ridge   2022-09-04 09:42:20.22  UTC}{gfz2022riez}
  \degree{ -0.85}N \degree{-21.76}E 10.0~km
  (BAZ=\azimuth{217.51}, $\Delta$=\degree{ 55.46})}
\newcommand{\gfzAoogoEQpara}{%
  \eventlink{Mw~6.99 Luzon, Philippines   2022-07-27 00:43:25.36  UTC}{gfz2022oogo}
  \degree{ 17.50}N \degree{120.82}E 21.0~km
  (BAZ=\azimuth{ 61.70}, $\Delta$=\degree{ 91.06})}
\newcommand{\surfacewavefigures}[2]{
  \begin{figure*}
    {\includegraphics[trim=0 0 0 60,clip,width=\textwidth]{#1_Anton}}\par
    \includegraphics[trim=0 0 0 60,clip,width=\textwidth]{#1_Vorstollen}
    \caption{Surface wave data in the frequency band 0.05\,Hz to 0.1\,Hz for
      \csname #2EQpara\endcsname.
      Top: \Lanton{} (\azimuth{330}),
      bottom: \Lvorstollen{} (\azimuth{90}).
      DAS data are scaled (factors are given in the legend) to minimize the
      rms misfit with respect to the strainmeter waveform.
    }
    \label{fig:#2}
  \end{figure*}
}
\appendix
\setcounter{figure}{0}
\setcounter{table}{0}
\setcounter{section}{0}
\setcounter{equation}{0}
\renewcommand{\thefigure}{S\arabic{figure}}
\renewcommand{\thetable}{S\arabic{table}}
\renewcommand{\thesection}{S\arabic{section}}
\renewcommand{\theequation}{S\arabic{equation}}
\newcommand{\secref}[1]{\ref{#1} \qmarks{\nameref{#1}}}
\title{Electronic supplement to\\
\emph{Calibration of the strain amplitude recorded with DAS\\
using a strainmeter array}}
\author{Thomas Forbriger,
Jérôme Azzola,
Nasim Karamzadeh,\\
Emmanuel Gaucher,
Rudolf Widmer-Schnidrig,
Andreas Rietbrock}
\date{}
\maketitle
\begin{multicols}{2}
\section*{Introduction}
This electronic supplement introduces additional information and diagrams that are not essential to an understanding of the manuscript, but provide further insight into the data analysis and support the results of the study. 

The supplementary material focuses first on the recording environment. 
It details the position of the DAS read-out locations in the \Lanton{} and
\Lvorstollen{} and the
extent of the associated gauges on the floor
of the gallery (Sec.~\secref{sec:location}). 
It also details the characteristics of the fiber optic cables used in the
study (Sec.~\secref{sec:cablespec}).

The supplement gives then further details about the 19 earthquakes used to
carry out a waveform comparison in the main study. 
In Section~\secref{subsec:used_events}, earthquake parameters for all 19
events are detailed. We show the DAS and strainmeter surface wave
signals corresponding to each event, for read-out locations in \Lanton{} and
\Lvorstollen{} (Sec.~\secref{subsubsec:surface_waves}). 
For the two Kahramanmaraş earthquakes, we also include the body wave data
(Sec.~\secref{subsubsec:body_waves}). 
We complement the discussion in Section~\secref{sec:data} by 
details on the pre-processing of DAS data in section
\secref{subsec:noise:reduction}.
Section \secref{subsec:strainmeter:calibration} provides information on the
calibration of the BFO strainmeters and section
\secref{subsec:cavity:effects} adds considerations on strain-strain coupling.

Extended results are then introduced to support the main findings of the
study (Sec.~\secref{sec:extended:results}). 
They focus on the analysis of \Tstraintransfer\ (shortened to STR), which
quantifies how much of \Trockstrain\ is transferred to the signal recorded by
the IU, and on normalized correlation coefficients (shortened to NCC), which
quantify the waveform similarity between the two strain
measurements. 
We assess the correlation between STR and NCC and evaluate the symmetry of the
regression matrix, analyzing differences in fitting strainmeter data against
DAS data or the opposite. 
The correlation of NCC and STR to earthquake parameters, including
backazimuth and observed maximum strain, is also evaluated.  

Section~\secref{sec:seismometer:as:a:refernce} gives further insight into the
comparison of DAS recordings to strain signals computed from seismometer data.
Section~\secref{subsec:scaling:particle:velocity} details the procedure used
for the conversion from particle velocity to strain and
Sec.~\secref{subsec:regression:DAS:with:seismometer:data} compares the
regression coefficients computed when fitting strainmeter signals to DAS
signals with those obtained by fitting scaled seismometer data to DAS. 
Seismometer and strainmeter data are finally compared in
Sec.~\secref{subsec:regression:seismometer:and:strainmeter:data}. 

\subsection*{Pre-processing}
Signal pre-processing is carried out following the procedure 
described in the main paper.
All signals shown in the supplemental material are band-pass filtered between
0.05~Hz and 0.1~Hz, where the DAS signal-to-noise ratio is good enough and
strainmeter signals are free from parasitic components.

\section{Fiber Optic Cable setting}
\enlargethispage{11pt}
\subsection{Location of sensing points and gauges in the mine}
\label{sec:location}
\begin{figure*}
  \begin{center}
    {\includegraphics[clip,trim=10 10 10 50,width=\textwidth]{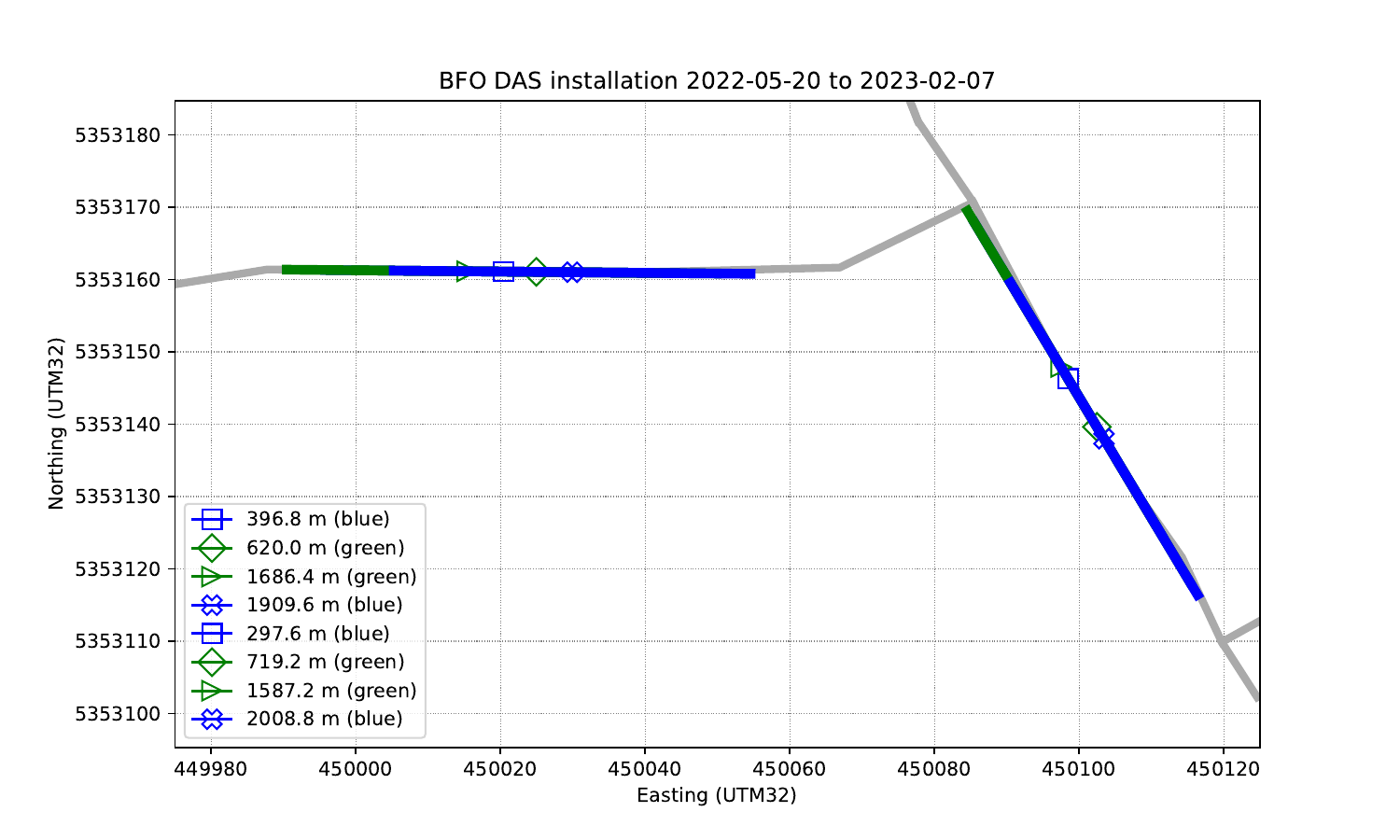}}
  \end{center}
  \caption{Floor map of the gallery of the mine with a zoom on the \Lanton\
  (70\,m long, azimuth  \azimuth{330}) and on the \Lvorstollen\ (80\,m long,
  azimuth \azimuth{90}). 
  It includes the extent of the gauge length, i.e. the
  optical fiber length over which optical dephasing is measured to produce a
  single sensing point located at the middle. The associated symbol shows the
  position of the measurement, as estimated by tap-tests. The figure includes
  the gauge lengths used for the study, considering the cable setup used in
  the mine from May 20, 2022 to February 7, 2023.}
  \label{fig:mapzoom1}
\end{figure*}

\begin{figure*}
  \begin{center}
    {\includegraphics[clip,trim=10 10 10 50,width=\textwidth]{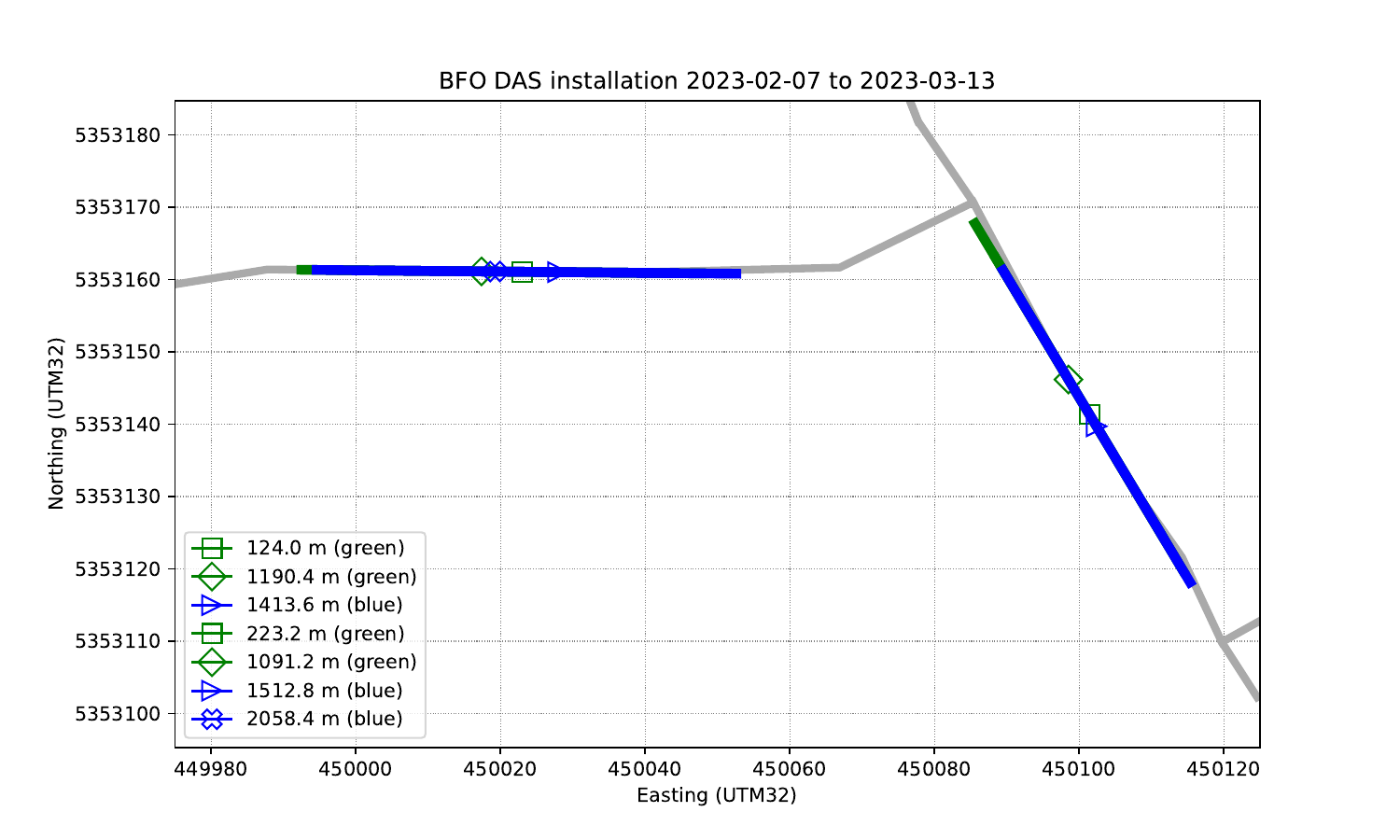}}
  \end{center}
  \caption{Same as previous figure, for the cable setup used in the mine from February 7 to March 13, 2023.}
  \label{fig:mapzoom2}
\end{figure*}

Table~\ref{tab:locations} details the position of the read-out locations
identified near the centers of \Lanton{} and \Lvorstollen{}. 
The read-out locations shown in the Figs.~\ref{fig:mapzoom1} and
\ref{fig:mapzoom2} are the ones used in the study. 

DAS strain is computed over so-called gauges, defined by the gauge length (GL)
parameter, which correspond to the length of optical fiber over which optical
dephasing is measured by the IU. 
The extent of the gauge on the cable plays therefore a significant role in
the measured strain.
The gauge-sections are displayed in detail in Figs.~\ref{fig:mapzoom1}
and \ref{fig:mapzoom2} for this reason.
Fig.~\ref{fig:mapzoom2} applies to the modified cable setup, after February 7,
2023. 
For each read-out location considered in the analysis on \Lanton{} and
\Lvorstollen{}, the physical location of the sensing point is represented as a
symbol with the cable offset indicated in the legend, and the extent of the
gauge is represented as a colored line. 

The read-out locations were selected because of their position, near the center
of \Lanton{} and \Lvorstollen{}. 
The figures show the small distance between the read-out locations positioned
in each section of interest. 
It shows also that the associated gauges cover these sections with a
consistent azimuth. 

\subsection{Specifications of optical fibers}
\label{sec:cablespec}
The analysis explores the capacities of two fiber optic cables with different
characteristics. 
We use on one hand a standard flexible
telecommunication cable with loose-tube fibers embedded in gel (blue cable on the diagrams).
It is distributed by Prysmian Group.
The print on the cable is:\\
      \texttt{DRAKA UC FIBRE I/O CT LSHF 3.0kN 4 SM2D
      G.652.D/SM7A1 BendBright 60011347 15 20506485 009201015}.

On the other hand, we use a second cable that has a stiff jacket containing
tight-buffered fibers (green cable in the diagrams).
It is distributed by Solifos AG and specified as a
`BRUsens DAS \& DTS \& Communication Hybrid' cable.
For distributed acoustic sensing, we use two 2.4\,mm simplex elements with
tight-buffered single mode fiber with Aramid, glass strain relief and plastic
outer sheath.
The print on the cable is:\\
      \texttt{BRUSENS ACOUSTIC TEMPERATURE HYBRID BSAH
      2SMF+2MMF+10SMF FIBRE OPTIC CABLE 00241904}.
\end{multicols}
\clearpage

\begin{table*}
  \tbl{Read-out locations along the fiber near the centers of \Lanton\ and
  \Lvorstollen\ (see Figs.~\ref{fig:mapzoom1} and \ref{fig:mapzoom2}).
  The locations are specified by the linear distance from the IU
  (interrogation unit) along the fiber route.
  \label{tab:locations}}{
\begin{tabular}{rlrr}
    \colrule
    \multicolumn{4}{l}{\textbf{\Lanton\ (azimuth \azimuth{330})}}\\
    \multicolumn{4}{l}{cables covered with
    sand and sandbags}\\
    & & \multicolumn{2}{c}{UTM32N}\\
    location / m & cable & easting / m & northing / m\\
    \cmidrule(r){1-1} \cmidrule(lr){2-2} \cmidrule(lr){3-3}
    \cmidrule(l){4-4}
    396.8 & loose-tube & 450102.6 & 5353139.4 \\
    620.0 & tight-buffered & 450098.3 & 5353146.5 \\
    1686.4 & tight-buffered & 450101.6 & 5353141.0 \\
    1909.6 & loose-tube & 450099.3 & 5353144.9 
    \trowsep
    \multicolumn{4}{l}{\textbf{\Lvorstollen\ (azimuth \azimuth{90})}}\\
    & & \multicolumn{2}{c}{UTM32N}\\
    location / m & cable & easting / m & northing / m\\
    \cmidrule(r){1-1} \cmidrule(lr){2-2} \cmidrule(lr){3-3}
    \cmidrule(l){4-4}
    297.6 & loose-tube & 450028.3 & 5353161.0 \\
    719.2 & tight-buffered & 450017.0 & 5353161.1 \\
    1587.2 & tight-buffered & 450023.4 & 5353161.1 \\
    2008.8 & loose-tube & 450022.0 & 5353161.1 \trowsep
    \multicolumn{4}{l}{\textbf{\Lanton\ (azimuth \azimuth{330}) after 2023-02-07}}\\
    & & \multicolumn{2}{c}{UTM32N}\\
    location / m & cable & easting / m & northing / m\\
    \cmidrule(r){1-1} \cmidrule(lr){2-2} \cmidrule(lr){3-3}
    \cmidrule(l){4-4}
    124.0 & tight-buffered & 450101.4 & 5353141.4 \\
    1190.4 & tight-buffered & 450098.5 & 5353146.2 \\
    1413.6 & loose-tube & 450102.4 & 5353139.7 \trowsep
    \multicolumn{4}{l}{\textbf{\Lvorstollen\ (azimuth \azimuth{90}) after 2023-02-07}}\\
    & & \multicolumn{2}{c}{UTM32N}\\
    location / m & cable & easting / m & northing / m\\
    \cmidrule(r){1-1} \cmidrule(lr){2-2} \cmidrule(lr){3-3}
    \cmidrule(l){4-4}
    223.2 & tight-buffered & 450023.0 & 5353161.1 \\
    1091.2 & tight-buffered & 450017.4 & 5353161.1\\
    1512.8 & loose-tube & 450027.9 & 5353161.0\\
    2058.4 & loose-tube & 450019.3 & 5353161.1 \trowsep
    \multicolumn{4}{l}{\textbf{\tunnellocation{Drum}}}\\
    location / m & cable &&\\
    \cmidrule(r){1-1} \cmidrule(l){2-3} 
    857.5 & \multicolumn{2}{l}{tight-buffered, drum start} \\
    1150.0 & \multicolumn{2}{l}{tight-buffered, drum center} \\
    1442.5 & \multicolumn{2}{l}{tight-buffered, drum end} \trowsep
    \multicolumn{4}{l}{\textbf{\tunnellocation{Drum} after 2023-02-07}}\\
    location / m & cable &&\\
    \cmidrule(r){1-1} \cmidrule(l){2-3} 
    367.5 & \multicolumn{2}{l}{tight-buffered, drum start} \\
    660.0 & \multicolumn{2}{l}{tight-buffered, drum center} \\
    952.5 & \multicolumn{2}{l}{tight-buffered, drum end} \\
    \botrule
  \end{tabular}
  }{}
\end{table*}
\clearpage
\begin{multicols}{2}
\section{Analyzed data}
\label{sec:data}
\subsection{Earthquakes used in the analysis}
\label{subsec:used_events}
Table~\ref{tab:UsedEvents} details the seismic events used in the analysis
with the main source parameters, which are taken from the GEOFON catalog.
From the GEOFON catalog \citep{quinteros2020} we find 84 earthquakes with
moment magnitude larger than 6 in the recording time period.
For 21 of them, we measure a maximum strain amplitude larger than 1\,nstrain in a
visual inspection of strainmeter data, which is considered large enough to
provide a sufficient signal-to-noise ratio in the DAS data.
For 19 of these events, DAS data is available and allows an analysis.
The largest amplitudes are found in the surface wave train.
For body waves, the horizontal strain amplitude decreases as the ray incidence becomes steeper.

The table details the backazimuth (BAZ) and epicentral distance ($\Delta$)
that are given with respect to BFO (\degree{48.33}N, \degree{8.33}E), as well
as the maximum strain amplitude associated with surface wave signals,
$A_{\text{max}}$. 
The latter is visually read from the strainmeter data and should
be understood as a proxy of the actual maximum signal amplitude in the
analysis.

\subsection{Strain waveforms}
\label{subsec:waveforms}
  In Figs.~\ref{fig:gfzBdswx} to \ref{fig:turkey:BWB:Vorstollen}
  we display waveforms of the analyzed data.
  All signals are consistently filtered with Butterworth high-pass
  (0.05~Hz, 4th order) and low-pass (0.1~Hz, 4th order)
  filters.
  Where STS-2 data is shown, the seismometer response is equalized to the
  response of the strain signals.
  Traces are shifted vertically for better visibility.

  The DAS signals from the tight-buffered cable (green)
  and the loose-tube cable (blue) 
  are taken with a gauge length of 50~m at a location
  in the center of the \Lanton\ and the \Lvorstollen, respectively
  (see Table~\ref{tab:locations} and
  Figs.~\ref{fig:mapzoom1} and \ref{fig:mapzoom2}).
  Linear strain in azimuth \azimuth{330} of \Lanton\ and azimuth \azimuth{90}
  of \Lvorstollen{} is obtained from the BFO strainmeter array like explained
  in the main paper.

\subsubsection{Surface wave signals}
\label{subsubsec:surface_waves}
Figures~\ref{fig:gfzBdswx} to ~\ref{fig:gfzAoogo} detail the surface wave
signals of the 19 events used in the manuscript to compare the 
DAS recorded \Tfiberstrain\ with the strainmeter recorded \Trockstrain. 
The figures focus on an ad-hoc waveform comparison of surface waves for linear
strain in the direction of \Lanton{} (top) and \Lvorstollen{} (bottom). 
Details about the analyzed frequency band, the event origin time and beam
parameters are included in the caption.  

\subsubsection{Body wave signals}
\label{subsubsec:body_waves}
For the large amplitude signals of the two main shocks (Mw~7.7 and Mw~7.6) of
the Kahramanmaraş earthquake sequence, we also analyze the body wave signals.
The body-wave signals of the 
Mw~7.7 Pazarcık earthquake are displayed for the \Lanton\ in
Fig.~\ref{fig:turkey:BWA:Anton} and in Fig.~\ref{fig:turkey:BWA:Vorstollen}
for the \Lvorstollen.
The surface wave signals are shown in Fig.~\ref{fig:gfzBcnwr}.
The corresponding diagrams for the Mw~7.6 Ekinözü earthquake are
shown in Figs.~\ref{fig:turkey:BWB:Anton}, \ref{fig:turkey:BWB:Vorstollen},
and \ref{fig:gfzBcoos}.

\subsection{Reduction of coherent noise}
\label{subsec:noise:reduction}
The Febus A1-R recordings contain a significant component of coherent noise at
frequencies below 0.5~Hz.
\citet[][their section~2.6 Optical Noise]{lindsey2020} attribute this type
of noise to vibrations of the optoelectronic system.

  Other studies use the average over all recorded channels to capture the
  coherent component of noise.
  This relies on the earthquake signal not being coherent (because of
  propagation delay and varying orientation of the fiber), such that it
  more or less cancels out in the average and the earthquake signal in a
  single channel is not deteriorated by subtracting the average over all
  channels.

  We follow a different approach.
  The idea behind using the reference coil on the drum is, that this section
  of the fiber is sufficiently decoupled from the rock and hence does not
  pick up a signal of the teleseismic earthquakes.
  For this reason it can be subtracted without affecting the amplitude of
  the earthquake signal.

  In Figs.~\ref{fig:noisered:P} and \ref{fig:noisered:all}
  we demonstrate this with the recordings of the Mw~7.7 Pazarcık
  earthquake.
  This is the earthquake which produced the by far largest strain amplitudes
  in the set of analyzed earthquakes.
  Even for the large amplitude surface waves, no signature of the earthquake
  signal is present in the recording from the reference coil (center diagram
  in Fig.~\ref{fig:noisered:all}).
  If a fraction of the earthquake signal should be hidden in the coherent
  noise on the reference drum, its amplitude is reduced to below 1.5\,per
  cent of the full signal.

  The P-wave signal of the earthquake becomes only apparent after
  this type of correction is applied,
  as shown in Fig.~\ref{fig:noisered:P}.

\subsection{Calibration of the strainmeters}
\label{subsec:strainmeter:calibration}
The strainmeters operated at BFO are Invar-wire strainmeters of the design by
\citet{king1976}.
\citet{agnew1986} and \citet{zuern2012} discuss instruments of this type
and their properties.
These instruments are primarily designed to record very-long period signals,
such as tidal strain or Earth's free oscillations.
For the analysis of tidal strain amplitudes instrument calibration is of
utmost importance.
The original calibration of the instruments was based on separate
calibrations of several components, which then were put in series.
In unfortunate cases the individual calibration inaccuracies added up
constructively, which resulted in an unacceptable inaccuracy for the entire
strainmeter.

For this reason an in-situ calibration mechanism for these instruments was
developed at BFO, based on so-called 'Crapaudines', which provides an accuracy
of about 2~per cent.
They were originally used by
\citet{verbaandert1959etalonnage,verbaandert1962etalonnage,verbaandert1963etalonnage}
in the calibration of tiltmeters.
This application is described by \citet[][pages~161 to 166]{melchior1966},
\citet[][his section~8.8]{melchior1978} and 
\citet[][his section~4.1.5]{agnew1986}.
\citet{melchior1966} calls them 'Expandable bearing plates of Verbaandert'.
\citet{agnew1986} calls them 'distensible support of Verbaandert'.
The 'Crapaudines' themselves are calibrated interferometrically
\citep{verbaandert1959etalonnage,verbaandert1962etalonnage}
and are used to impose a well defined motion on the fixed end of the wire in
the strainmeter.
The calibration factors of a few 'Crapaudines' used at BFO were 
confirmed by an independent interferometric measurement.

Because calibrations disturb the long-term tidal recordings, the stability of
the calibration is occasionally tested by a comparison against theoretical
tidal strain \citep{longman1959}.
This way we ensure a calibration accuracy and stability of about 5~per cent.
\citet{zuern2015} documented an episode of unstable calibration during the
2011 Tōhoku earthquake and how this was resolved.
They were able to obtain strain amplitudes for radial mode $_{0}$S$_{0}$,
which were consistent with Earth's radius and the observed radial surface
motion within 15~per cent after the 2004 Sumatra-Andaman earthquake (their
table~2).

The STR cannot be more accurate than the strainmeter calibration.
The variation of STR between locations and cable type, however, cannot be due
to limited calibration accuracy.
This is because the differences exceed the level of 5~per cent and because
the STR for the blue cable is larger in the \Lanton{} than in
\Lvorstollen{}, while it is the opposite for the green cable, which cannot
be caused by a calibration bias in principle.

\subsection{Distortion of the strain field due to local heterogeneity}
\label{subsec:cavity:effects}
Strain measurements and tilt measurements both are affected by local
heterogeneity, sometimes called 'cavity effects'.
This was first reported by \citet{king1973} and \citet{baker1973} 
for measurements of tidal tilt and limits the usefulness of tidal strain and
tilt amplitudes in the investigation of Earth's elastic properties.
These effects are strongest at free surfaces of the chambers (cavities) in
which instruments are installed
and near the free surface of the Earth, which can have a
significant topography.
The free-surface condition lets components of the stress tensor vanish and
this enforces specific linear combinations for the components of the
deformation tensor through the stress-strain relation.
\citet{harrison1976} investigated this by finite-element modeling. 

All strain measurements are affected by such strain-strain coupling.
This applies to DAS recordings as well as to conventional strainmeters.
In a more general sense the coupling issues between rock and fiber could also
be summarized under these effects of local heterogeneity.
Likewise particle motion is affected by free-surface conditions
\citep{kennett1991}.
When comparing DAS recorded strain with signals from reference instruments,
these effects generally could limit the usefulness of the observations, but
are commonly ignored in the respective studies (except for cable coupling).
The problem is even more critical, when comparing DAS recorded strain with
particle motion recorded by seismometers, because the effect of the free
surface on both is different and most likely the plane wave assumption
(see section~\secref{sec:seismometer:as:a:refernce}) is violated in this way.
If DAS recorded strain is compared to recordings of a strainmeter, as is done
in the current study, a problem may arise from spatially varying strain-strain
coupling, if instruments are installed at a distance from each other.

The 10~m long Invar-wire strainmeters show little cavity effects due to their
installation in the center and along the symmetry axes of 60~m long tunnels.
For the DAS cables the cavity effects in a more general sense are addressed as
coupling issues controlled by installation and cable type.

Topography at BFO, however, alters the strain field as demonstrated by
\citet{emter1985} with finite element simulations.
In particular the strainmeter in azimuth \azimuth{60}
experiences significant strain-strain coupling through the local topography.
\citet[][their table~S2 in the supporting material]{zuern2015} list the
strain-strain coupling coefficients derived from tidal analysis.
Their analysis of the radial mode $_{0}$S$_{0}$ 
\citep[][their table~2]{zuern2015} suggests that not the same
coupling might be at work for this type of straining.
The factor 0.58 (for \azimuth{60})
given by 
\citet[][their table~S2 in the supporting material]{zuern2015} 
for tidal analysis is consistent with the factor of 0.67
derived in a 2D finite element analysis by 
\citet[][their figure~5]{emter1985}.
For strain in the azimuth \azimuth{90} of the \Lvorstollen{} we presumably
observe part of this topography effect.
The seismometer data systematically over-estimates the strain amplitude in
this azimuth, as discussed in
section~\secref{subsec:regression:DAS:with:seismometer:data}.

In the current experiment we use the Invar-wire strainmeter in a comparison
with DAS data recorded at a distance of about 350~m.
If the strain-strain coupling would vary along this distance, it would
contribute to the amplitude difference seen between DAS recordings and
strainmeter recordings.
The difference shown by \citet[][their Fig. 5]{emter1985}, with strain
amplitude increasing towards the tunnel entrance near the free surface,
however, is opposite to the observed difference.
Because the section of the \Lvorstollen, which is used in the current
analysis, is at distance of about 200~m to the entrance the expected
strain-strain coupling would be almost the same as for the strainmeters,
according to \citet[][their Fig. 5]{emter1985}.

For strain recordings with cemented fibers in the \Lvorstollen{}
\citet{forbriger2024} recently found a STR of about 1 with respect to the
strainmeters for four different IUs, independently.
We take this as a confirmation that the DAS cable and the strainmeters
experience the same topography-effect, indeed.

\end{multicols}
\clearpage
\begin{sidewaystable*}
    \centering
  \tbl{Events used in the analysis.
  Source parameters are taken from the GEOFON catalog.
  backazimuth (BAZ) and epicentral distance ($\Delta$) are given with respect
  to BFO (\degree{48.33}N, \degree{8.33}E).
  Maximum amplitude $A_{\text{max}}$ is the largest amplitude in the surface
  wave train and is from visual inspection of the strainmeter signals.}{
\begin{tabular}{lrllrrrrrr}
    \colrule
    ID & Mw & origin time & region & 
    \multicolumn{1}{l}{latitude} & 
    \multicolumn{1}{l}{longitude} & 
    \multicolumn{1}{l}{depth} & 
    \multicolumn{1}{l}{BAZ} & 
    \multicolumn{1}{l}{$\Delta$} & 
    \multicolumn{1}{l}{$A_{\text{max}}$}  \\
    & & & & & & 
    \multicolumn{1}{l}{/ km}
    & &  & 
    \multicolumn{1}{l}{/ nstrain} \\
    \cmidrule(r){1-1} 
    \cmidrule(lr){2-2} 
    \cmidrule(lr){3-3}
    \cmidrule(lr){4-4}
    \cmidrule(lr){5-7}
    \cmidrule(lr){8-9}
    \cmidrule(l){10-10}
  \texttt{\eventlink{gfz2023dswx}{gfz2023dswx}}
    &  6.79 &  2023-02-23 00:37:39.01  & Tajikistan-Xinjiang Border Region 
    & \degree{ 38.06}N & \degree{ 73.29}E & 10.0
    & \azimuth{  77.24} & \degree{ 47.06} &   20.0 \\
  \texttt{\eventlink{gfz2023doqy}{gfz2023doqy}}
    &  6.34 &  2023-02-20 17:04:30.09  & Turkey 
    & \degree{ 36.11}N & \degree{ 35.93}E & 10.0
    & \azimuth{ 110.83} & \degree{ 23.63} &    8.0 \\
  \texttt{\eventlink{gfz2023coos}{gfz2023coos}}
    &  7.59 &  2023-02-06 10:24:49.96  & Turkey 
    & \degree{ 38.11}N & \degree{ 37.22}E & 10.0
    & \azimuth{ 105.26} & \degree{ 23.24} &  130.0 \\
  \texttt{\eventlink{gfz2023cnwr}{gfz2023cnwr}}
    &  7.69 &  2023-02-06 01:17:34.97  & Turkey 
    & \degree{ 37.23}N & \degree{ 37.05}E & 10.0
    & \azimuth{ 107.24} & \degree{ 23.64} &  250.0 \\
  \texttt{\eventlink{gfz2023apzg}{gfz2023apzg}}
    &  7.58 &  2023-01-09 17:47:35.03  & Banda Sea 
    & \degree{ -7.11}N & \degree{129.98}E & 108.1
    & \azimuth{  69.87} & \degree{115.92} &    1.5 \\
  \texttt{\eventlink{gfz2022wxsg}{gfz2022wxsg}}
    &  6.06 &  2022-11-23 01:08:16.89  & Turkey 
    & \degree{ 40.97}N & \degree{ 31.05}E & 10.0
    & \azimuth{ 106.06} & \degree{ 17.69} &    3.0 \\
  \texttt{\eventlink{gfz2022wvyo}{gfz2022wvyo}}
    &  7.01 &  2022-11-22 02:03:07.65  & Solomon Islands 
    & \degree{ -9.78}N & \degree{159.57}E & 9.7
    & \azimuth{  41.48} & \degree{134.41} &    2.5 \\
  \texttt{\eventlink{gfz2022wpnl}{gfz2022wpnl}}
    &  6.83 &  2022-11-18 13:37:09.2  & Southwest of Sumatra, Indonesia 
    & \degree{ -4.77}N & \degree{100.90}E & 10.0
    & \azimuth{  91.10} & \degree{ 95.16} &    2.0 \\
  \texttt{\eventlink{gfz2022wcnj}{gfz2022wcnj}}
    &  7.28 &  2022-11-11 10:48:48.57  & Tonga Islands Region 
    & \degree{-19.27}N & \degree{-172.38}E & 44.5
    & \azimuth{   1.38} & \degree{150.73} &    3.0 \\
  \texttt{\eventlink{gfz2022vpri}{gfz2022vpri}}
    &  6.08 &  2022-11-04 10:02:51.02  & Gulf of California 
    & \degree{ 28.13}N & \degree{-112.15}E & 10.0
    & \azimuth{ 310.50} & \degree{ 86.94} &    1.2 \\
  \texttt{\eventlink{gfz2022uoko}{gfz2022uoko}}
    &  6.78 &  2022-10-20 11:57:11.66  & South of Panama 
    & \degree{  7.66}N & \degree{-82.32}E & 10.0
    & \azimuth{ 275.73} & \degree{ 84.68} &    1.5 \\
  \texttt{\eventlink{gfz2022sovf}{gfz2022sovf}}
    &  6.79 &  2022-09-22 06:16:08.89  & Michoacan, Mexico 
    & \degree{ 18.33}N & \degree{-102.94}E & 15.0
    & \azimuth{ 297.89} & \degree{ 89.69} &    5.0 \\
  \texttt{\eventlink{gfz2022skgc}{gfz2022skgc}}
    &  7.59 &  2022-09-19 18:05:08.76  & Near Coast of Michoacan, Mexico 
    & \degree{ 18.44}N & \degree{-103.01}E & 20.0
    & \azimuth{ 298.02} & \degree{ 89.65} &   19.0 \\
  \texttt{\eventlink{gfz2022shod}{gfz2022shod}}
    &  6.92 &  2022-09-18 06:44:16.55  & Taiwan 
    & \degree{ 23.20}N & \degree{121.35}E & 10.0
    & \azimuth{  57.83} & \degree{ 86.89} &   10.0 \\
  \texttt{\eventlink{gfz2022sggk}{gfz2022sggk}}
    &  6.46 &  2022-09-17 13:41:20.35  & Taiwan 
    & \degree{ 23.11}N & \degree{121.29}E & 10.0
    & \azimuth{  57.93} & \degree{ 86.92} &    2.0 \\
  \texttt{\eventlink{gfz2022rufw}{gfz2022rufw}}
    &  7.51 &  2022-09-10 23:46:59.95  & Papua New Guinea Region 
    & \degree{ -6.25}N & \degree{146.48}E & 100.0
    & \azimuth{  53.88} & \degree{124.91} &    2.5 \\
  \texttt{\eventlink{gfz2022rjqx}{gfz2022rjqx}}
    &  6.60 &  2022-09-05 04:52:19.87  & Sichuan, China 
    & \degree{ 29.64}N & \degree{102.16}E & 10.0
    & \azimuth{  66.71} & \degree{ 70.75} &    2.0 \\
  \texttt{\eventlink{gfz2022riez}{gfz2022riez}}
    &  6.88 &  2022-09-04 09:42:20.22  & Central Mid-Atlantic Ridge 
    & \degree{ -0.85}N & \degree{-21.76}E & 10.0
    & \azimuth{ 217.51} & \degree{ 55.46} &    2.5 \\
  \texttt{\eventlink{gfz2022oogo}{gfz2022oogo}}
    &  6.99 &  2022-07-27 00:43:25.36  & Luzon, Philippines 
    & \degree{ 17.50}N & \degree{120.82}E & 21.0
    & \azimuth{  61.70} & \degree{ 91.06} &    2.5 \\
    \botrule
  \end{tabular}
  }{}\par
\label{tab:UsedEvents}
\end{sidewaystable*}
\clearpage

\surfacewavefigures{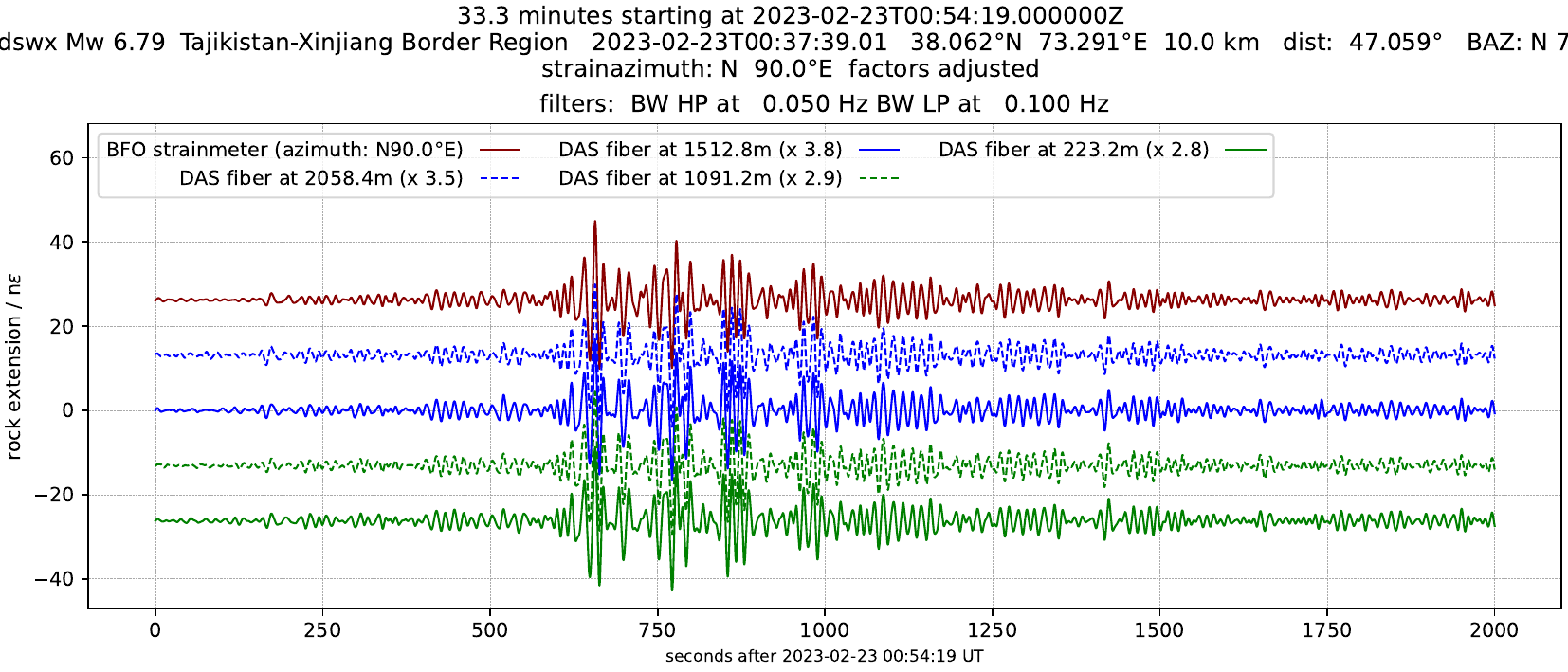}{gfzBdswx}
\surfacewavefigures{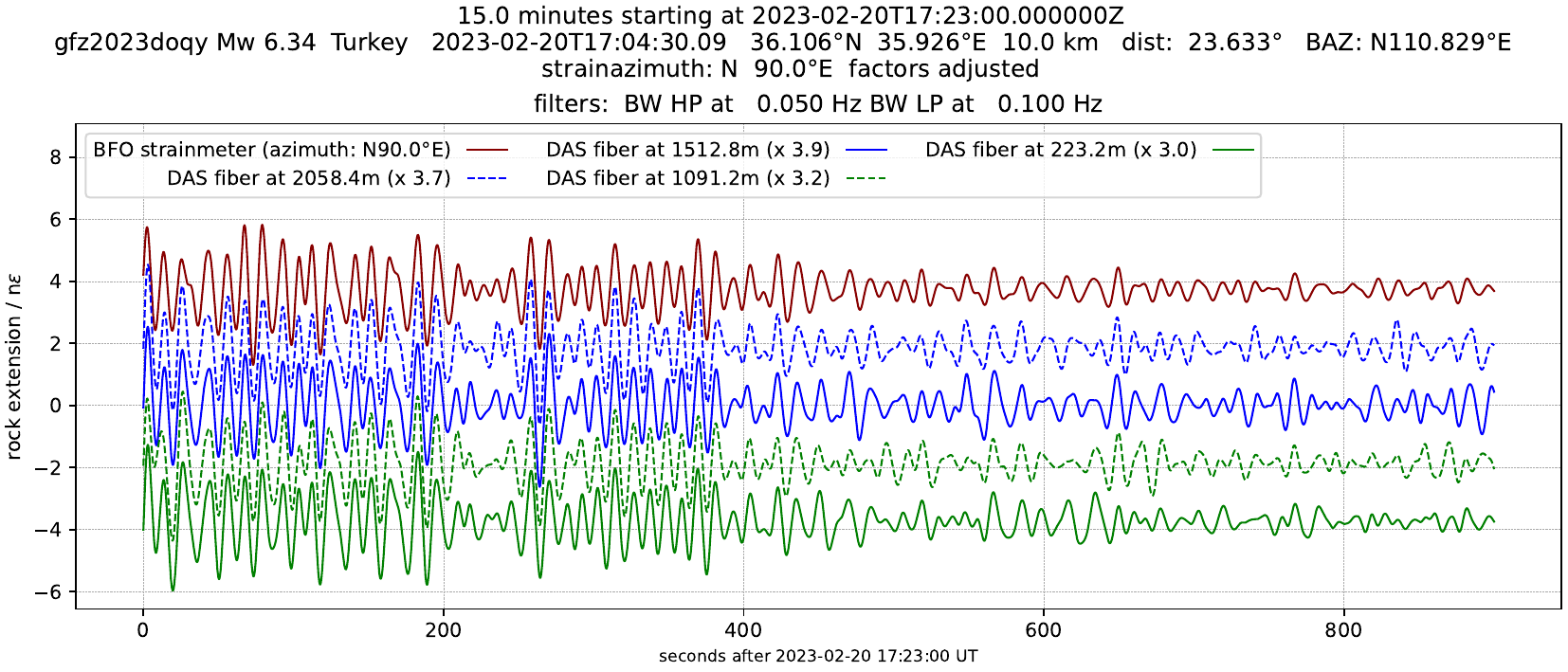}{gfzBdoqy}
\surfacewavefigures{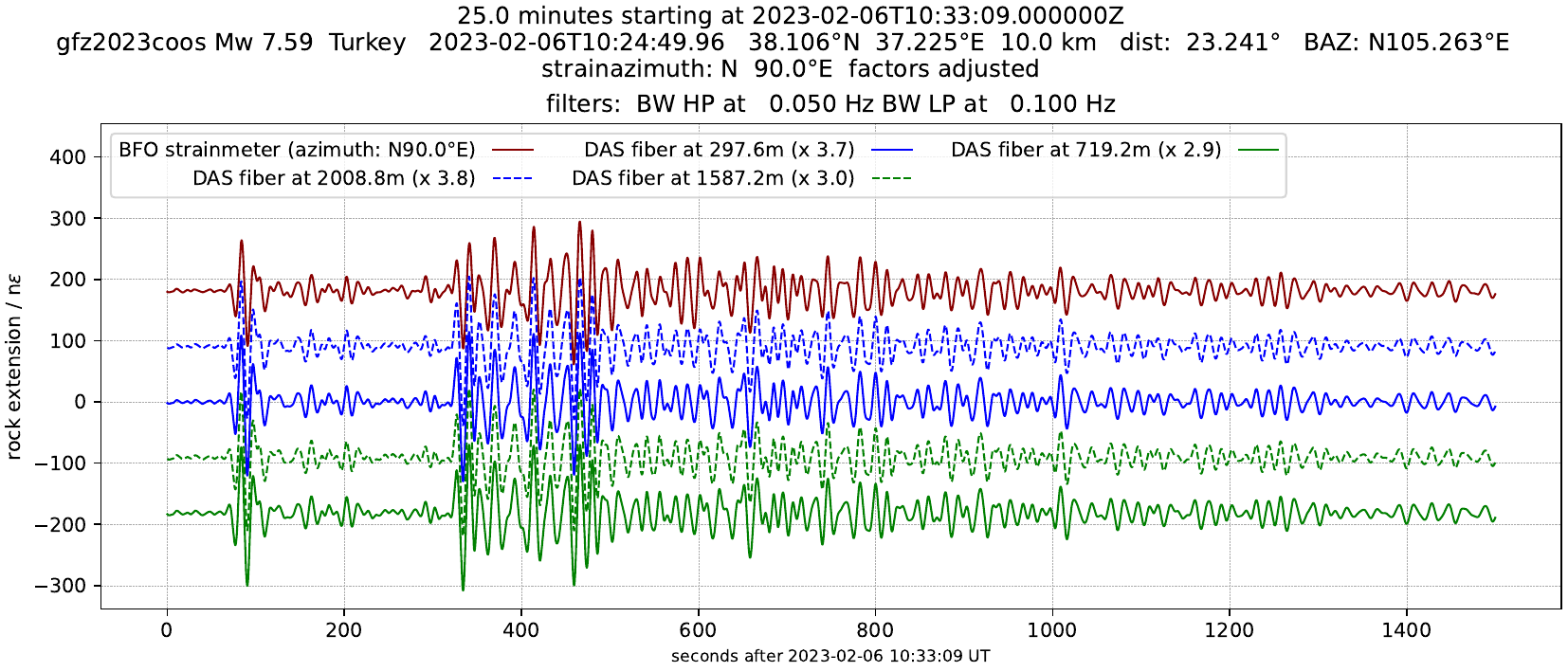}{gfzBcoos}
\surfacewavefigures{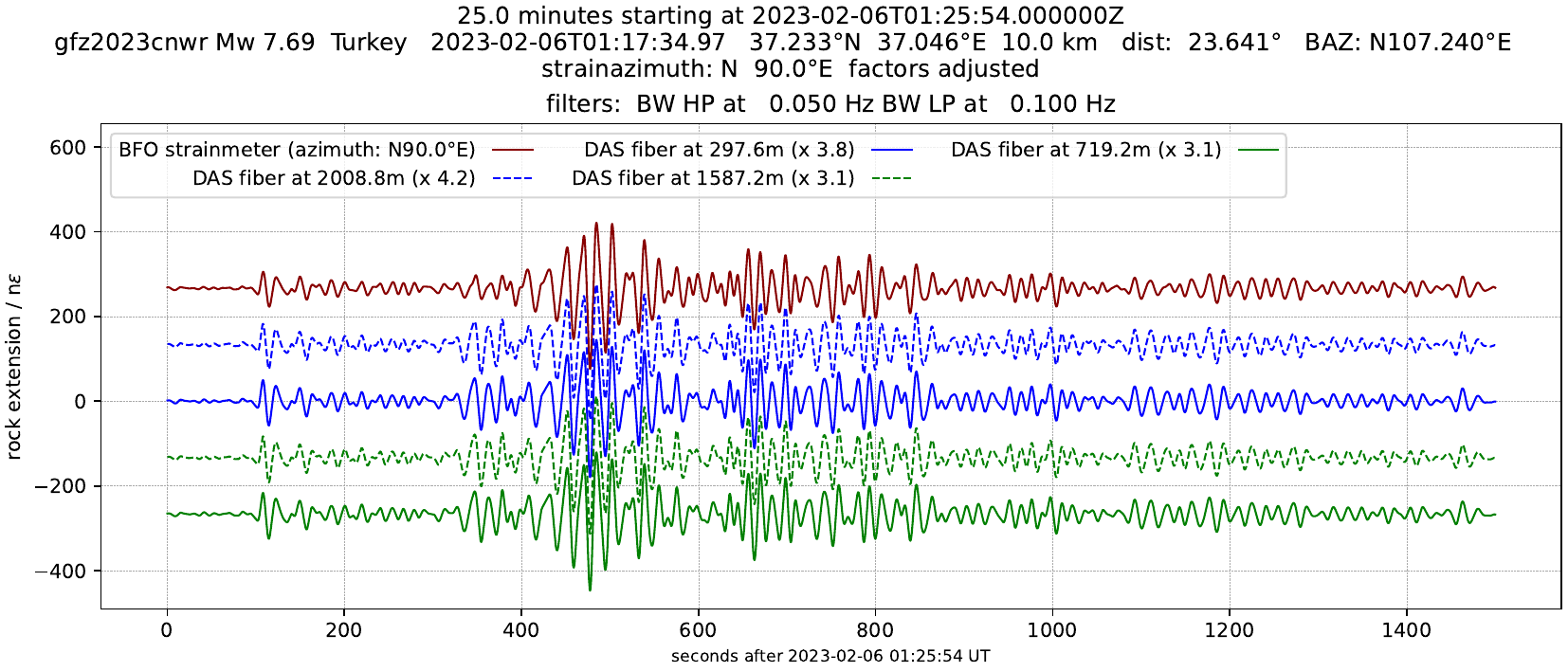}{gfzBcnwr}
\surfacewavefigures{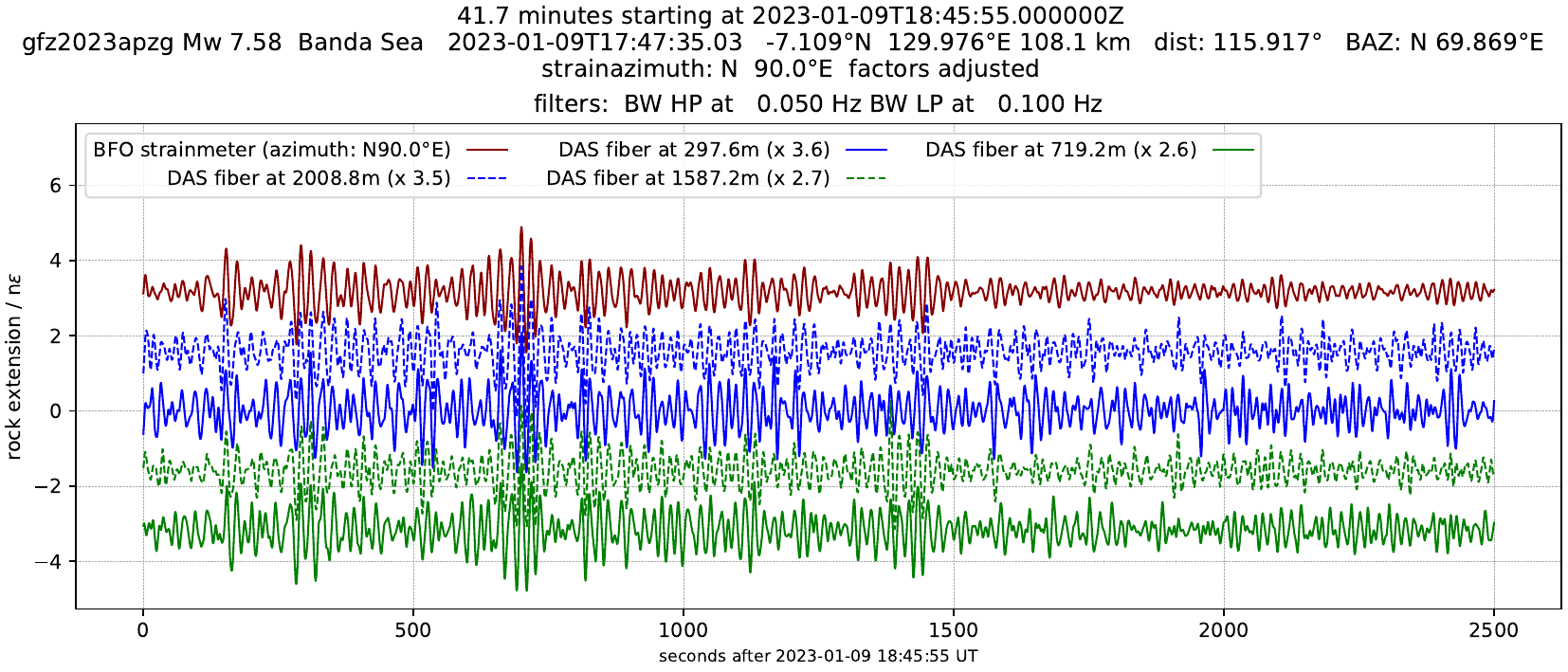}{gfzBapzg}
\surfacewavefigures{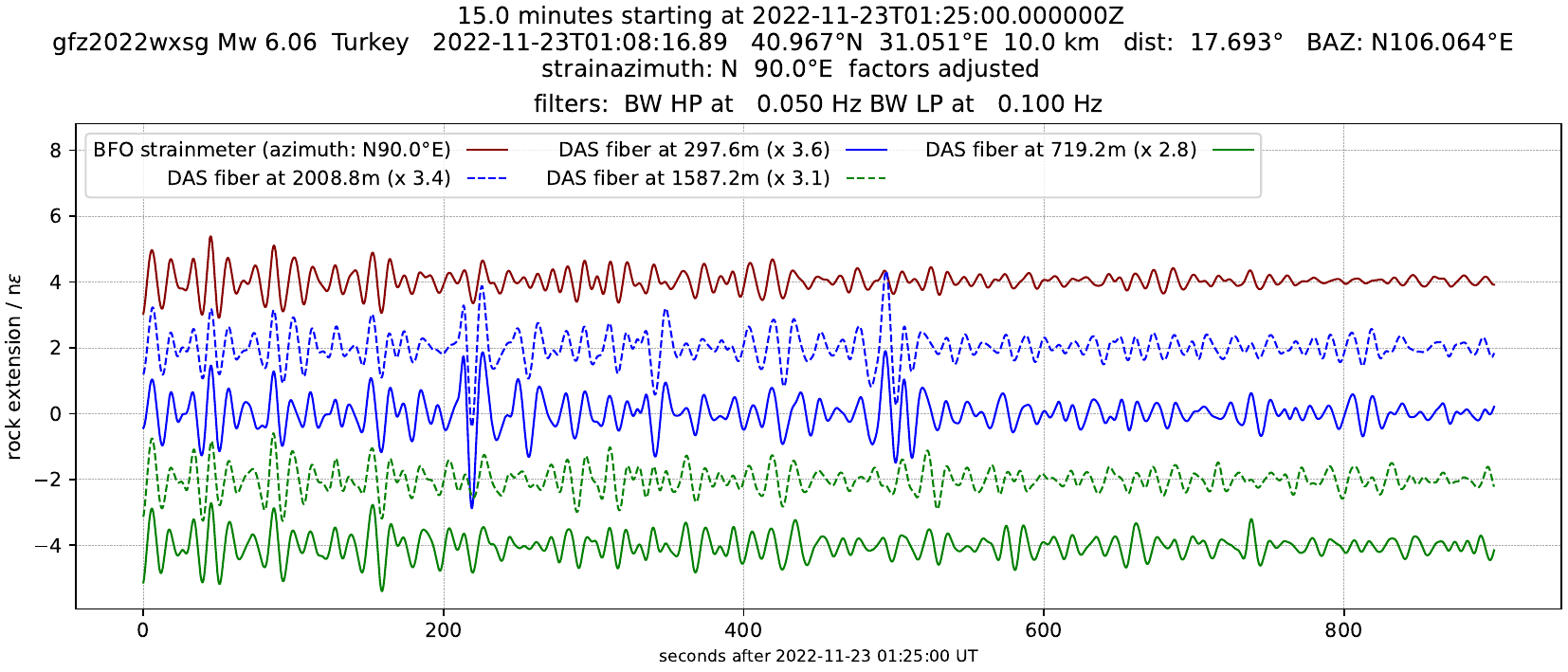}{gfzAwxsg}
\surfacewavefigures{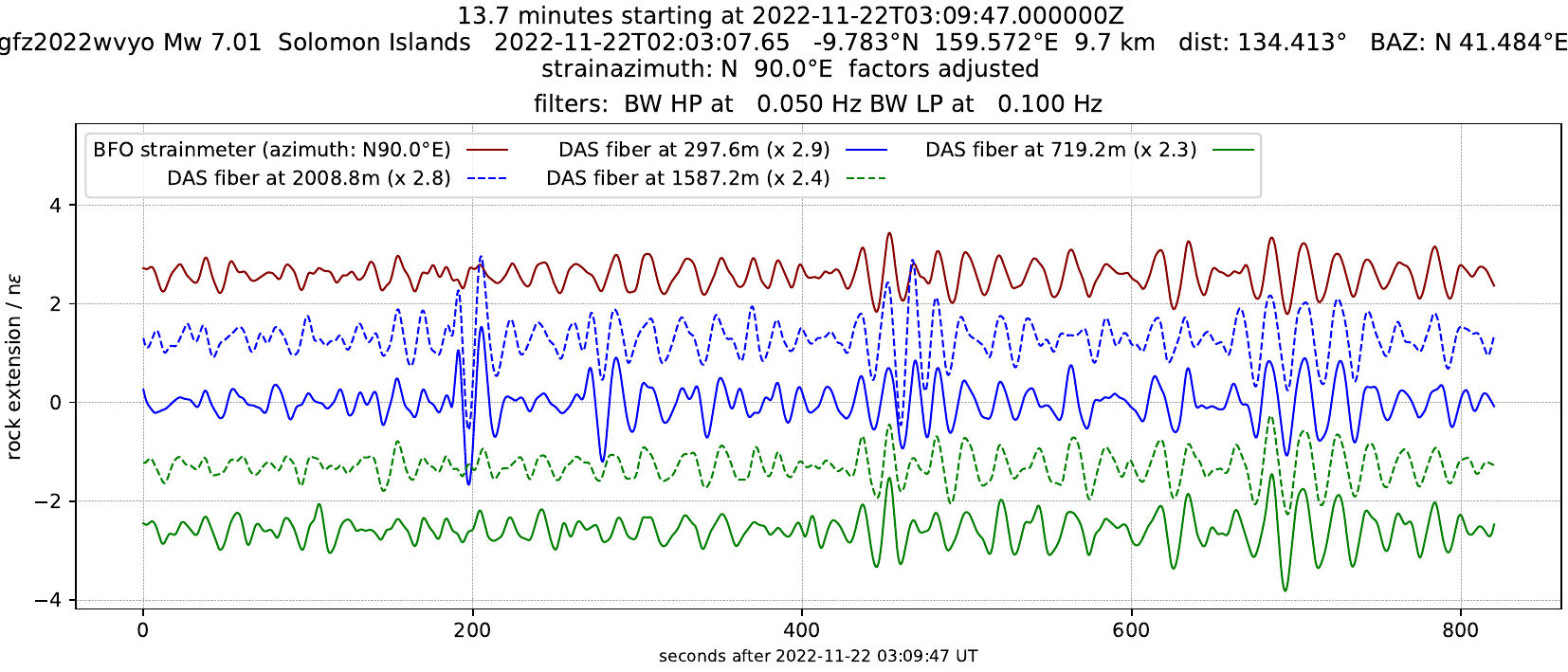}{gfzAwvyo}
\surfacewavefigures{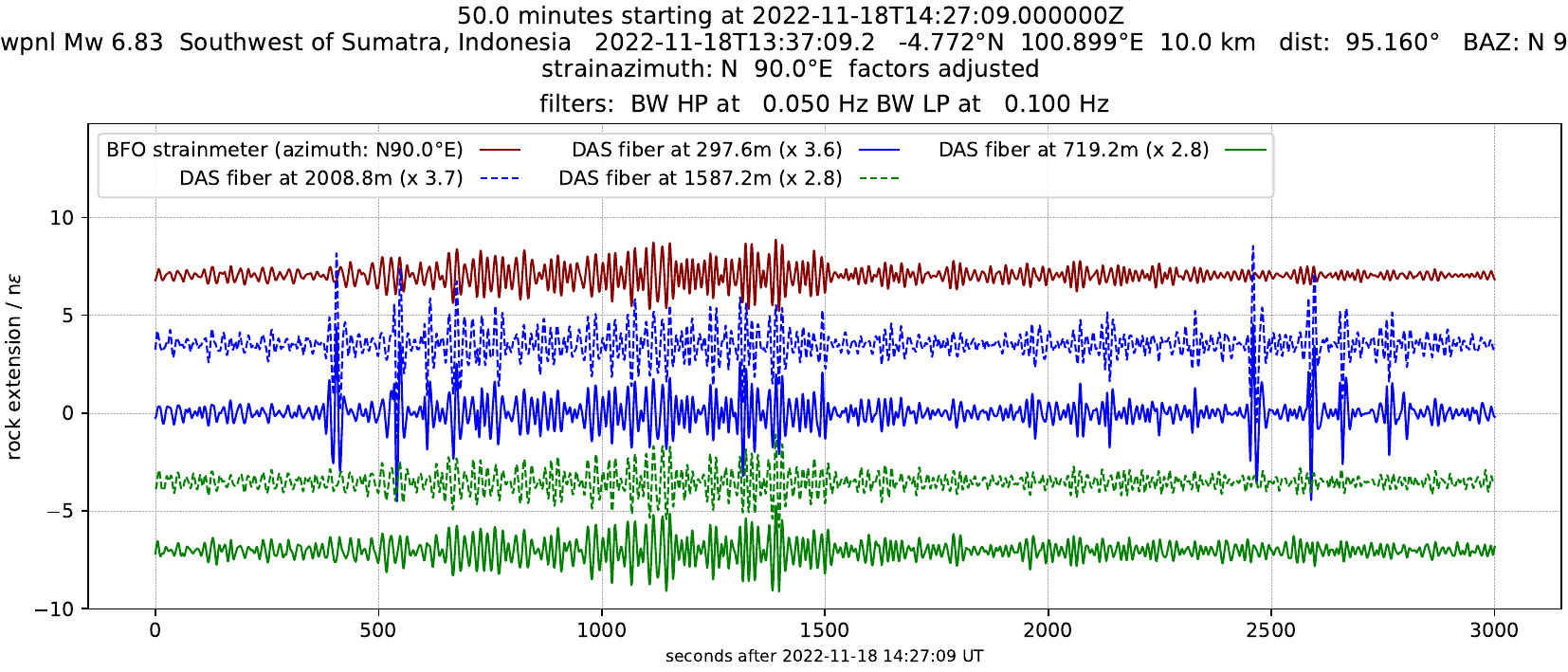}{gfzAwpnl}
\surfacewavefigures{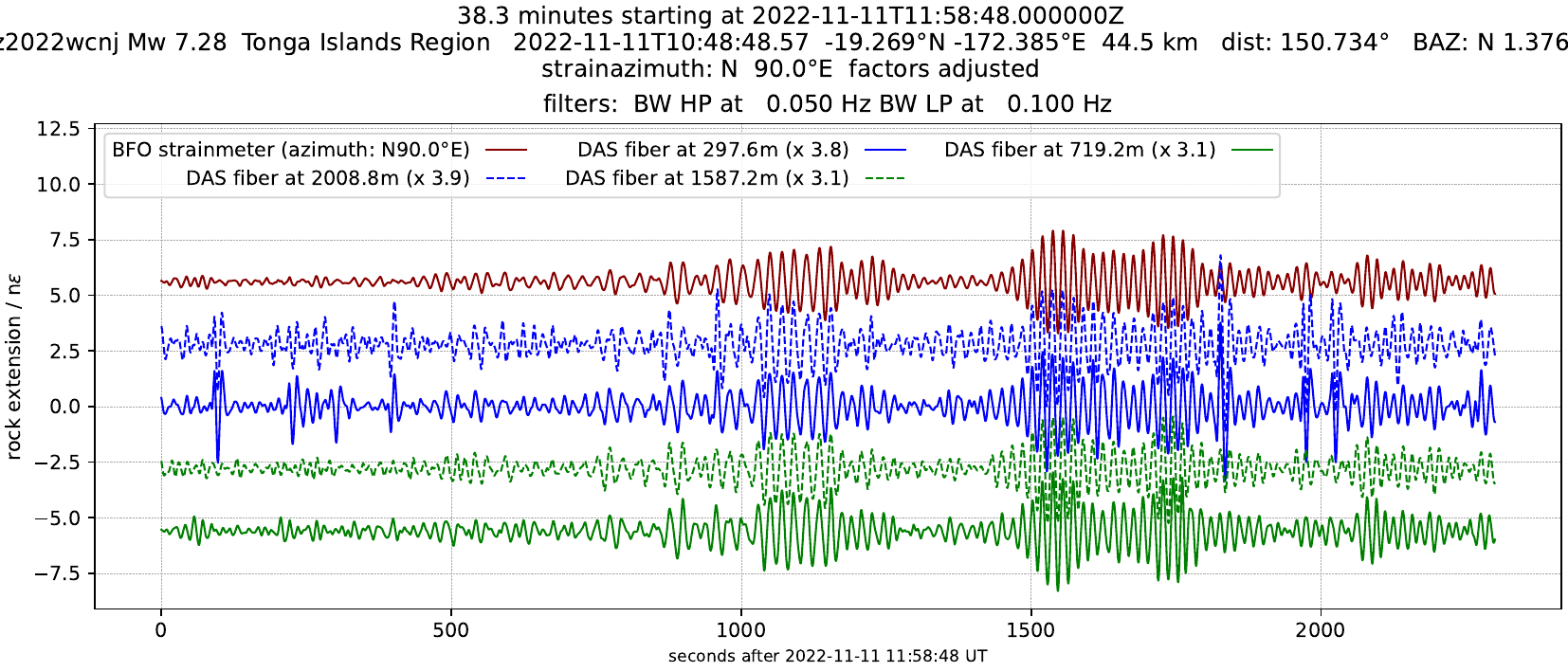}{gfzAwcnj}
\surfacewavefigures{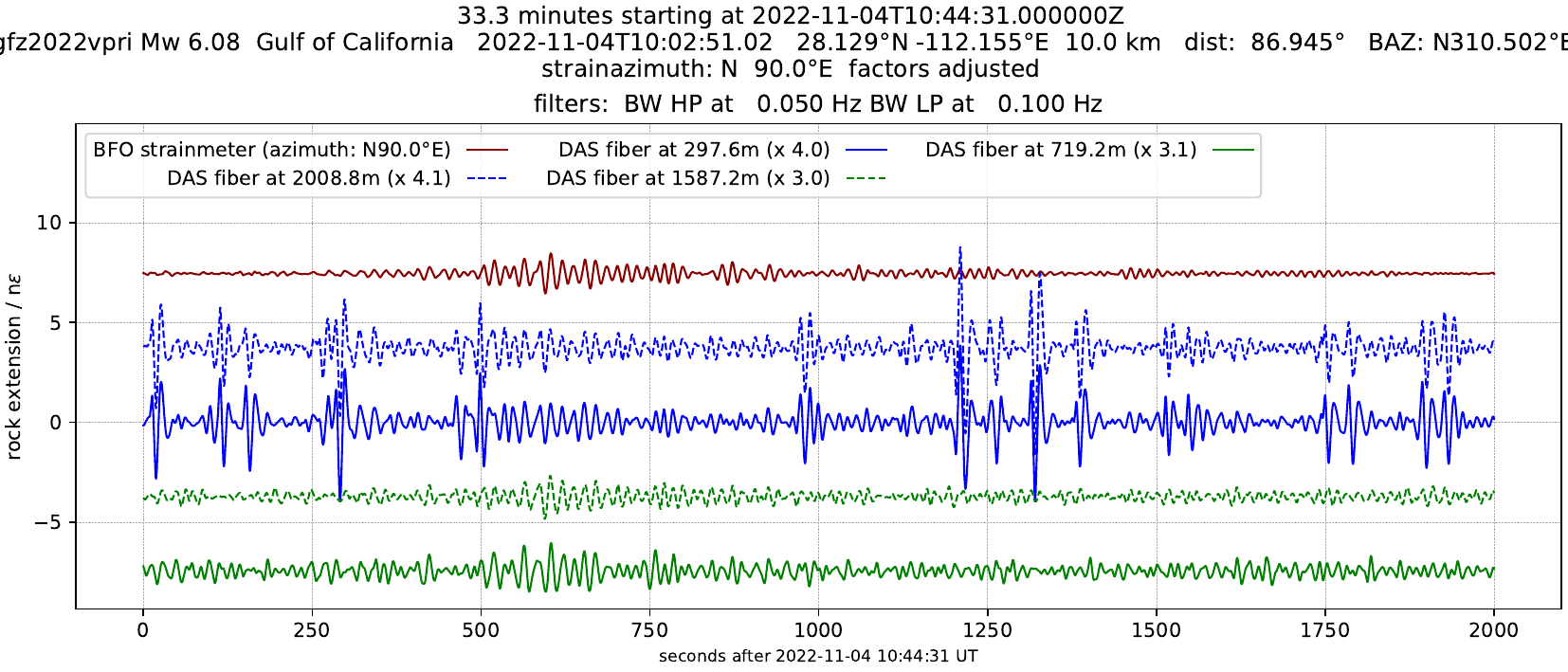}{gfzAvpri}
\surfacewavefigures{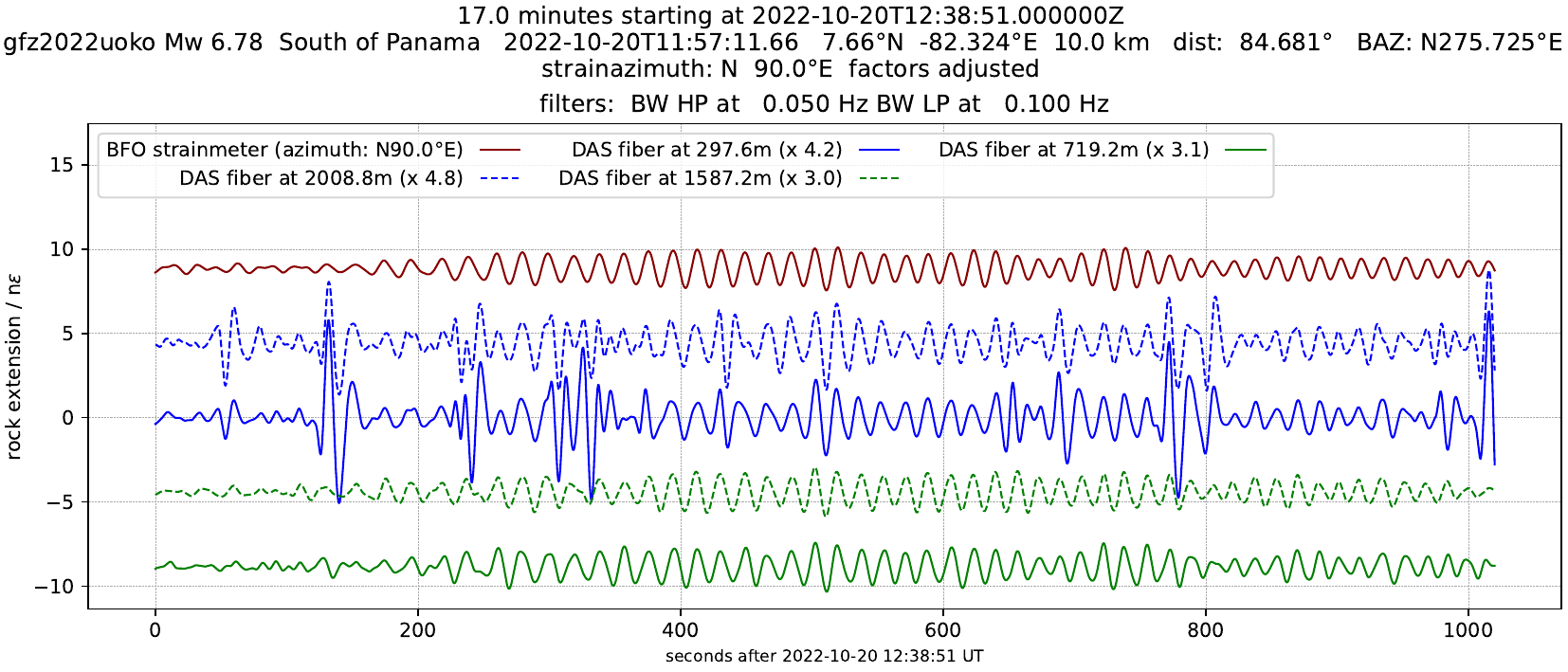}{gfzAuoko}
\surfacewavefigures{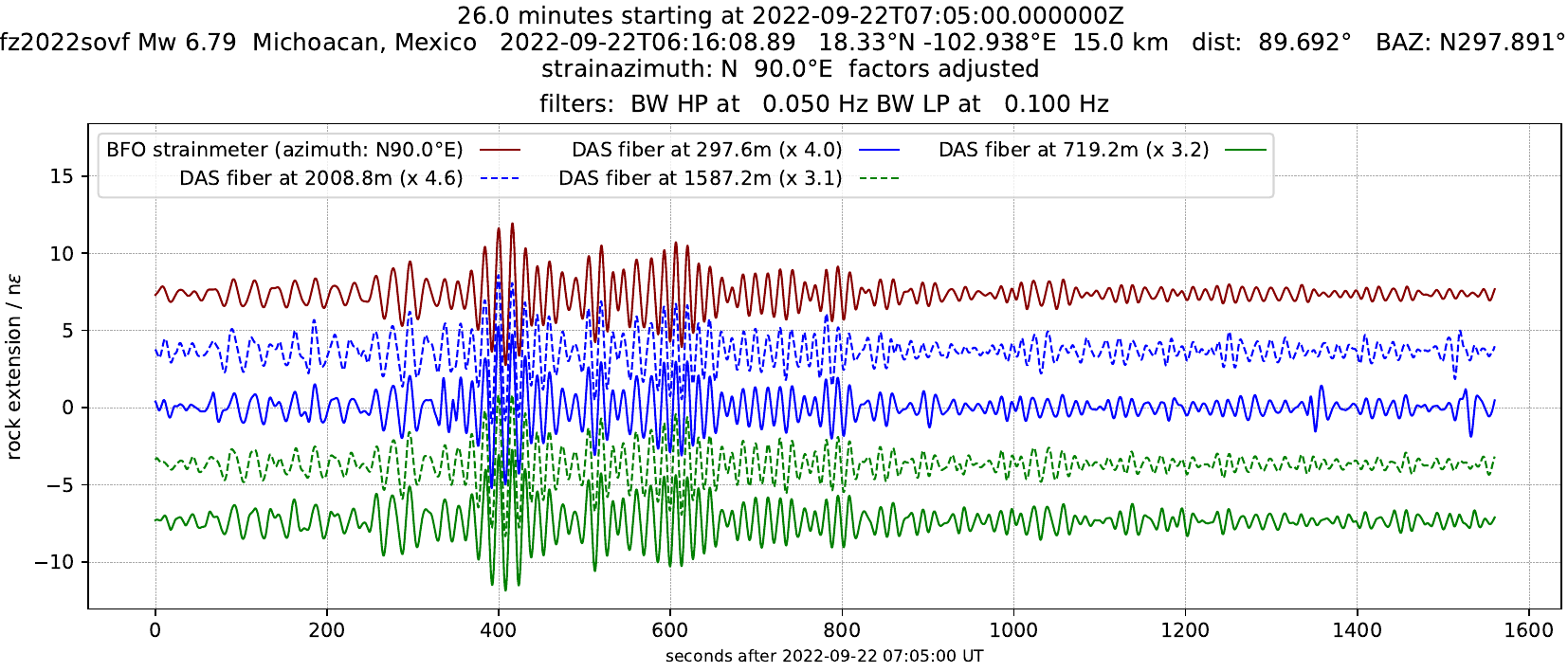}{gfzAsovf}
\surfacewavefigures{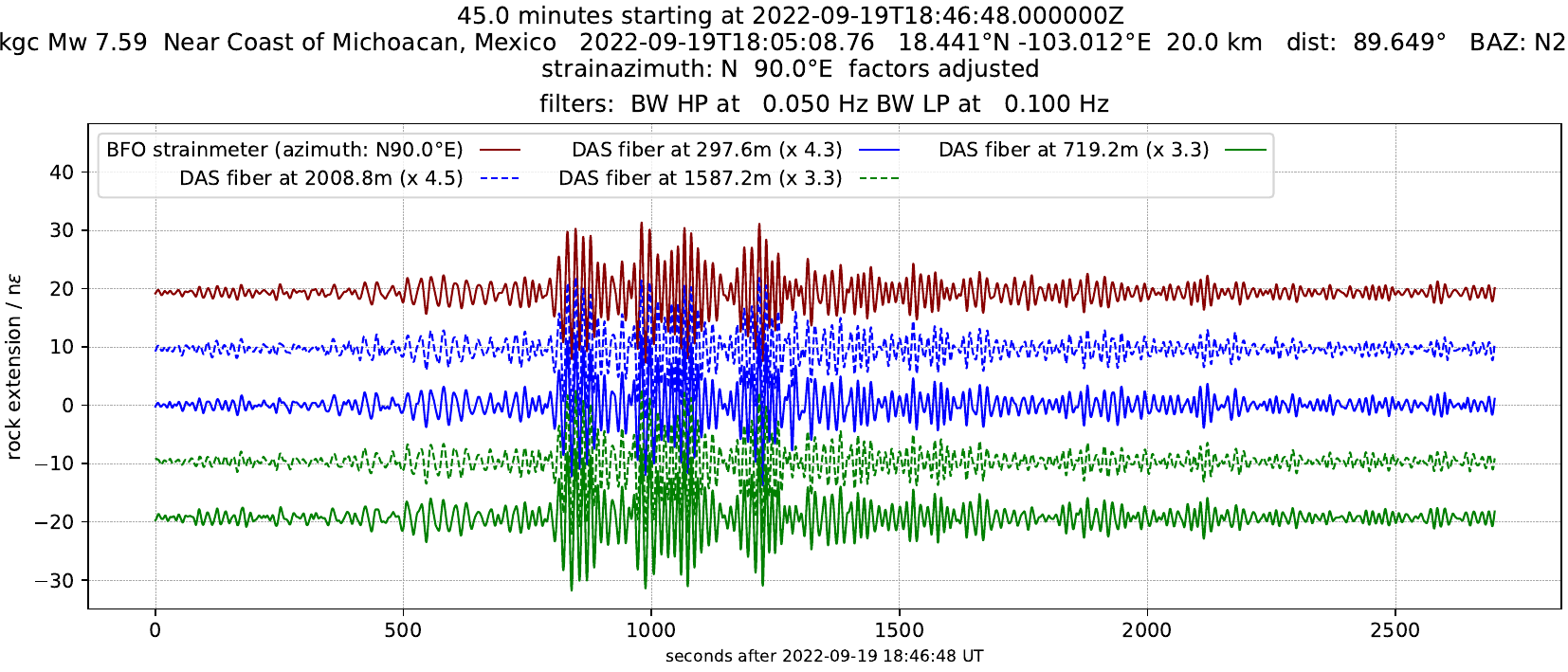}{gfzAskgc}
\surfacewavefigures{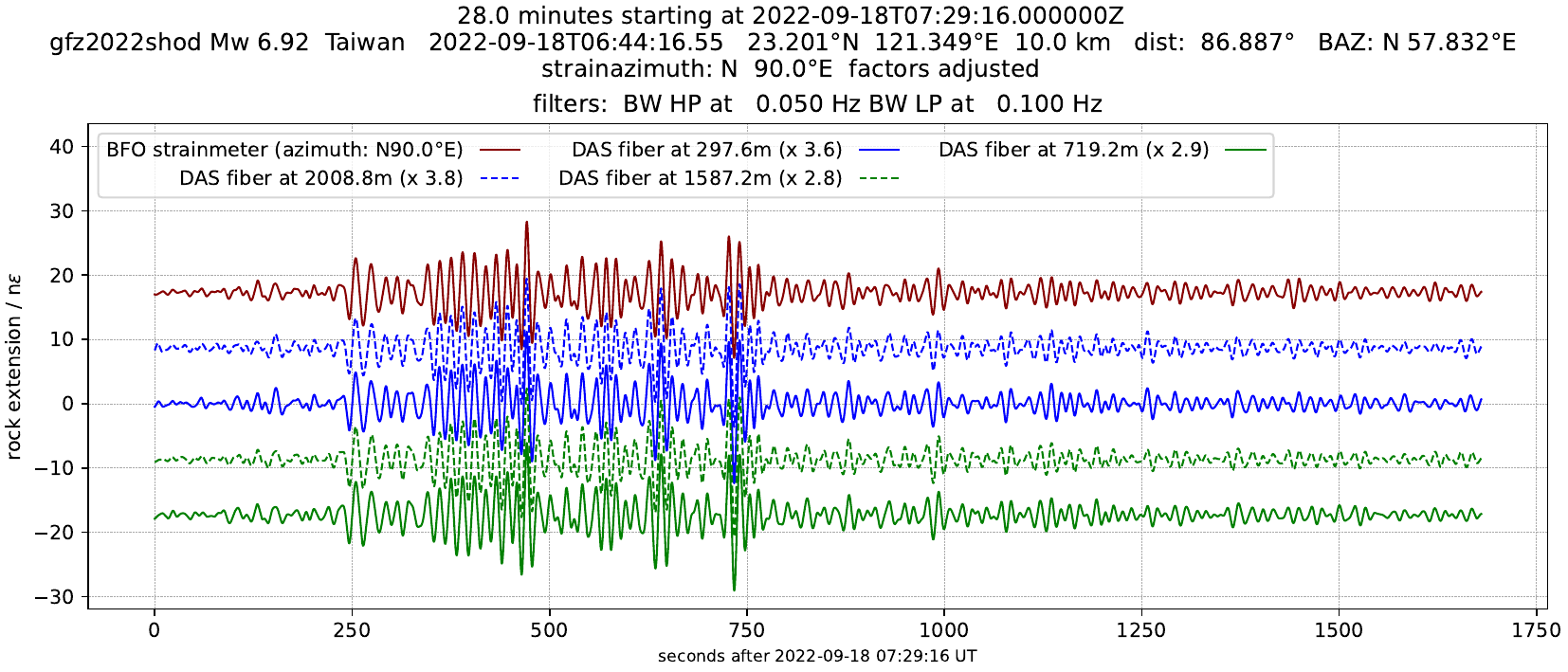}{gfzAshod}
\surfacewavefigures{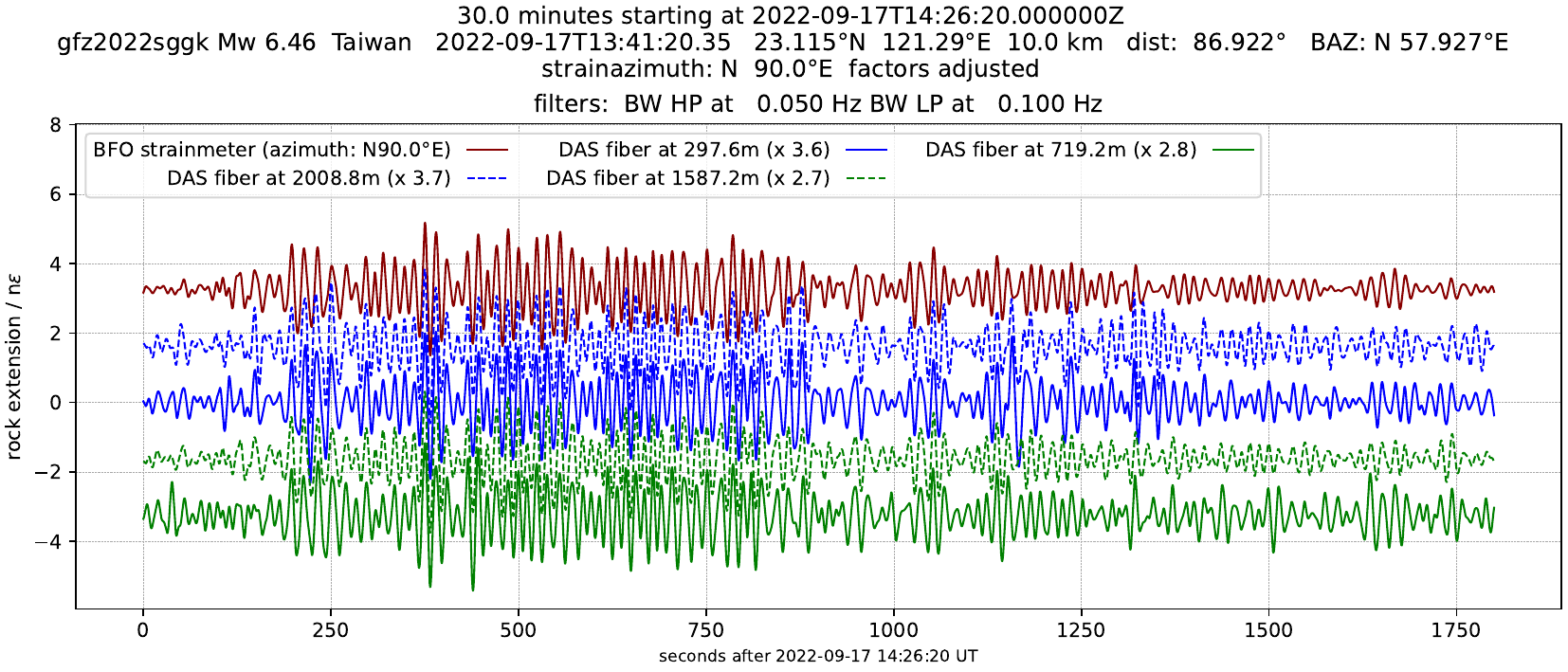}{gfzAsggk}
\surfacewavefigures{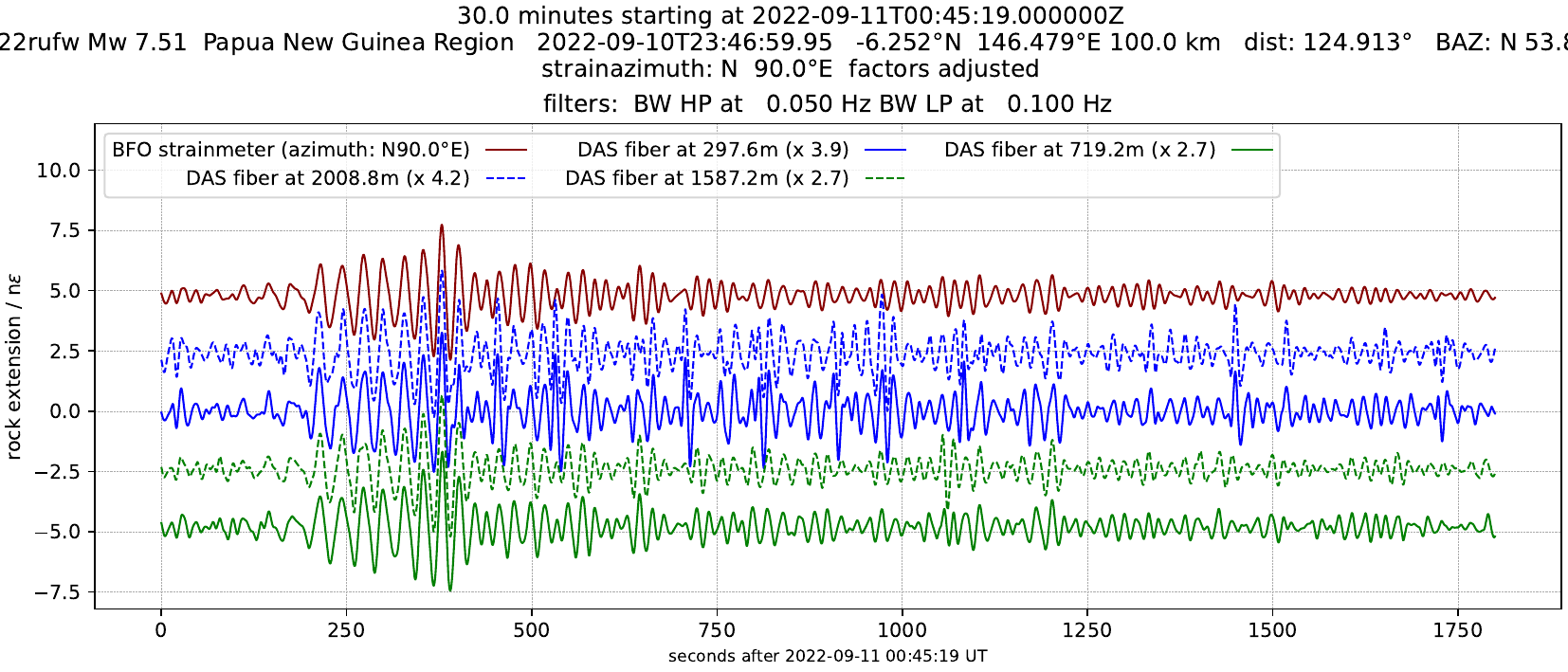}{gfzArufw}
\surfacewavefigures{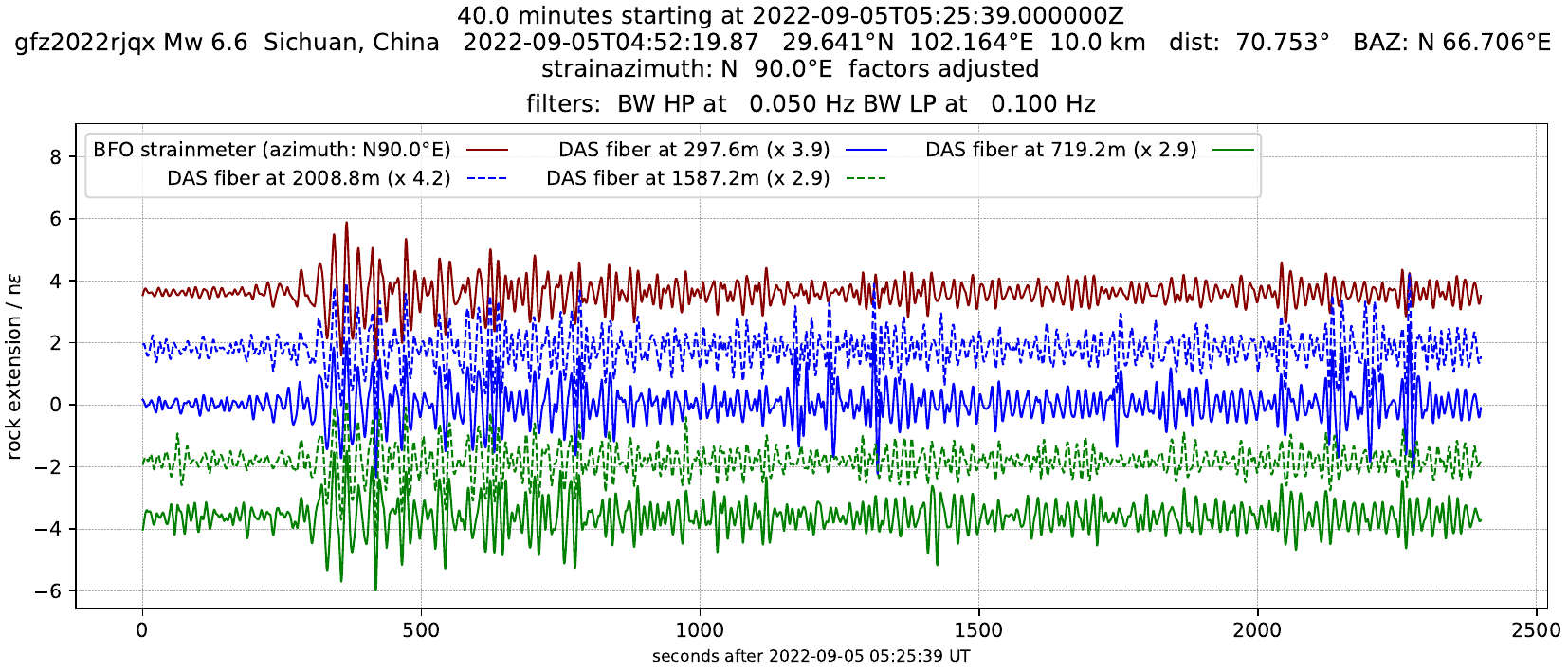}{gfzArjqx}
\surfacewavefigures{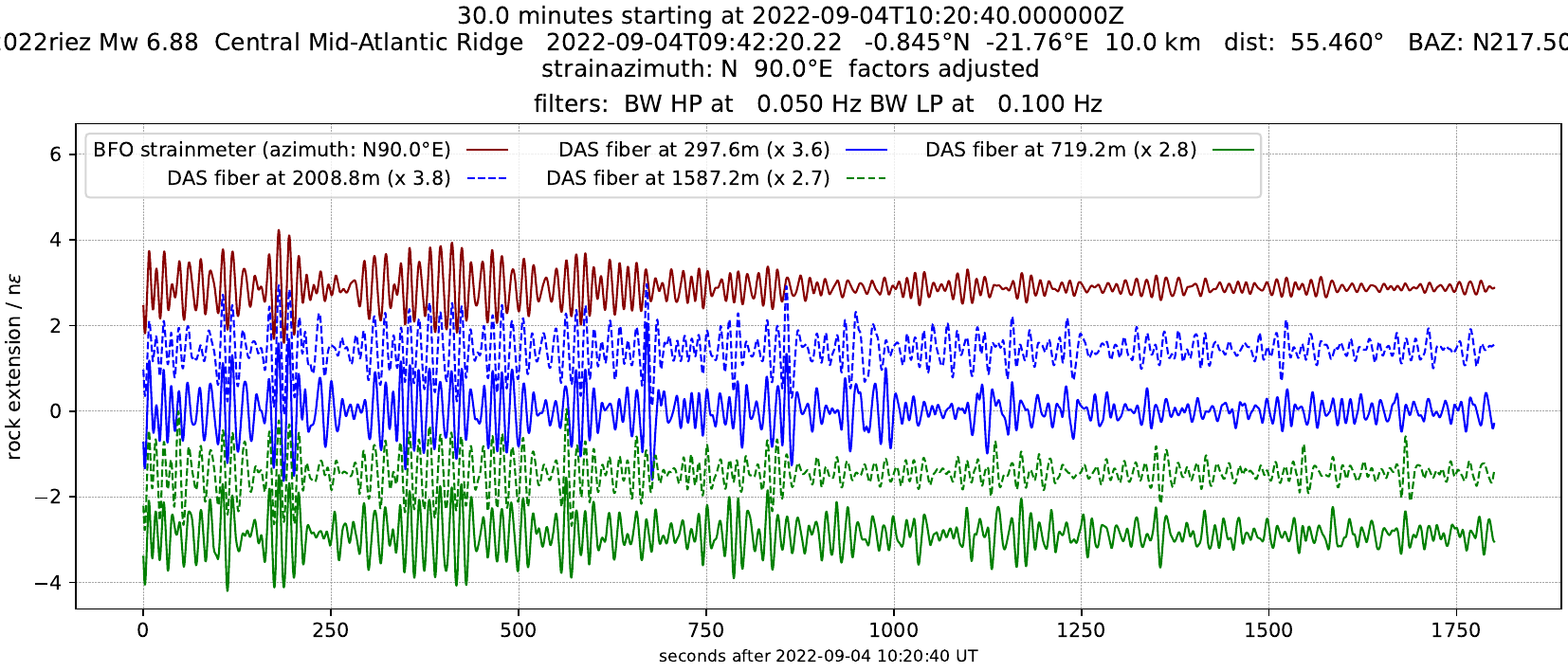}{gfzAriez}
\surfacewavefigures{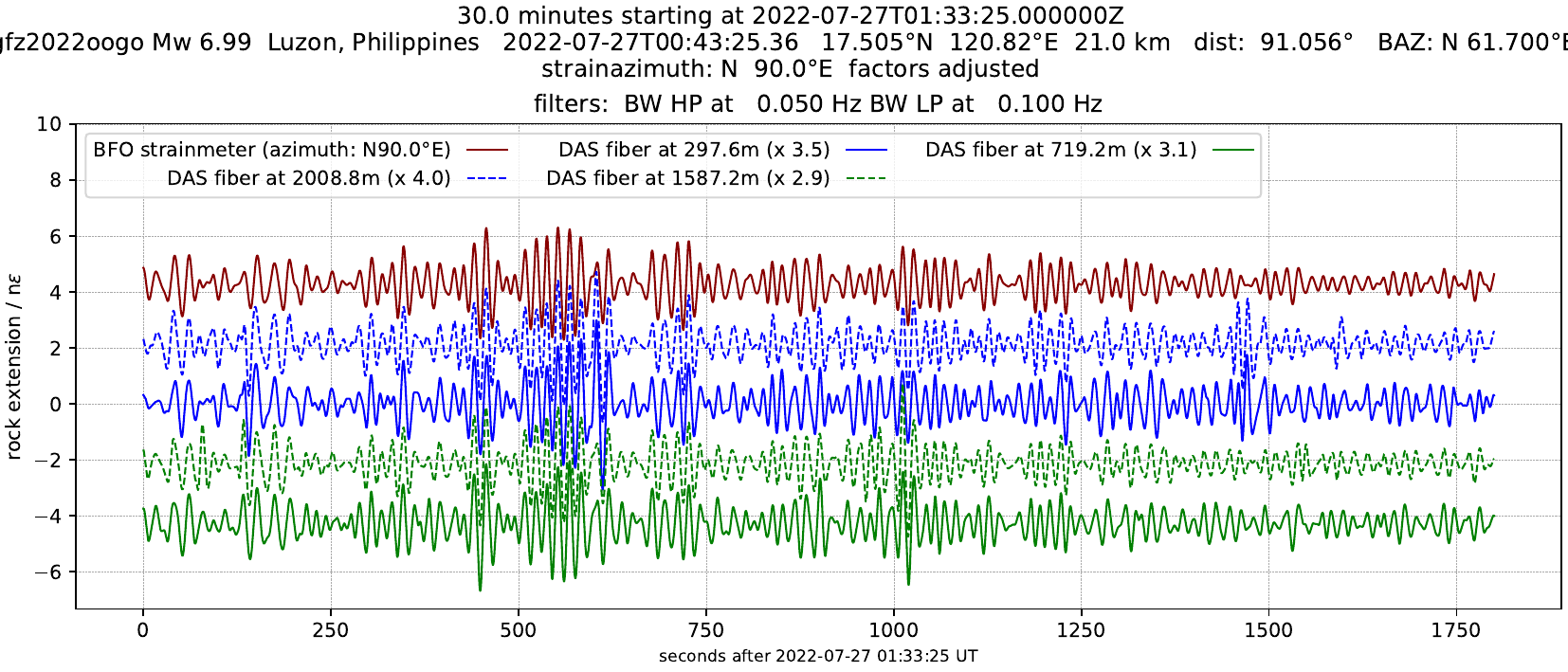}{gfzAoogo}


\begin{figure*}
{\includegraphics[clip,trim=10 10 10 50,width=\textwidth]{Turkey77_Anton_Pwaveform}}
{\includegraphics[clip,trim=10 10 10 50,width=\textwidth]{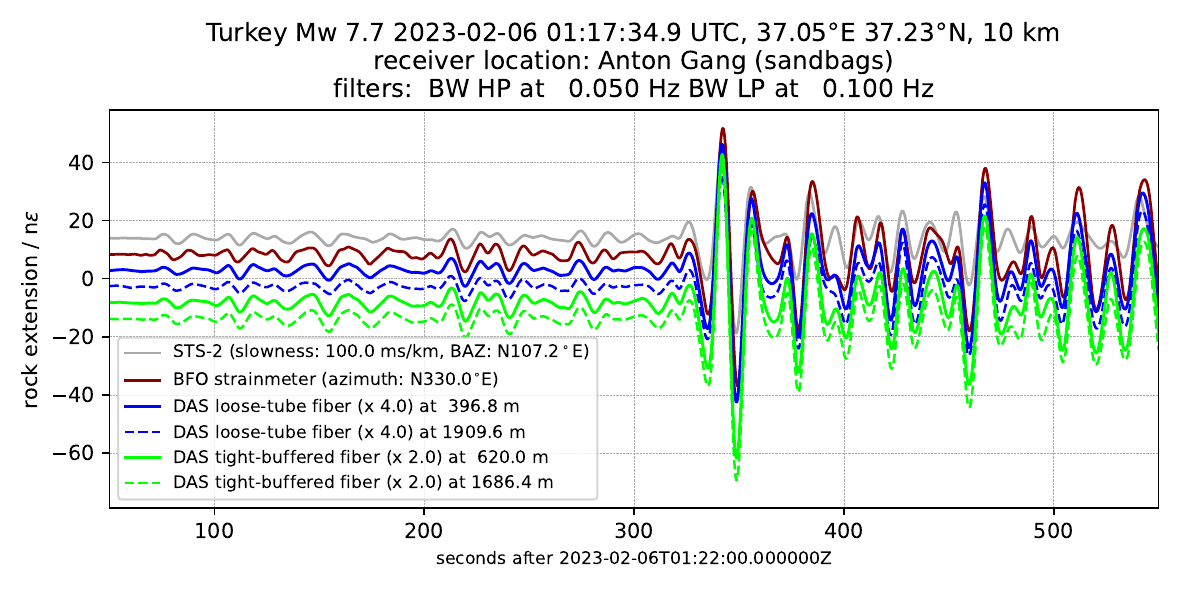}}
  \caption{Strain waveforms as recorded for the azimuth (\azimuth{330}) of the
  \Lanton\ for the first of the main shocks of the Kahramanmaraş earthquakes: 
  \gfzBcnwrEQpara.
  Top: P-waves, bottom: P- and S-waves.
  First arrival times according to raytracing (obspy.taup.TauPyModel) with
  IASP91 are:
  P: 2023-02-06 01:22:46.44 UT and
  S: 2023-02-06 01:27:22.57 UT.
  P-waves arrive with a minimum slowness of 
  $82.3\,\text{ms}\,\text{km}^{-1}$ and a maximum slowness of
  $94.2\,\text{ms}\,\text{km}^{-1}$.
  The value range for the S-waves is
  $146.0\,\text{ms}\,\text{km}^{-1}$ to
  $214.5\,\text{ms}\,\text{km}^{-1}$.
  The seismometer data are scaled by \eqref{eq:supp:plane:wave:strain}.
  A band-pass from 0.05~Hz to 0.1~Hz is applied to all signals.
  The seismometer data was equalized to this filter response.
  }
\label{fig:turkey:BWA:Anton}
\end{figure*}
\begin{figure*}
{\includegraphics[clip,trim=10 10 10 50,width=\textwidth]{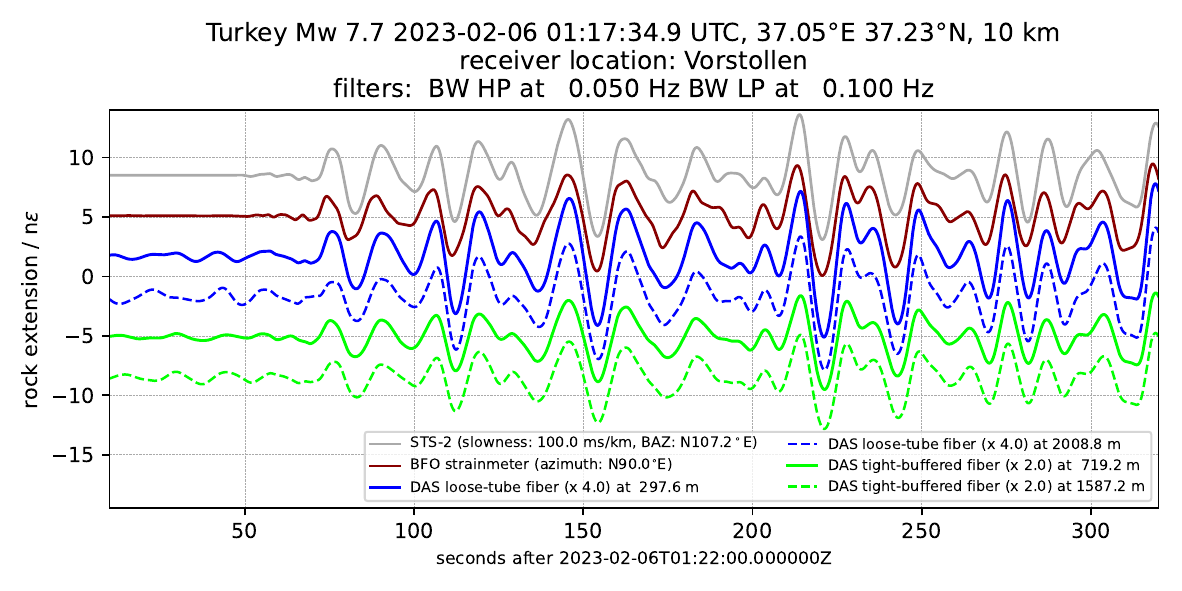}}
{\includegraphics[clip,trim=10 10 10 50,width=\textwidth]{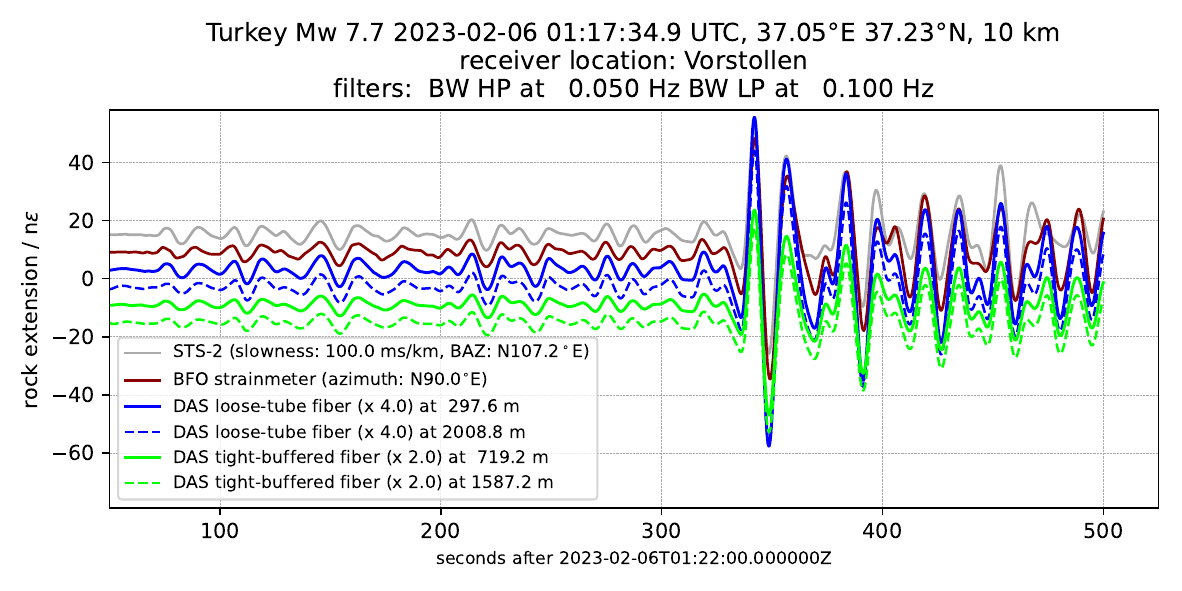}}
  \caption{Strain waveforms as recorded for the azimuth (\azimuth{90}) of the
  \Lvorstollen\ for the first of the main shocks of the Kahramanmaraş
  earthquakes: \gfzBcnwrEQpara.
  Top: P-waves, bottom: P- and S-waves.
  First arrival times according to raytracing (obspy.taup.TauPyModel) with
  IASP91 are:
  P: 2023-02-06 01:22:46.44 UT and
  S: 2023-02-06 01:27:22.57 UT.
  P-waves arrive with a minimum slowness of 
  $82.3\,\text{ms}\,\text{km}^{-1}$ and a maximum slowness of
  $94.2\,\text{ms}\,\text{km}^{-1}$.
  The value range for the S-waves is
  $146.0\,\text{ms}\,\text{km}^{-1}$ to
  $214.5\,\text{ms}\,\text{km}^{-1}$.
  The seismometer data are scaled by \eqref{eq:supp:plane:wave:strain}.
  A band-pass from 0.05~Hz to 0.1~Hz is applied to all signals.
  The seismometer data was equalized to this filter response.
  }
\label{fig:turkey:BWA:Vorstollen}
\end{figure*}
\begin{figure*}
{\includegraphics[clip,trim=10 10 10 50,width=\textwidth]{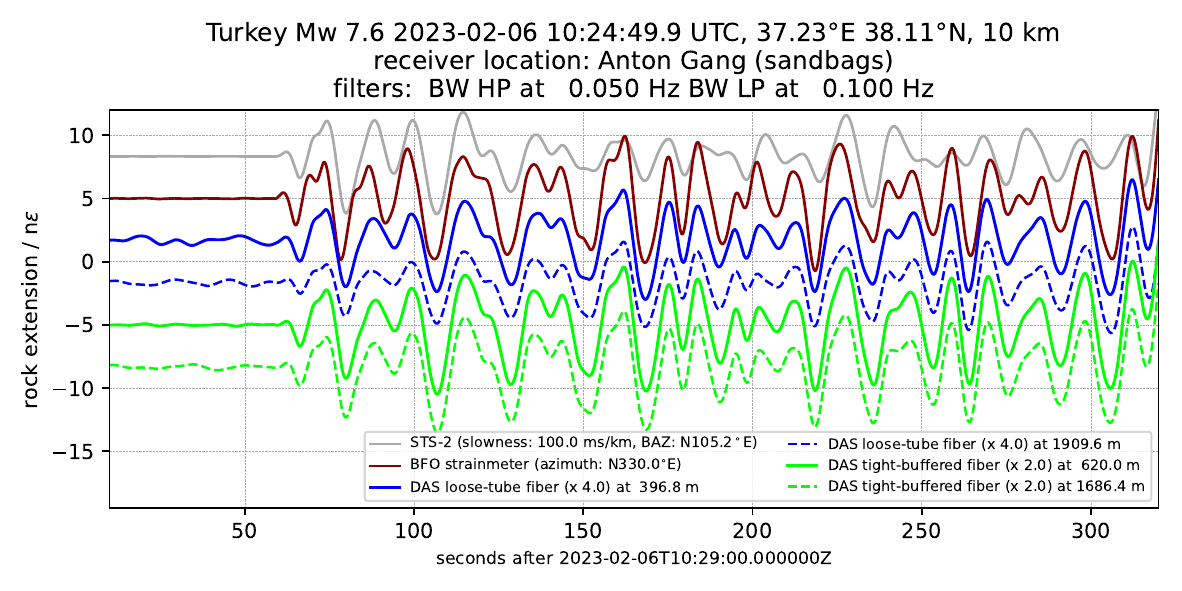}}
{\includegraphics[clip,trim=10 10 10 50,width=\textwidth]{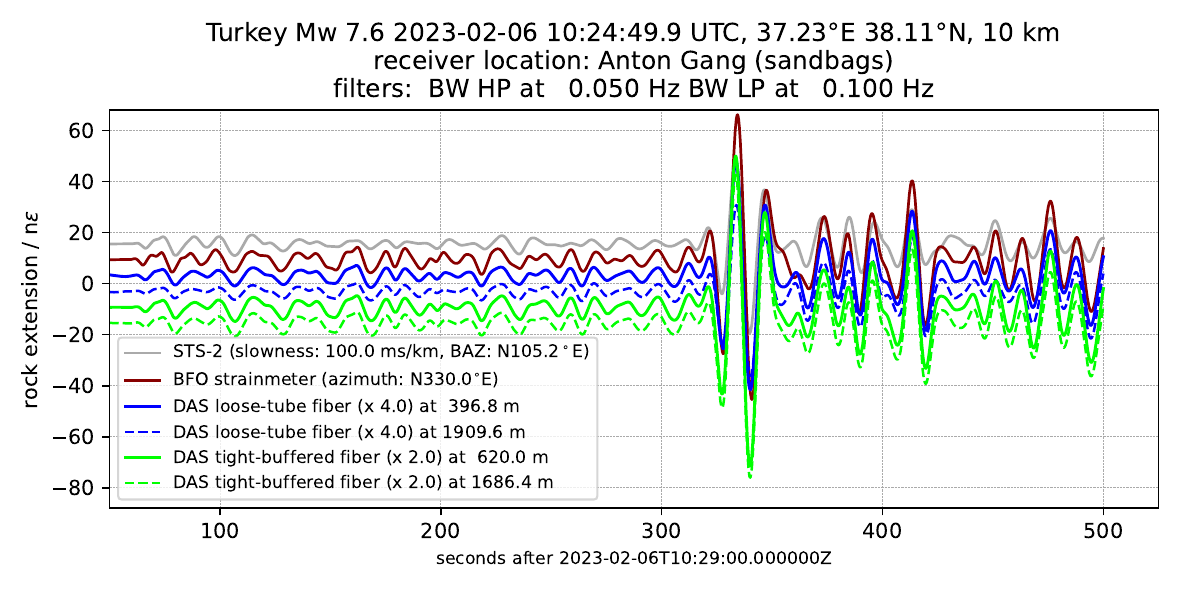}}
  \caption{Strain waveforms as recorded for the azimuth (\azimuth{330}) of the
  \Lanton\ for the first of the main shocks of the Kahramanmaraş
  earthquakes: \gfzBcoosEQpara.
  Top: P-waves, bottom: P- and S-waves.
  First arrival times according to raytracing (obspy.taup.TauPyModel) with
  IASP91 are:
  P: 2023-02-06 10:29:57.35 UT and
  S: 2023-02-06 10:34:27.98 UT.
  P-waves arrive with a minimum slowness of 
  $82.4\,\text{ms}\,\text{km}^{-1}$ and a maximum slowness of
  $94.7\,\text{ms}\,\text{km}^{-1}$.
  The value range for the S-waves is
  $146.7\,\text{ms}\,\text{km}^{-1}$ to
  $214.7\,\text{ms}\,\text{km}^{-1}$.
  The seismometer data are scaled by \eqref{eq:supp:plane:wave:strain}.
  A band-pass from 0.05~Hz to 0.1~Hz is applied to all signals.
  The seismometer data was equalized to this filter response.
  }
\label{fig:turkey:BWB:Anton}
\end{figure*}
\begin{figure*}
{\includegraphics[clip,trim=10 10 10 50,width=\textwidth]{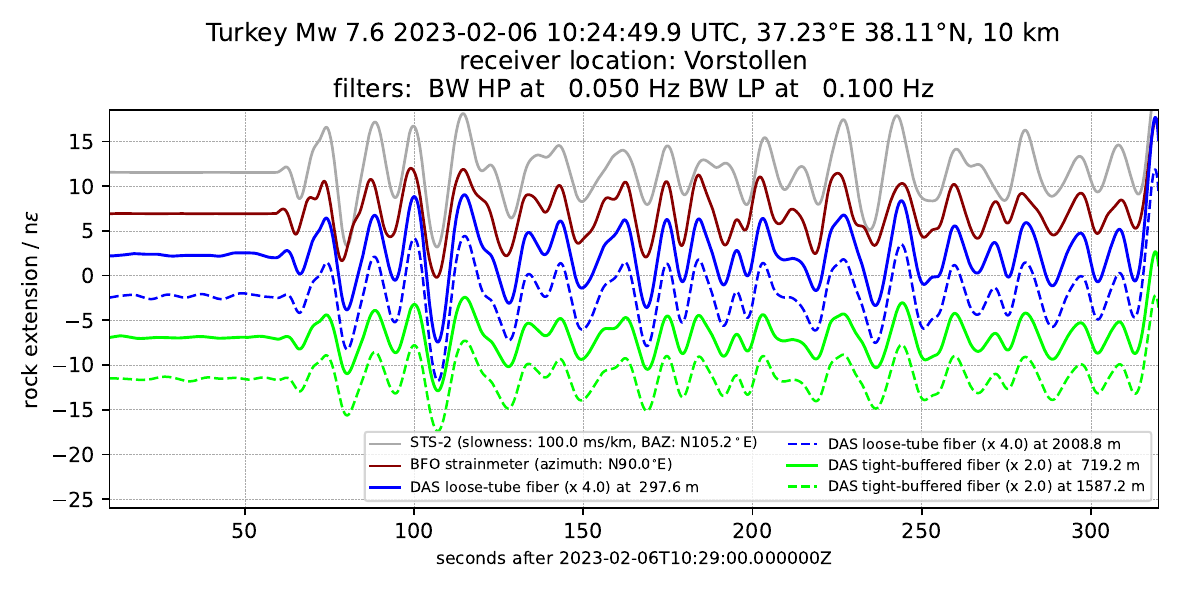}}
{\includegraphics[clip,trim=10 10 10 50,width=\textwidth]{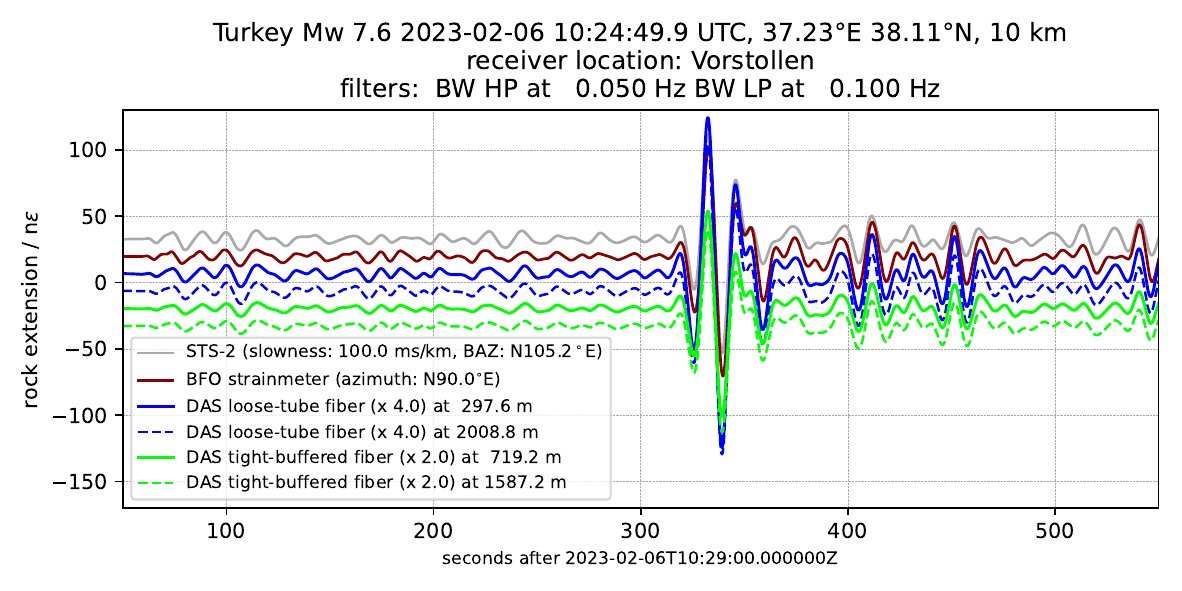}}
  \caption{Strain waveforms as recorded for the azimuth (\azimuth{90}) of the
  \Lvorstollen\ for the first of the main shocks of the Kahramanmaraş
  earthquakes: \gfzBcoosEQpara.
  Top: P-waves, bottom: P- and S-waves.
  First arrival times according to raytracing (obspy.taup.TauPyModel) with
  IASP91 are:
  P: 2023-02-06 10:29:57.35 UT and
  S: 2023-02-06 10:34:27.98 UT.
  P-waves arrive with a minimum slowness of 
  $82.4\,\text{ms}\,\text{km}^{-1}$ and a maximum slowness of
  $94.7\,\text{ms}\,\text{km}^{-1}$.
  The value range for the S-waves is
  $146.7\,\text{ms}\,\text{km}^{-1}$ to
  $214.7\,\text{ms}\,\text{km}^{-1}$.
  The seismometer data are scaled by \eqref{eq:supp:plane:wave:strain}.
  A band-pass from 0.05~Hz to 0.1~Hz is applied to all signals.
  The seismometer data was equalized to this filter response.
  }
\label{fig:turkey:BWB:Vorstollen}
\end{figure*}
\begin{figure*}
  \begin{center}
     \includegraphics[trim=0 0 0 24,clip,width=0.85\textwidth]{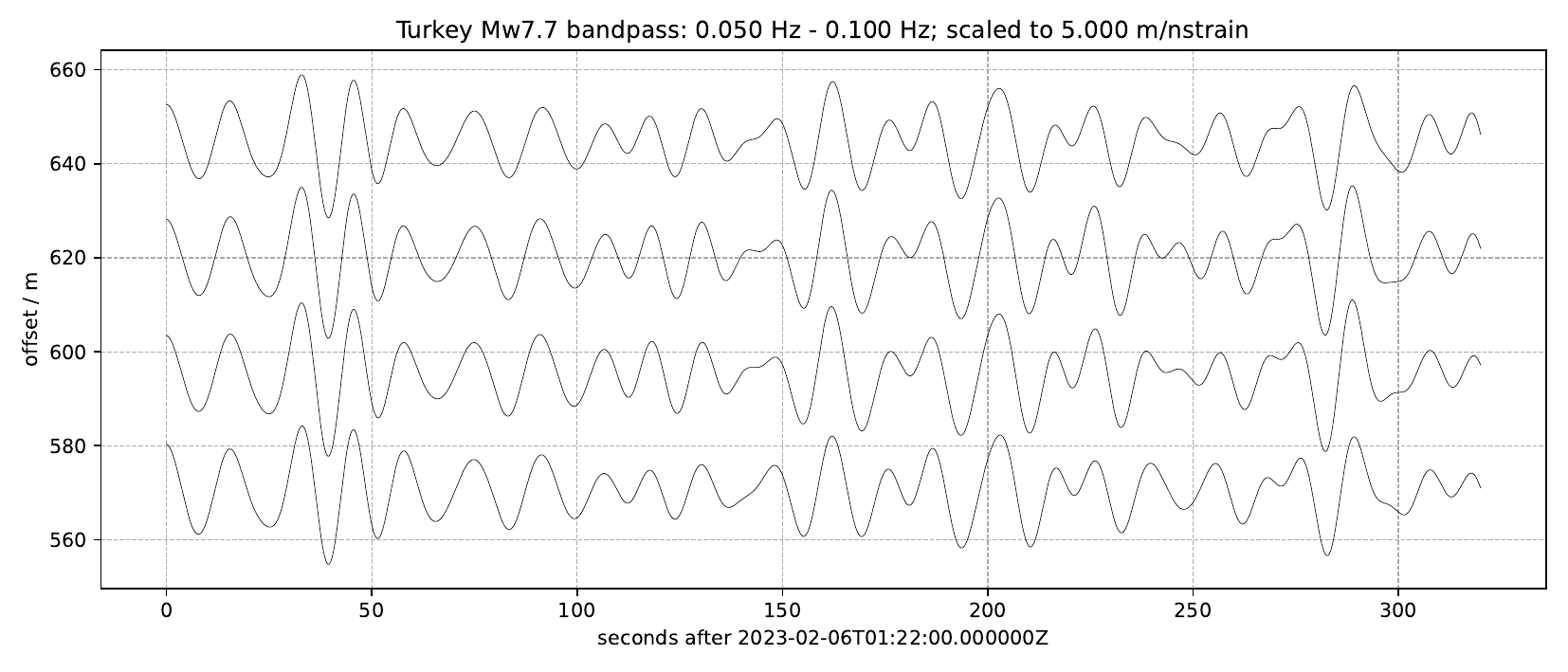}
     \par
     \includegraphics[trim=0 0 0 24,clip,width=0.85\textwidth]{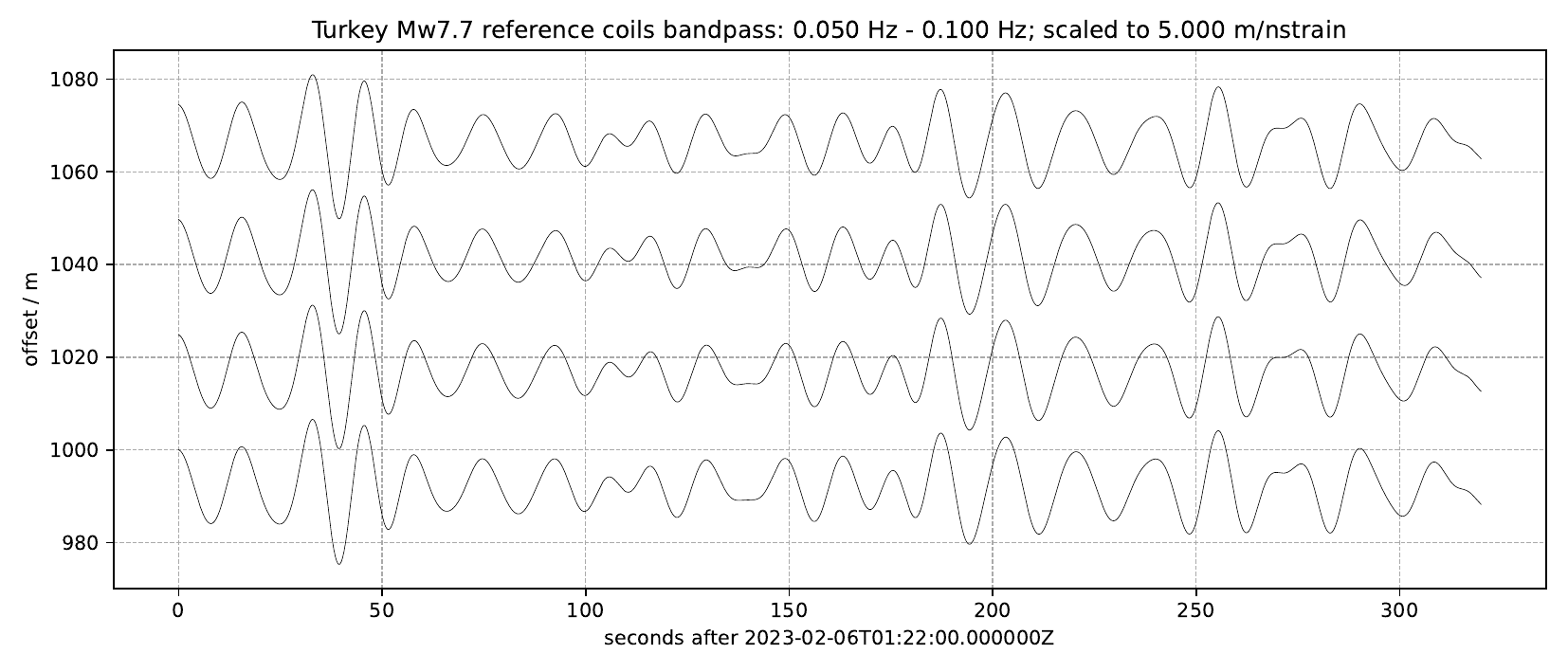}
     \par
     \includegraphics[trim=0 0 0 24,clip,width=0.85\textwidth]{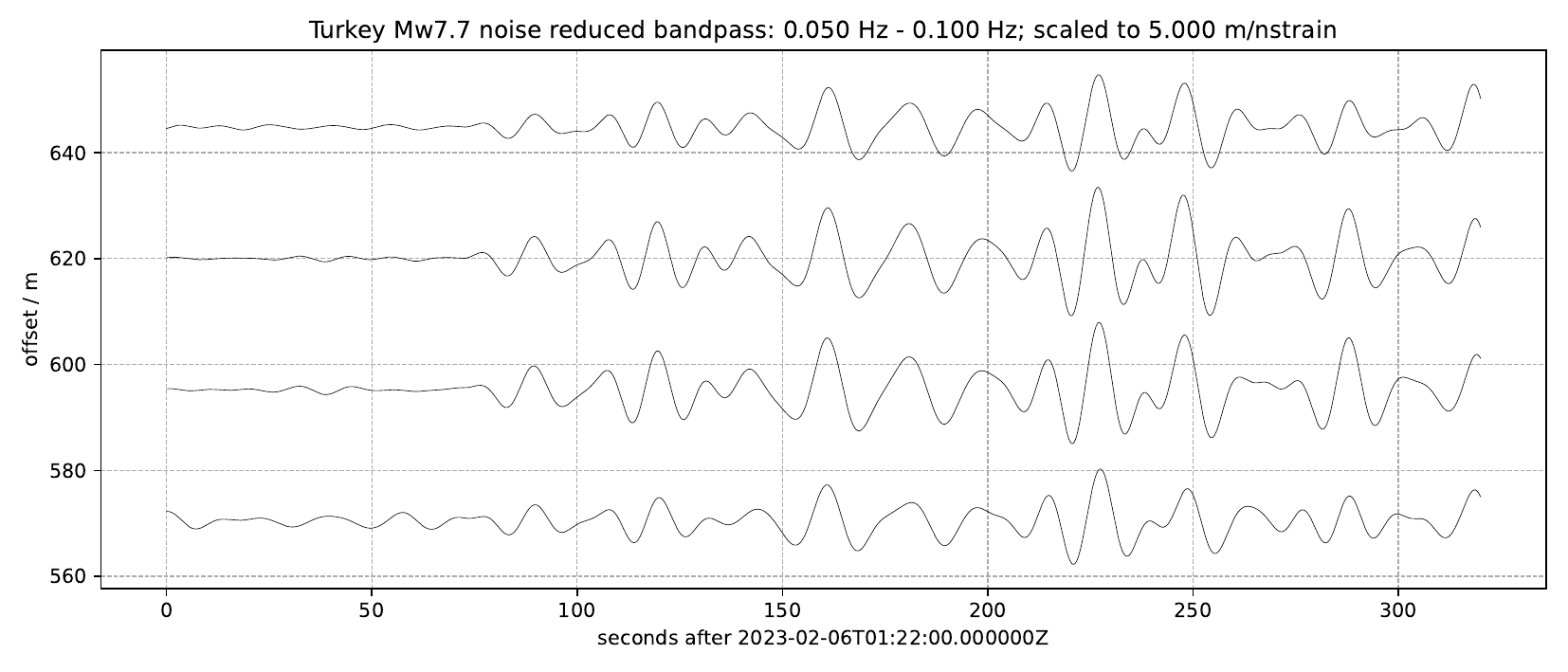}
     \par
  \end{center}
  \caption{Demonstration of the reduction of coherent (common-mode) noise.
    The P-waves from the Mw~7.7
    Pazarcık earthquake arrive in the displayed time window.
    All signals are filtered to the frequency band of 0.05~Hz to 0.1~Hz
    and are scaled to the same 5~m/nstrain.
    Top: 
      recorded waveforms for channels between 570~m and 670~m.
    Center:
      recorded waveforms for channels between 1000~m and 1100~m (on the
        reference drum).
    Bottom:
      recorded waveforms for channels between 570~m and 670~m after the
        average of the recorded waveforms between 1000~m and 1100~m has been
        removed.
    The channel at 620\,m is one of the two channels from the tight-buffered
    cable in the \Lanton.\\
    Only after subtracting the signal from the reference coil, the P-wave
    arrival (at 70\,s on the time scale) becomes apparent.
  }
  \label{fig:noisered:P}
\end{figure*}
\begin{figure*}
  \begin{center}
     \includegraphics[trim=0 0 0 24,clip,width=0.85\textwidth]{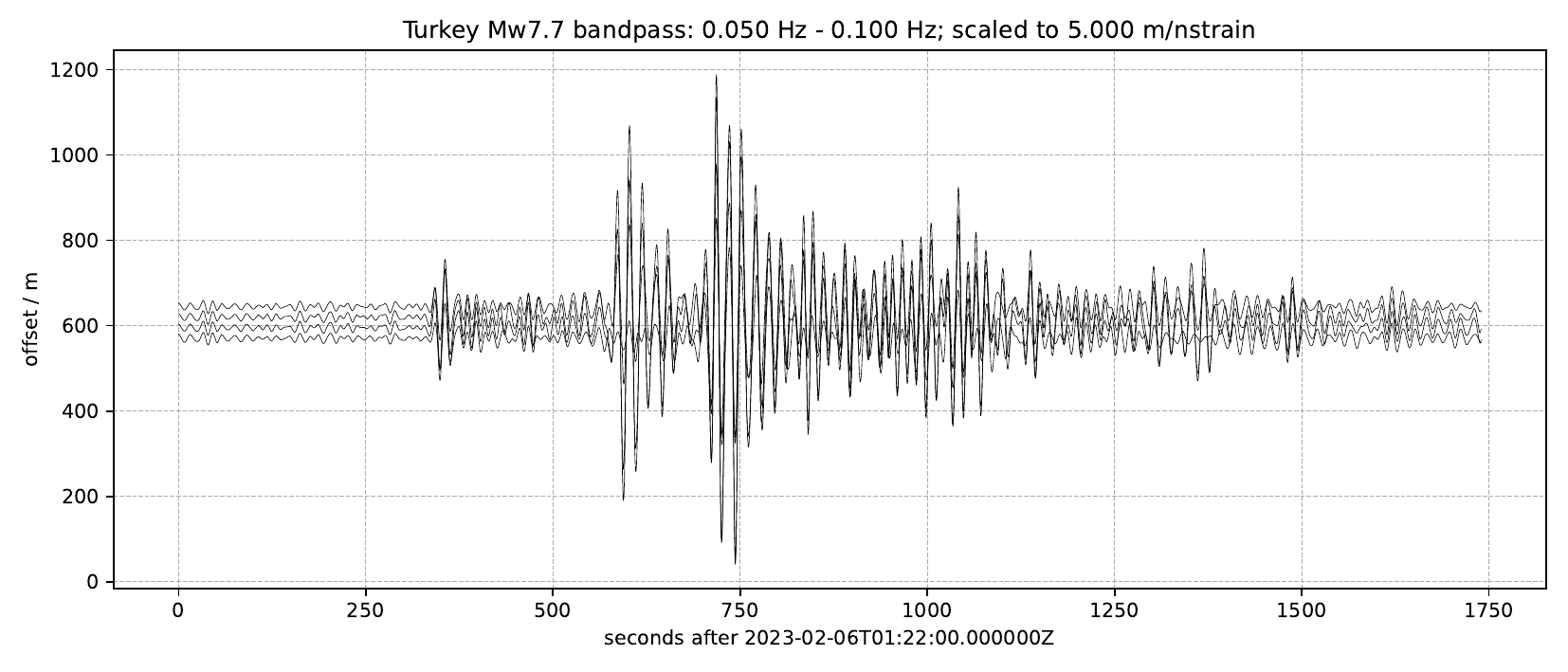}
     \par
     \includegraphics[trim=0 0 0 24,clip,width=0.85\textwidth]{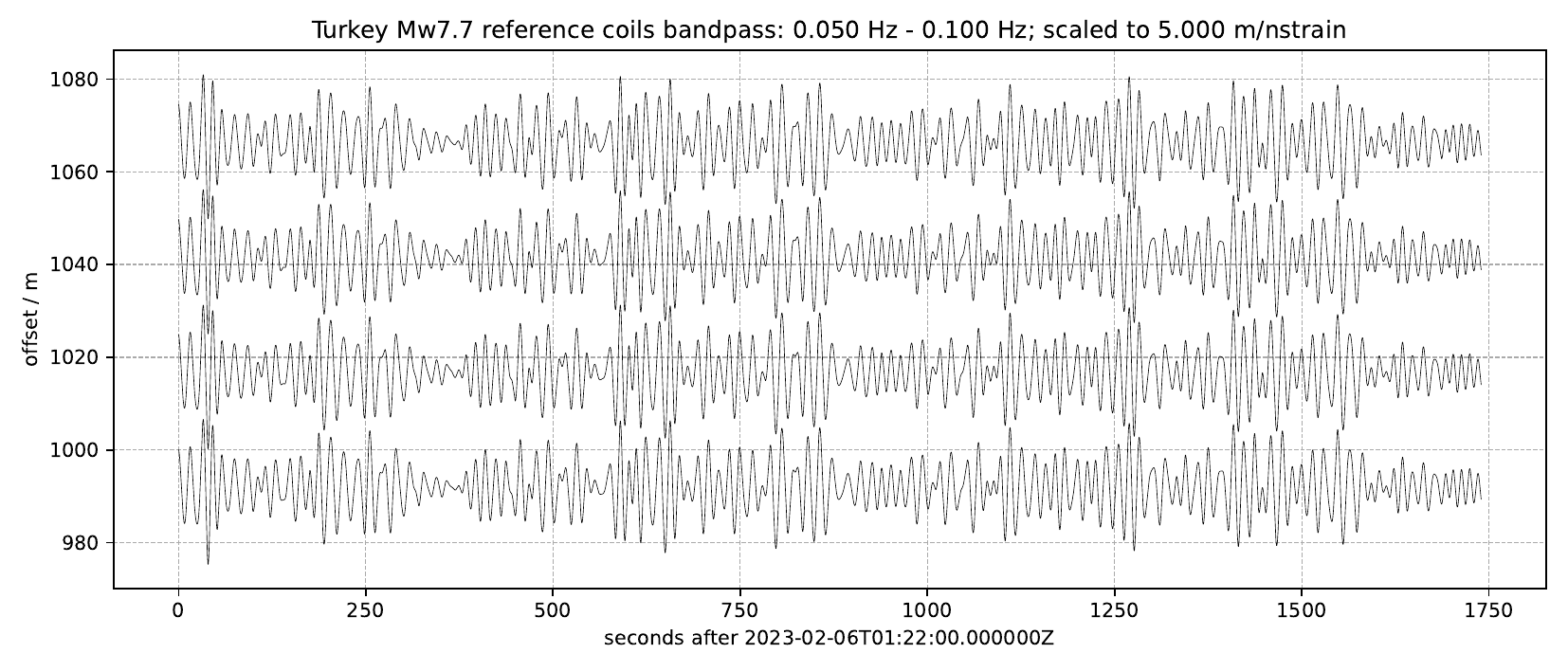}
     \par
     \includegraphics[trim=0 0 0 24,clip,width=0.85\textwidth]{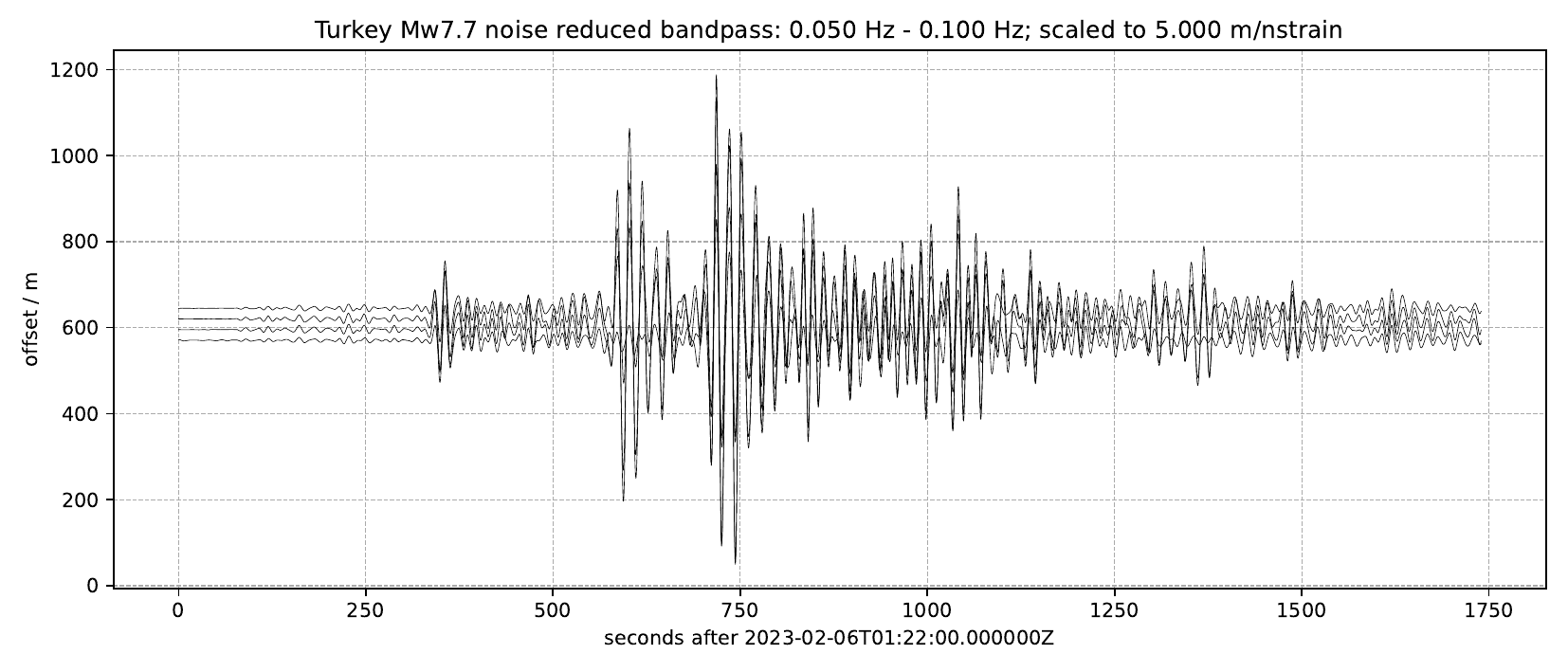}
     \par
  \end{center}
  \caption{Demonstration of the reduction of coherent (common-mode) noise.
    The wave-train (including P-, S-, and surface waves) of the Mw~7.7
    Pazarcık earthquake arrives in the displayed time window.
    All signals are filtered to the frequency band of 0.05~Hz to 0.1~Hz
    and are scaled to the same 5~m/nstrain, while the total axes
    range differs for the three panels.
    Top: 
      recorded waveforms for channels between 570~m and 670~m.
    Center:
      recorded waveforms for channels between 1000~m and 1100~m (on the
        reference drum).
    Bottom:
      recorded waveforms for channels between 570~m and 670~m after the
        average of the recorded waveforms between 1000~m and 1100~m has been
        removed.
    The channel at 620\,m is one of the two channels from the tight-buffered
    cable in the \Lanton.\\
    The surface waves of the Mw~7.7 Pazarcık earthquake are the largest
    amplitude strain signals in the entire analyzed data.
    No signature of this earthquake signal, 
    which has a signal-to-noise ratio of about 60
    with respect to the coherent noise in the raw recording shown 
    in the top panel, is apparent in the recording from the reference drum.
  }
  \label{fig:noisered:all}
\end{figure*}

\clearpage
\begin{multicols}{2}
\section{Additional diagrams to complement the main results}
\label{sec:extended:results}
\begin{figure*}
  \begin{tabular}{rr}
  {\includegraphics[trim=60 60 40 90,clip,width=0.48\textwidth]{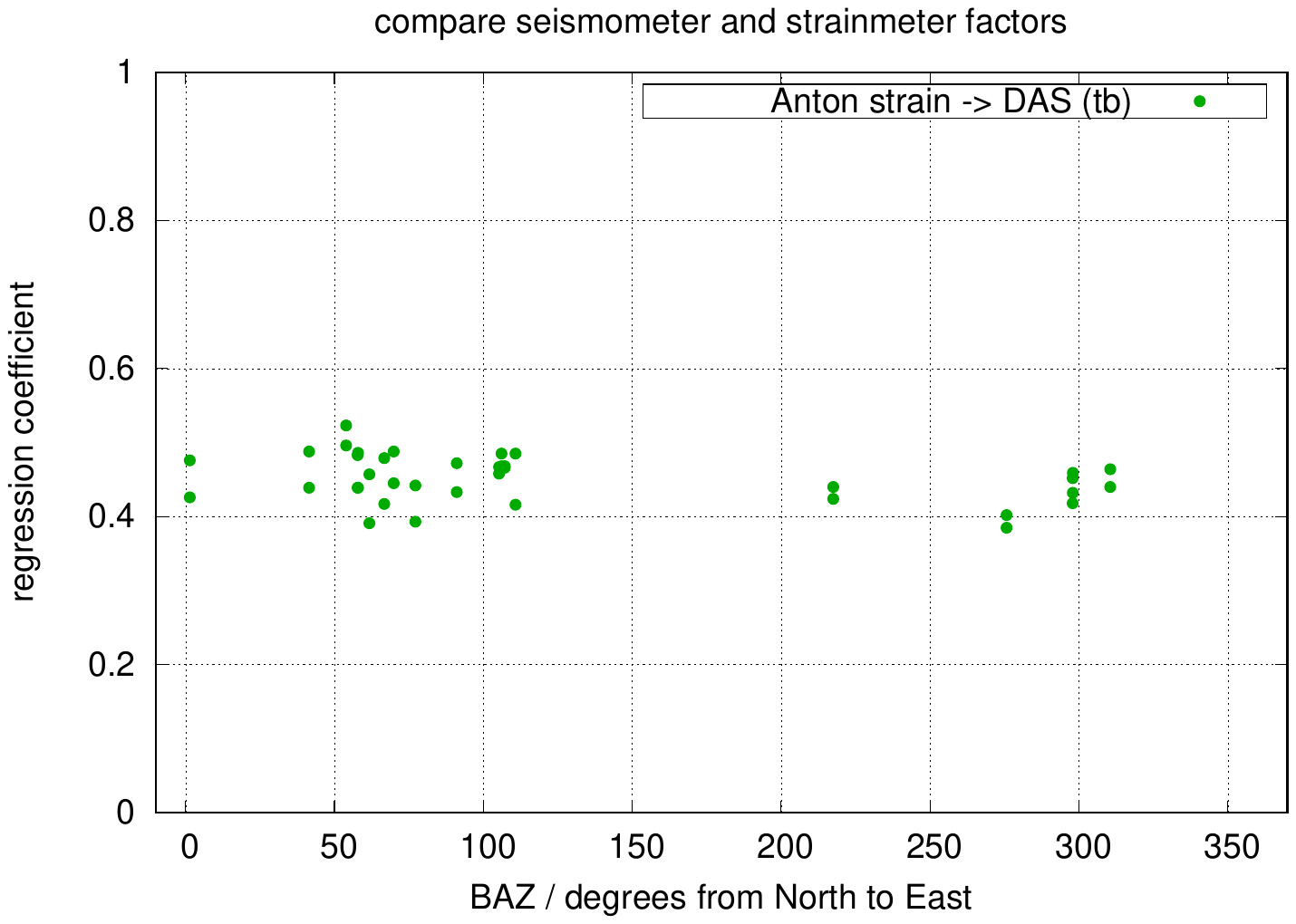}}
    &
  {\includegraphics[trim=60 60 40 90,clip,width=0.48\textwidth]{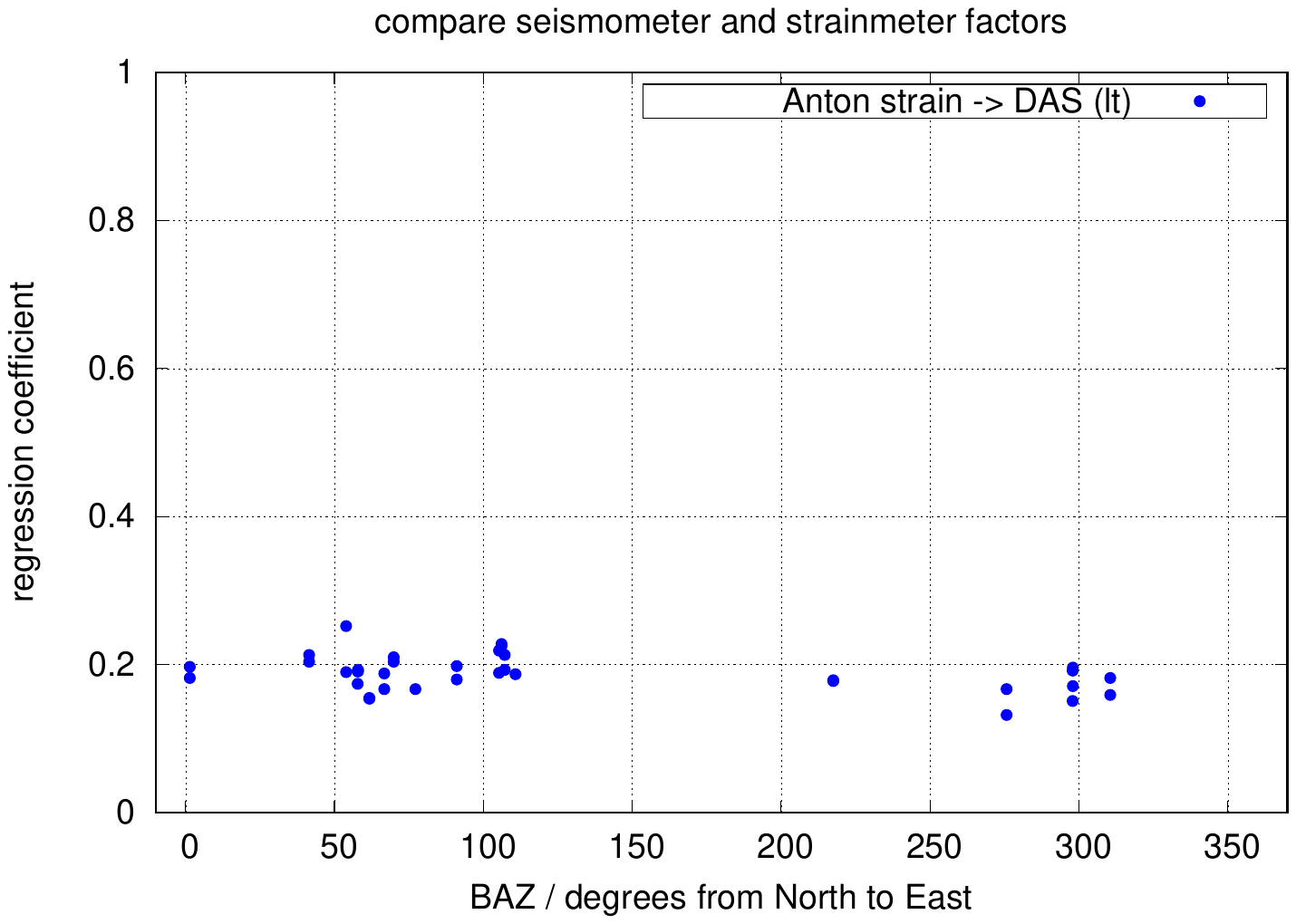}}\\
  {\includegraphics[trim=60 60 40 90,clip,width=0.48\textwidth]{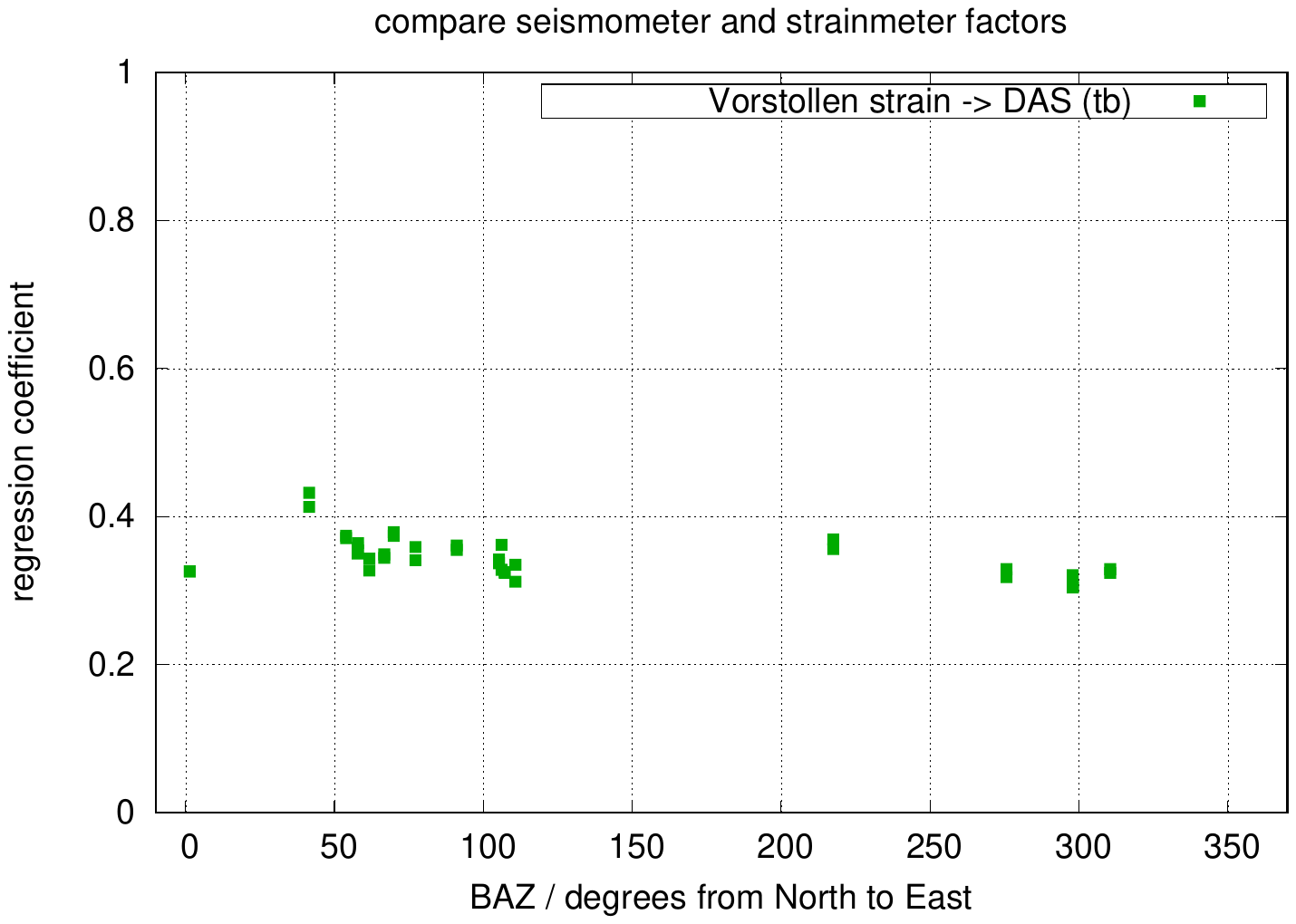}}
    &
  {\includegraphics[trim=60 60 40 90,clip,width=0.48\textwidth]{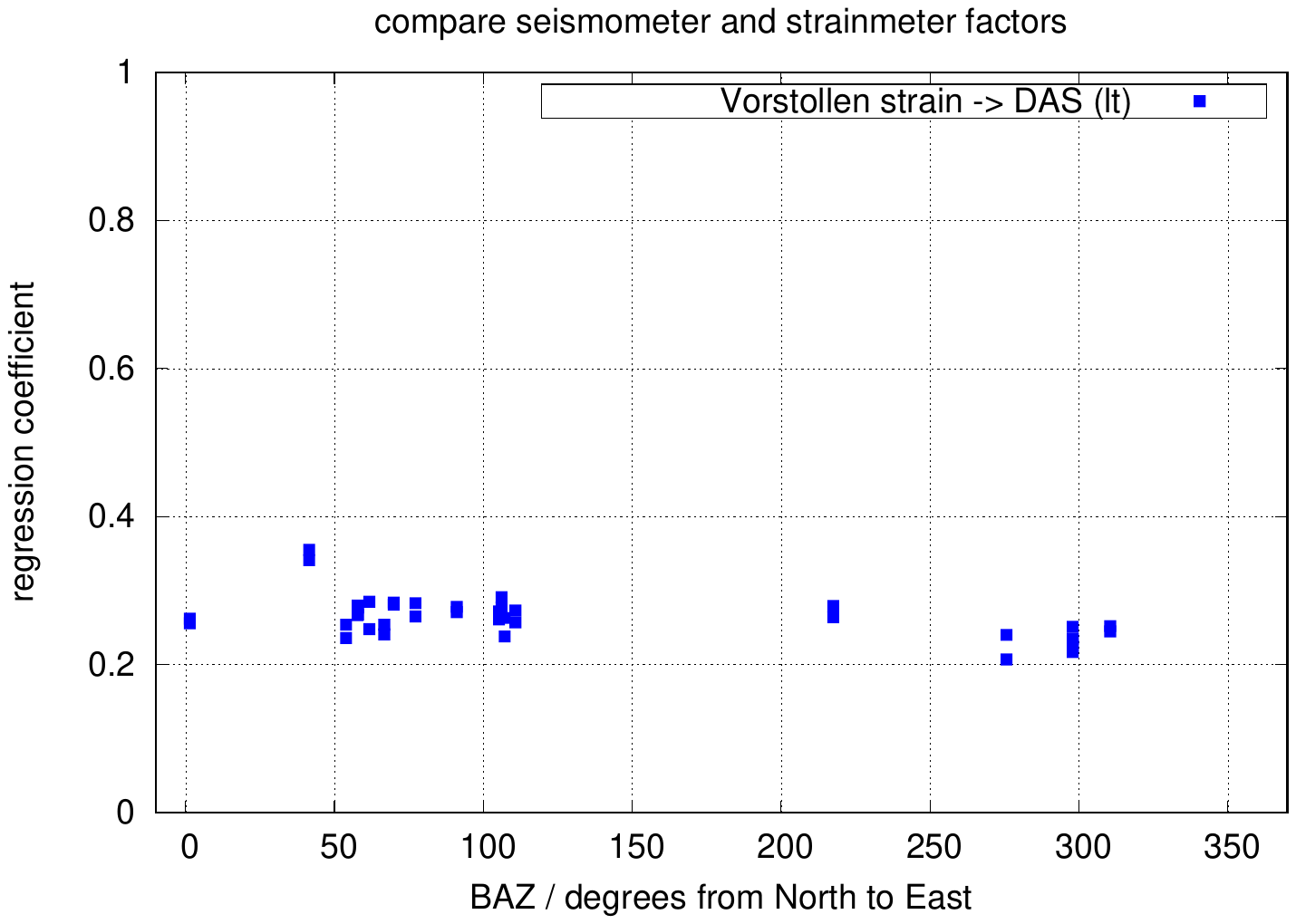}}
  \end{tabular}
  \caption{\Tstraintransfer\
  (regression coefficients for fitting strainmeter data to the DAS data)
  plotted against the backazimuth.
  Results for both cables (left: tight-buffered, right: loose-tube) and both
  locations (top: \Lanton, bottom: \Lvorstollen) are displayed.
  }
  \label{fig:str:vs:baz}
\end{figure*}

\begin{figure*}
  \begin{center}
    {\includegraphics[clip,trim=50 60 50 95,width=\textwidth]{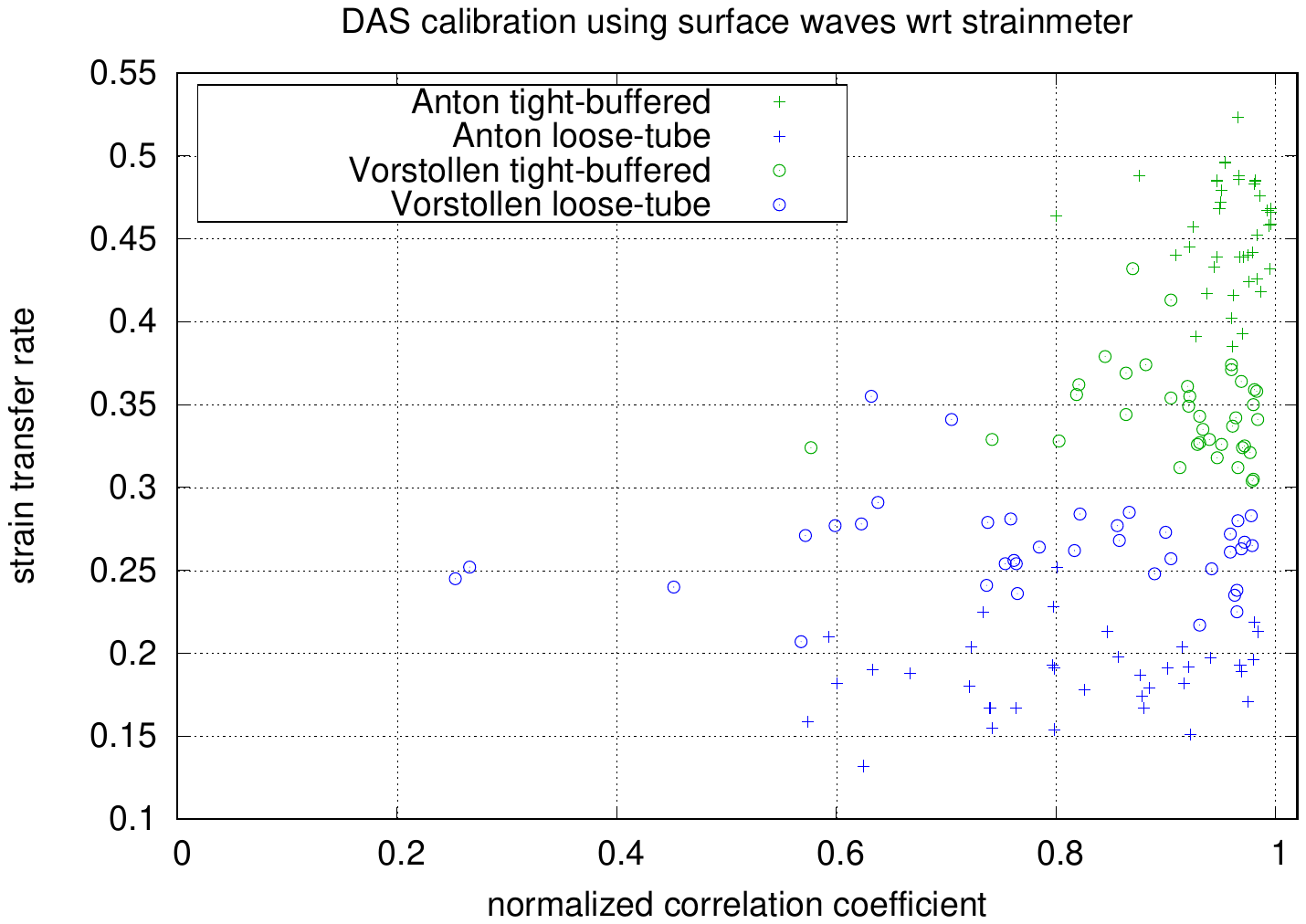}}
  \end{center}
  \caption{Measured \Tstraintransfer\ plotted against the normalized
  correlation coefficient for each of the 19 events studied in the analysis.
  The analysis focuses on surface waves and on the comparison between linear
  strain recorded by the DAS and strainmeter in a given azimuth. The
  measurements are carried out on sensing points situated along the \Lanton{}
  (crosses) and \Lvorstollen{} (circles) and in loose-tube (blue) and
  tight-buffered (green) cables.}
  \label{fig:STR_f_NCC}
\end{figure*}

Here we present some additional diagrams to support the main findings of the
study. 
In Fig.~\ref{fig:str:vs:baz} \Tstraintransfer\ (STR) is plotted against the
backazimuth (BAZ) of the respective event.
We observe no significant correlation between both parameters.
Fig.~\ref{fig:STR_f_NCC} shows STRs against NCCs. 
We find no correlation either. These results highlight that the
STRs predominantly depends on cable type and installation
conditions.

\begin{figure*}
  \begin{center}
    {\includegraphics[clip,trim=50 60 50 110,width=\textwidth]{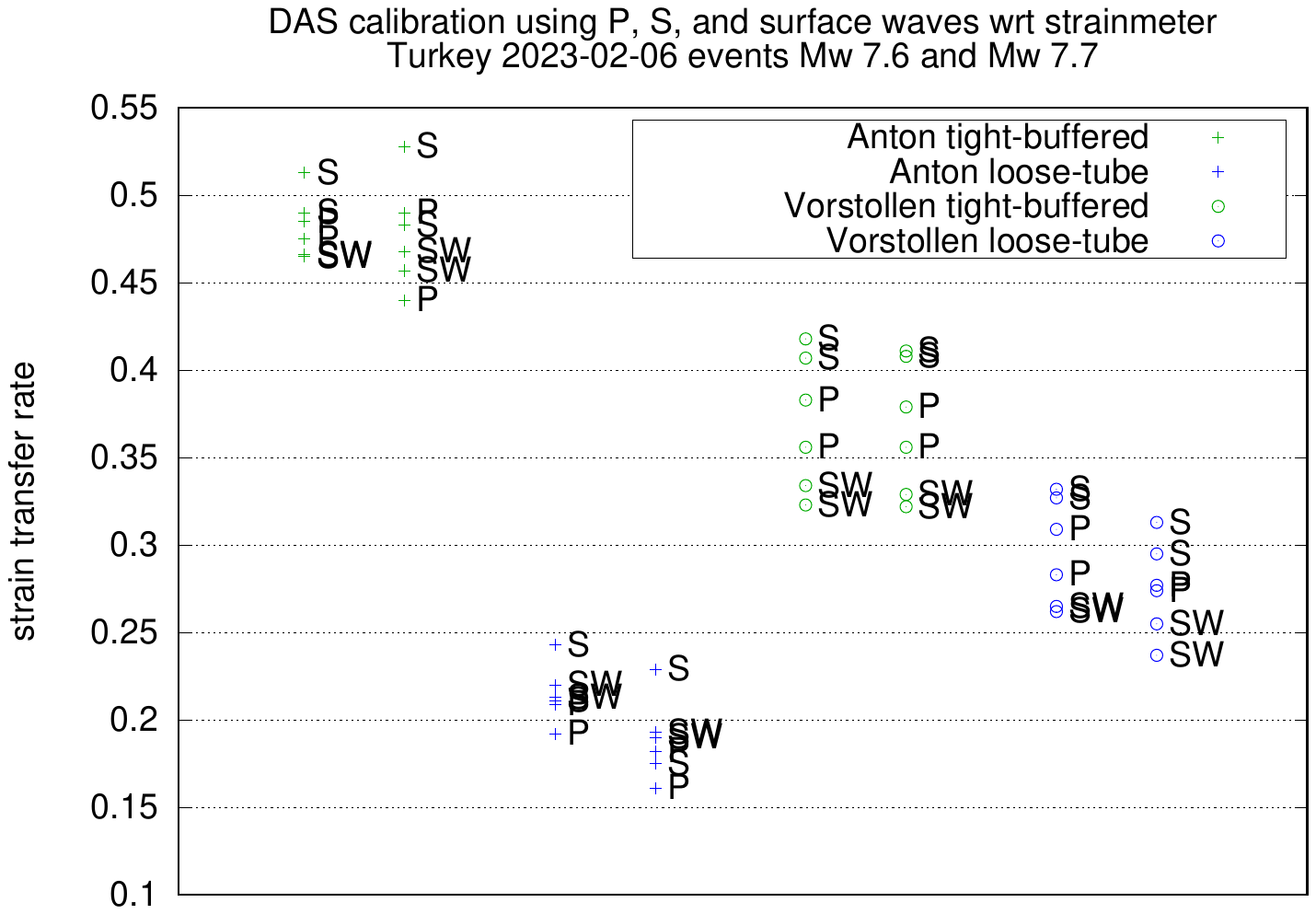}}
  \end{center}
  \caption{Measured \Tstraintransfer\ for 
  the main shocks (Mw~7.7 and Mw~7.6) of the
  Kahramanmaraş earthquake sequence on February 6th 2023.
  The analysis focuses on P, S and surface waves (SW), on the comparison
  between linear strain recorded by the strainmeter versus the DAS, in a given
  azimuth. 
  For each phase, the measurements are carried out on sensing points situated
  along the \Lanton{} (crosses) and \Lvorstollen{} (circles) and in 
  loose-tube (blue)
  and tight-buffered (green) cables. 
  The corresponding phases are indicated near the data points.}
  \label{fig:turkey_STR}
\end{figure*}

\begin{figure*}
  \begin{center}
    {\includegraphics[clip,trim=50 60 50 110,width=\textwidth]{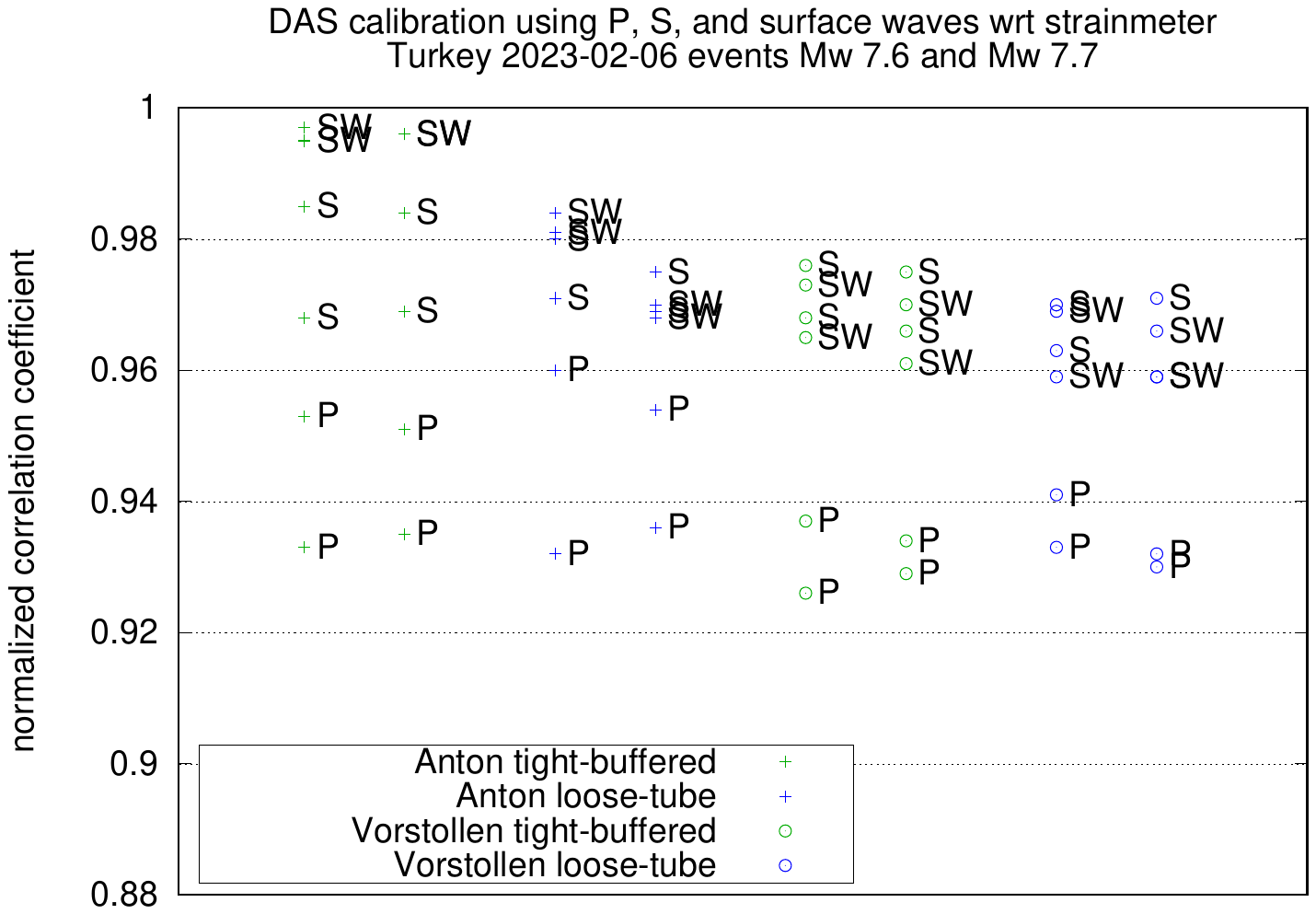}}
  \end{center}
  \caption{Normalized correlation coefficients for 
  the main shocks (Mw~7.7 and Mw~7.6) of the
  Kahramanmaraş earthquake sequence on February 6th 2023.
  The analysis focuses on P, S and surface waves (SW), on the comparison
  between linear strain recorded by the strainmeter versus the DAS, in a given
  azimuth. 
  For each phase, the measurements are carried out on sensing points situated
  along the \Lanton{} (crosses) and \Lvorstollen{} (circles) and in 
  loose-tube (blue)
  and tight-buffered (green) cables. 
  The corresponding phases are indicated
  near the data points.}
  \label{fig:turkey_NCC}
\end{figure*}

For the main shocks (Mw~7.7 and Mw~7.6) of the Kahramanmaraş earthquake
sequence on February 6th 2023, we analyze the body waves in addition to the
surface waves.
Fig.~\ref{fig:turkey_STR} shows STRs and
Fig.~\ref{fig:turkey_NCC} displays NCCs.
For these large amplitude signals, we observe large signal-to-noise ratio, which implies a high waveform similarity. Hence, the NCCs generally are larger than 0.92 and are largest ($>0.99$) for the
surface wave signals recorded in the \Lanton\ with the tight-buffered cable.

\begin{figure*}
  \begin{center}
    {\includegraphics[clip,trim=50 60 50 95,width=\textwidth]{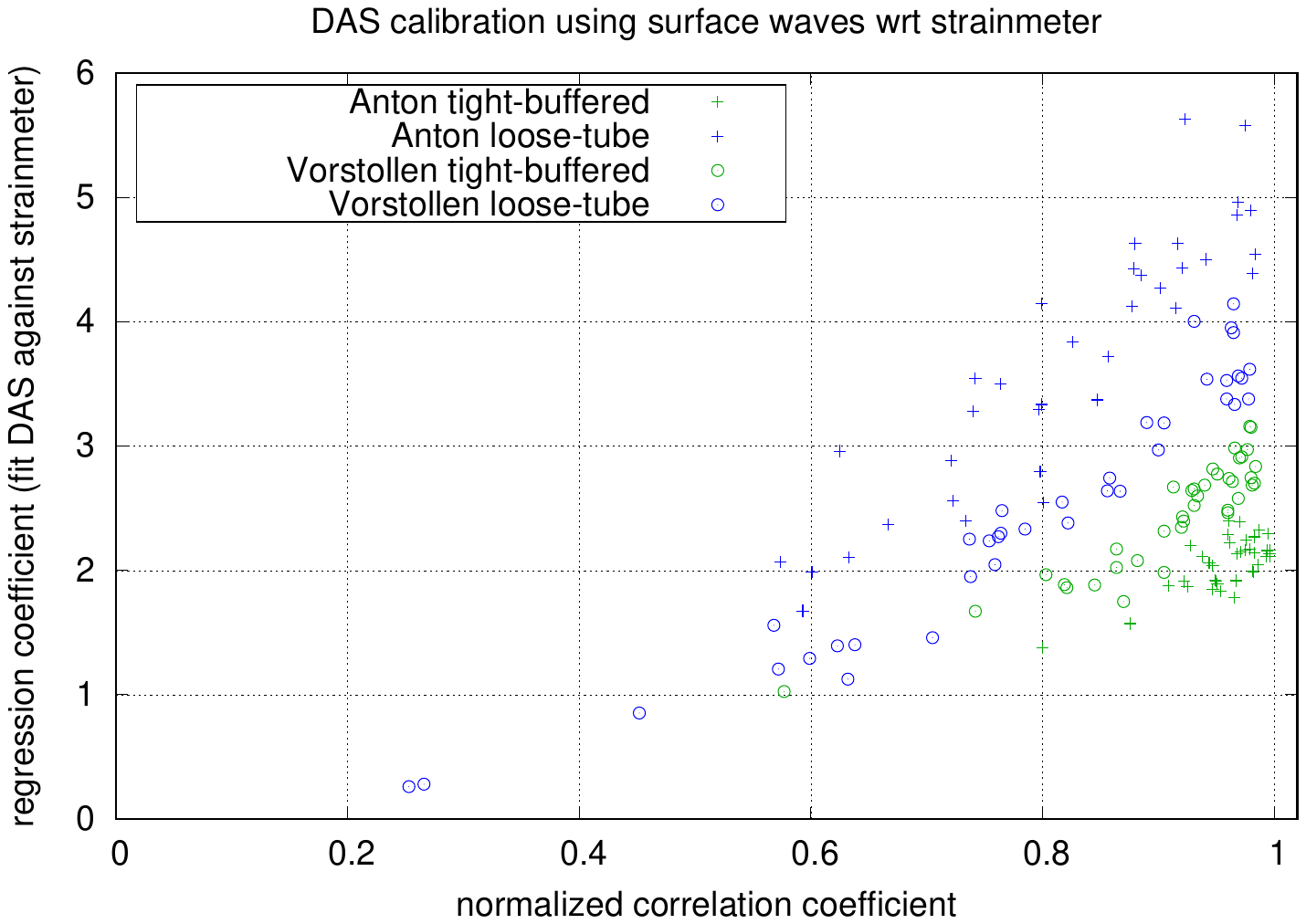}}
  \end{center}
  \caption{Regression coefficient for fitting DAS data to strainmeter
  waveforms plotted against the normalized correlation coefficient for each of
  the 19 events studied in the analysis. 
  The analysis focuses on surface waves and on the
  comparison between linear strain recorded by the DAS and strainmeter in a
  given azimuth. 
  The measurements are carried out on sensing points situated along the
  \Lanton{} (crosses) and \Lvorstollen{} (circles) and in loose-tube (blue) and
  tight-buffered (green) cables.}
  \label{fig:RC_f_NCC}
\end{figure*}

The \Tstraintransfer\ is not correlated to the normalized correlation coefficient, like shown in
Fig.~\ref{fig:STR_f_NCC}, because smaller values of NCC are due to noise in
the DAS data, not in the strainmeter data.
In this study, we fit strainmeter against DAS data in
order to determine the STR. 
This approach provides reasonable STR, even for heavily
disturbed DAS data. 
This is particularly the case in the \Lvorstollen, with the loose-tube cable,
for which NCC are globally small. 
Among the 19 events, illustrative examples of the situation
are shown in Figs.~\ref{fig:gfzAvpri}, \ref{fig:gfzAwxsg},
\ref{fig:gfzAwvyo}, and \ref{fig:gfzAuoko} (bottom panels), 

In the linear regression the noise in the regressor causes a bias for the
regression coefficients to smaller values.
If we use the DAS data as regressors in fitting the DAS waveform to the
strainmeter data, we find a correlation of the regression coefficients with
the NCC, which is demonstrated in Fig.~\ref{fig:RC_f_NCC}.
The signal-to-noise ratio for the strainmeter data is much better in all
cases, such that no bias is found in Fig.~\ref{fig:STR_f_NCC}, where the
strainmeter data is used as regressor.

\begin{figure*}
  \begin{center}
    {\includegraphics[clip,trim=50 60 50 95,width=\textwidth]{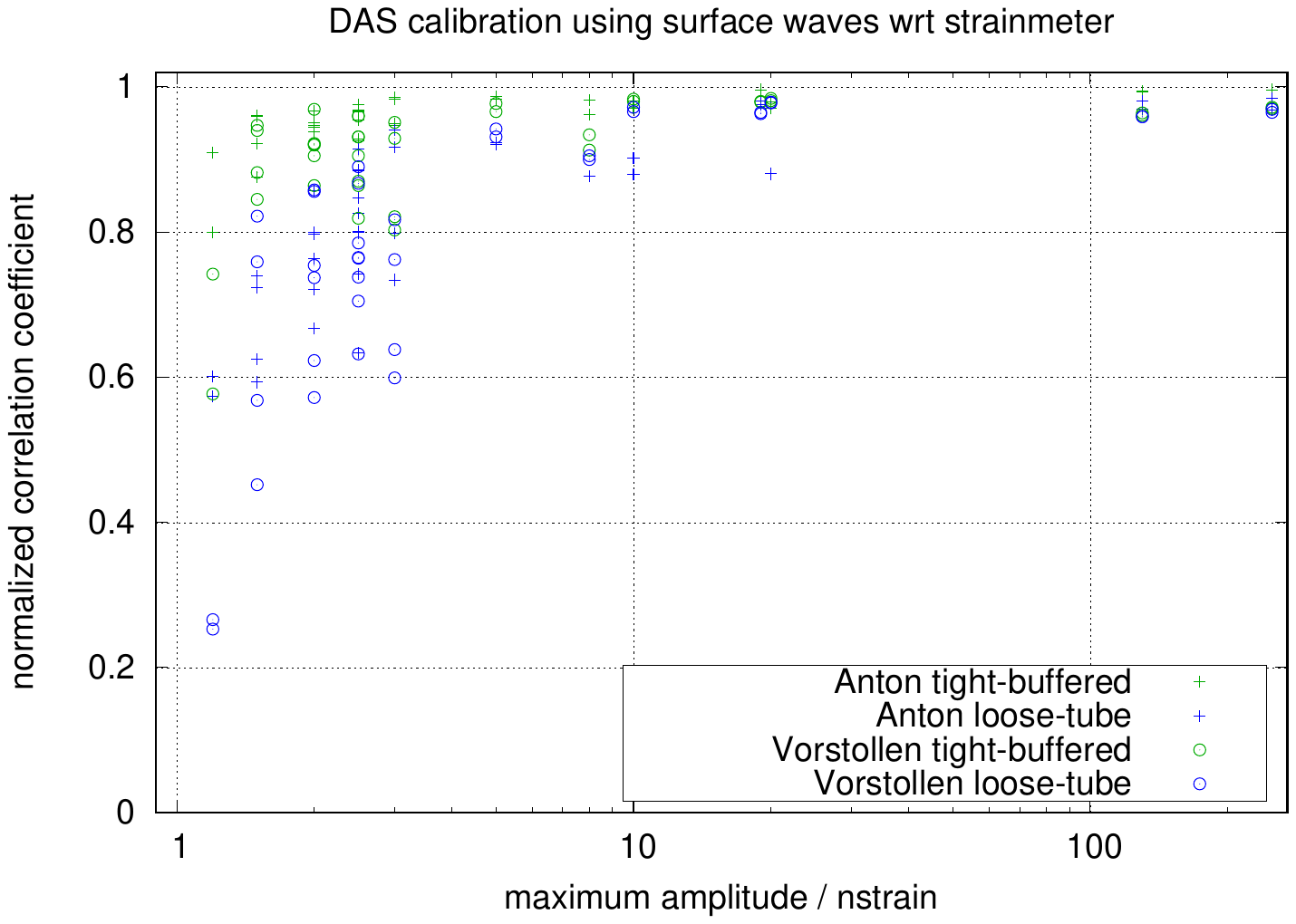}}
  \end{center}
  \caption{Normalized correlation coefficient plotted against the
  maximum signal
  amplitude for each of the 19 events studied in the analysis. 
  The values of maximum amplitude are those in listed in
  Table~\ref{tab:UsedEvents} and are given per event and not per actual time
  series and should be understood as a proxy for the actual signal amplitude.
  The analysis focuses on
  surface waves and on the comparison between linear strain recorded by the
  DAS and strainmeter in a given azimuth. The measurements are carried out on
  sensing points situated along the \Lanton{} (crosses) and \Lvorstollen{}
  (circles) and in loose-tube (blue) and tight-buffered (green) cables.}
  \label{fig:NCC_vs_maxAmp}
\end{figure*}

Fig.~\ref{fig:NCC_vs_maxAmp} displays a plot of NCC against the maximum strain
amplitude.
We observe a clear trend where smaller strain amplitudes are associated with lower NCC values.
For smaller strain amplitudes the noise in the DAS data takes a larger
fraction of the signal energy.
The effect is stronger for the loose-tube cable than for the tight-buffered
cable and is strongest in the \Lvorstollen, where cables are not embedded in
sand.

\end{multicols}
\clearpage

\clearpage
\begin{multicols}{2}
\section{Plane wave strain from seismometer recordings}
\label{sec:seismometer:as:a:refernce}
In this study, we fit waveforms of \Trockstrain\ recorded by the
array of Invar-wire strainmeters at BFO \citep{zuern2015} 
to \Tfiberstrain\ recorded by DAS.
The regression coefficient is the \Tstraintransfer{} (STR), a measure of how much of
the strain amplitude is picked up by the DAS fibers.
Strainmeters are rare installations and in the absence of these types of
sensors studies of DAS recorded strain occasionally compare the signals with
strain simulated from particle velocity recorded by seismometers, or convert
DAS signals to equivalent particle velocity recordings.
This conversion is possible in cases where the waveform represents a plane
wave of known phase velocity and azimuth of propagation.

In section~\secref{subsec:scaling:particle:velocity} we summarize the theory
behind this conversion.
The final conversion rule is given by \eqref{eq:supp:plane:wave:strain} and is
valid for a single, non-dispersive plane wave.

When using surface wave-data, which provide superior signal-to-noise ratio
because of their large amplitudes (compared to body-wave signals) the
limitations of the plane-wave approximation become significant.
The surface waves are dispersive and do not propagate with a single
phase-velocity and Love- and Rayleigh-waves are inseparably superimposed. 
Due to their propagation along the surface of the heterogeneous Earth's crust,
they have non-plane character and contain significant components, which do not
propagate along the backazimuth (BAZ) great circle.
These limitation become most obvious for BAZ being perpendicular to the
azimuth of measured linear strain.
The plane wave does not strain the material
perpendicular to the direction of
propagation and the value of the cosine in \eqref{eq:plane:wave:strain}
necessarily vanishes.
The recorded signals then entirely are composed by the non-plane components of
the wave, not propagating along the BAZ great circle.
The problem becomes already apparent for BAZ being close to perpendicular to
strain azimuth.
The seismometer recorded particle velocity contains as well these non-plane
components, which are at variance with the expected polarization of the plane
wave.
Thus the scaling in \eqref{eq:supp:plane:wave:strain} is not valid for them.
We discuss the consequences of this on the regression coefficients in
section~\secref{subsec:regression:DAS:with:seismometer:data}.
In section~\secref{subsec:regression:seismometer:and:strainmeter:data} we discuss
a direct comparison of strainmeter and seismometer data.

\subsection{Scaling rule for plane waves}
\label{subsec:scaling:particle:velocity}
For signals of single non-dispersive plane waves, strain can be simulated from
particle velocity recorded by a seismometer.
Consider that ground deformation is due to a non-dispersive plane wave, then
particle displacement at location $\vec{r}$ and time $t$ is
\begin{equation}
  \vec{u}(\vec{r},t)=\vec{U}\,f(\vec{s}\,\vec{r}-t),
\end{equation}
where $\vec{s}$ is the slowness-vector of the plane wave, $\vec{U}$ defines
the polarization and $f(t)$ is the shape of the wave (d'Alembert's solution to
the wave equation), which propagates in
direction of the slowness vector.

The linear strain in $x$-direction for this wave is
\begin{equation}
  \epsilon_{xx}(\vec{r},t)=
  \frac{\Sd}{\Sd x}\,u_x(\vec{r},t)
  =U_x\,s_x\,f^\prime(\vec{s}\,\vec{r}-t),
\end{equation}
where $f^\prime(t)$ is the derivative of $f(t)$ with respect to $t$
and $u_x$, $s_x$, and $U_x$ are the $x$-components of 
the slowness vector $\vec{s}$, of $\vec{u}$, and
of $\vec{U}$, respectively.
Likewise the $x$-component of particle velocity is 
\begin{equation}
  v_{x}(\vec{r},t)=
  \frac{\Sd}{\Sd t}\,u_x(\vec{r},t)
  =-U_x\,f^\prime(\vec{s}\,\vec{r}-t).
\end{equation}
Hence
\begin{equation}
  \frac{\epsilon_{xx}(\vec{r},t)}{v_{x}(\vec{r},t)}
  =-s_x.
\end{equation}

If the $x$-direction is the horizontal direction along the \Lanton\ or the
\Lvorstollen, given by azimuth $\AZI_{x}$ and $s_h$ is the horizontal
component of the plane wave slowness for propagation in a 1D structure, as can
be derived by ray-tracing, then
\begin{equation}
  s_x=s_h\,\cos\bigl(\AZI_{\text{BAZ}}-\AZI_{x}-180^{\circ}\bigr),
\end{equation}
where $\AZI_{\text{BAZ}}$ is the backazimuth of the source.
In this way we estimate the linear strain in $x$-direction
\begin{equation}
  \epsilon_{xx}(\vec{r},t)=
  -s_h\,\cos\bigl(\AZI_{\text{BAZ}}-\AZI_{x}-180^{\circ}\bigr)\,v_{x}(\vec{r},t)
  \label{eq:supp:plane:wave:strain}
\end{equation}
from the particle velocity $v_{x}(\vec{r},t)$ as recorded by the broad-band
seismometer.

\subsection{Regression with respect to seismometer data}
\label{subsec:regression:DAS:with:seismometer:data}
\begin{figure*}
\begin{center}
  \begin{minipage}{0.99\textwidth}
  \begin{tabular}{rr}
  {\includegraphics[trim=60 60 40 90,clip,width=0.48\textwidth]{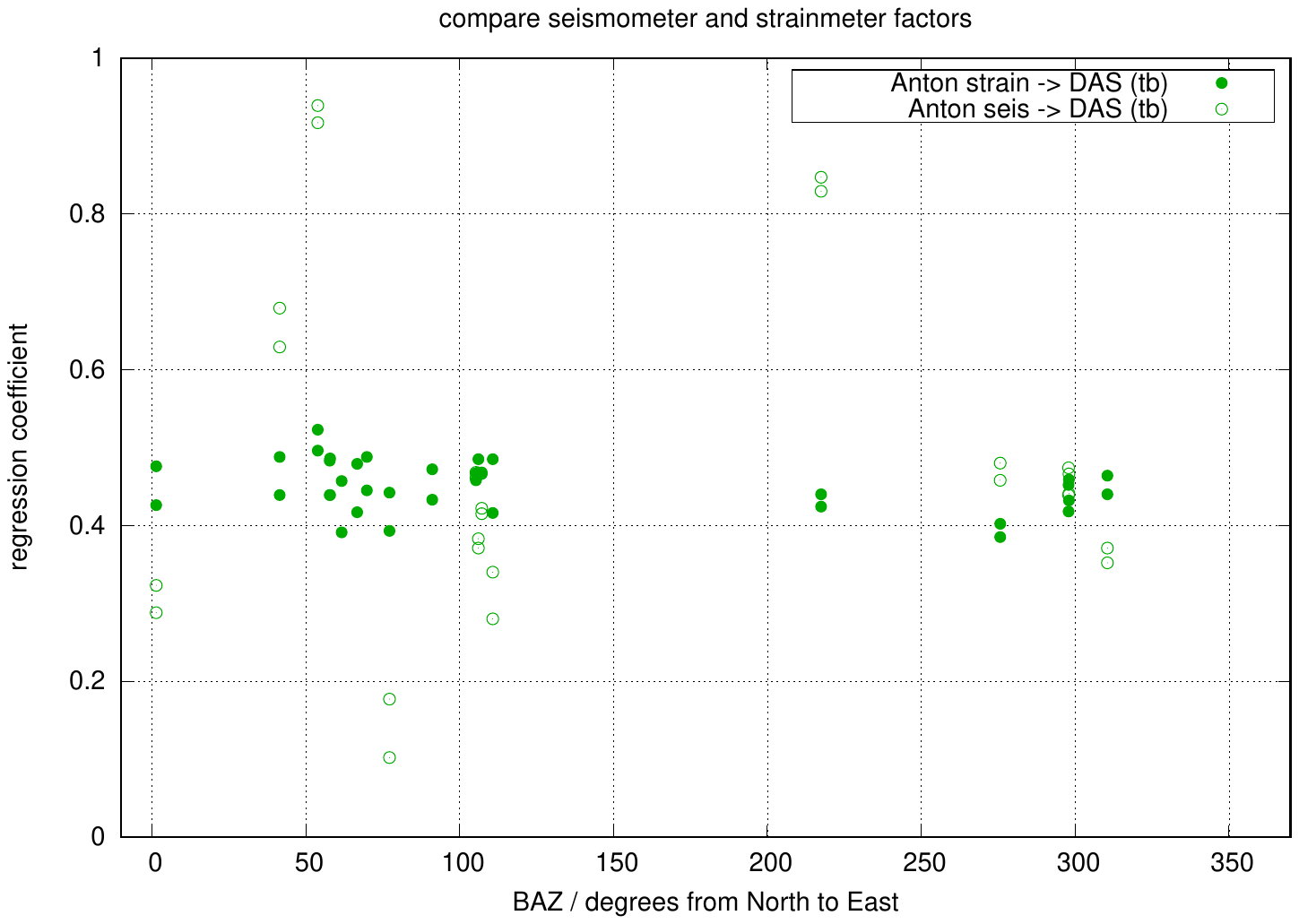}}
    &
  {\includegraphics[trim=60 60 40 90,clip,width=0.48\textwidth]{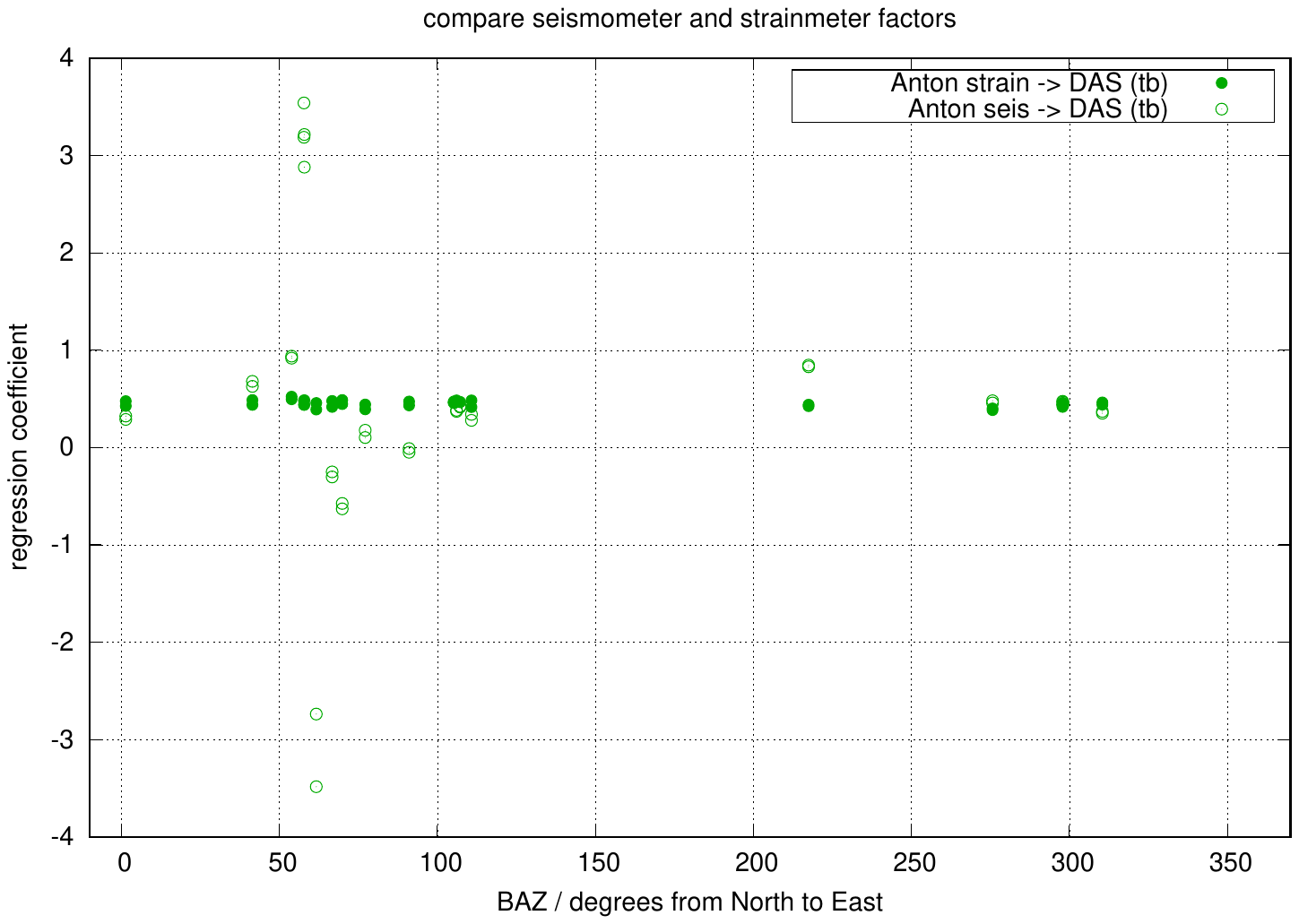}}\\
  {\includegraphics[trim=60 60 40 90,clip,width=0.48\textwidth]{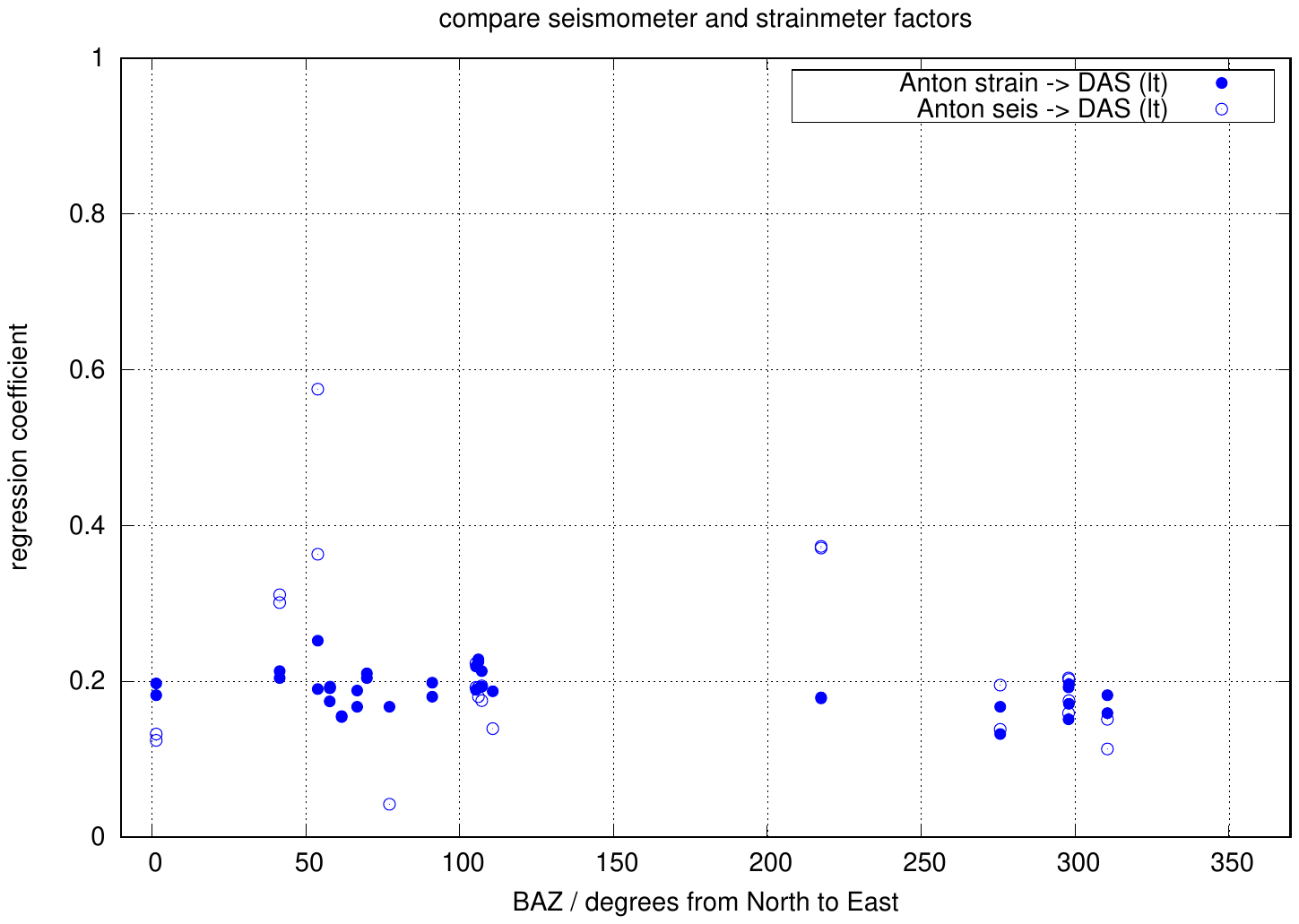}}
    &
  {\includegraphics[trim=60 60 40 90,clip,width=0.48\textwidth]{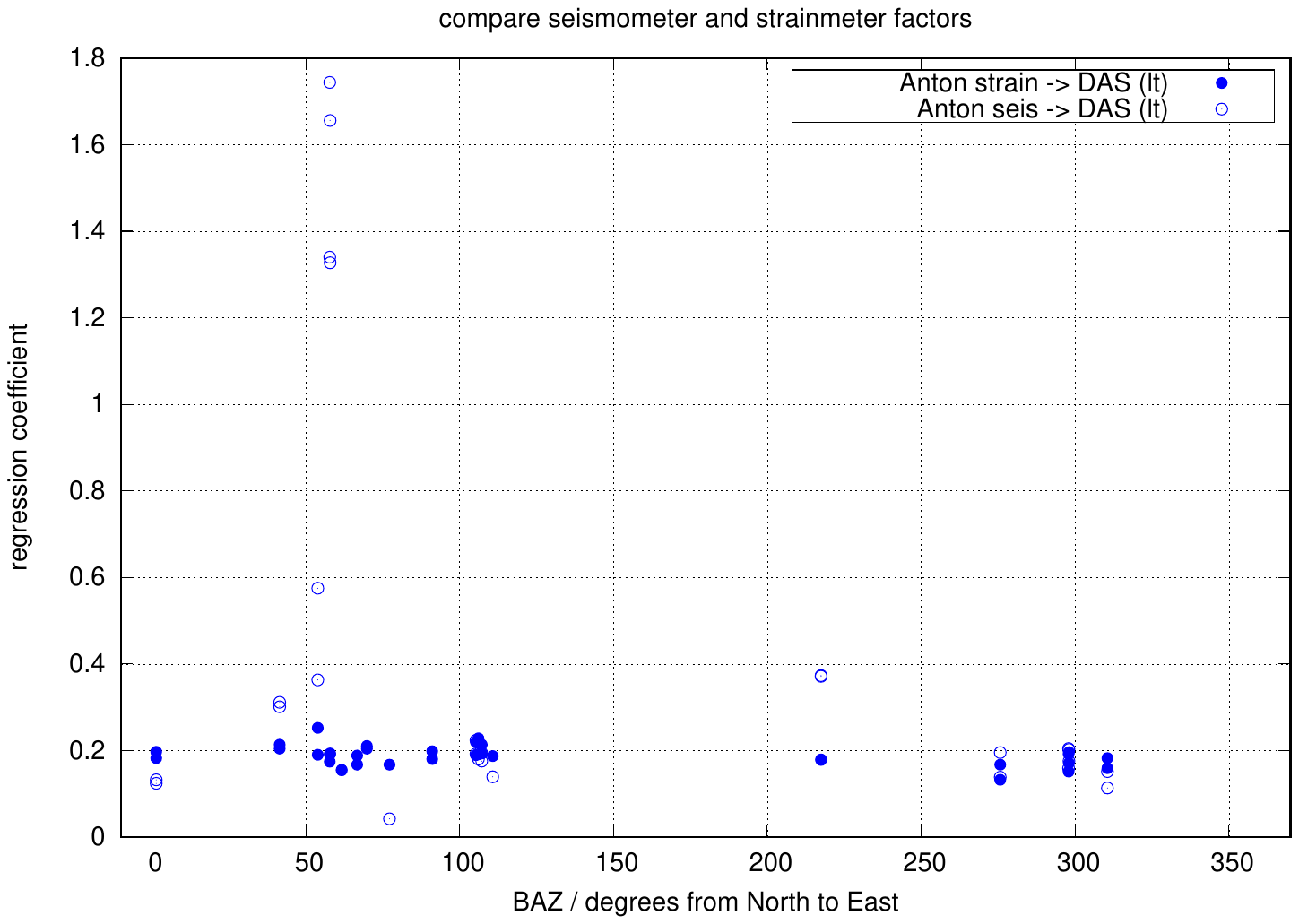}}\\
  {\includegraphics[trim=60 60 40 90,clip,width=0.48\textwidth]{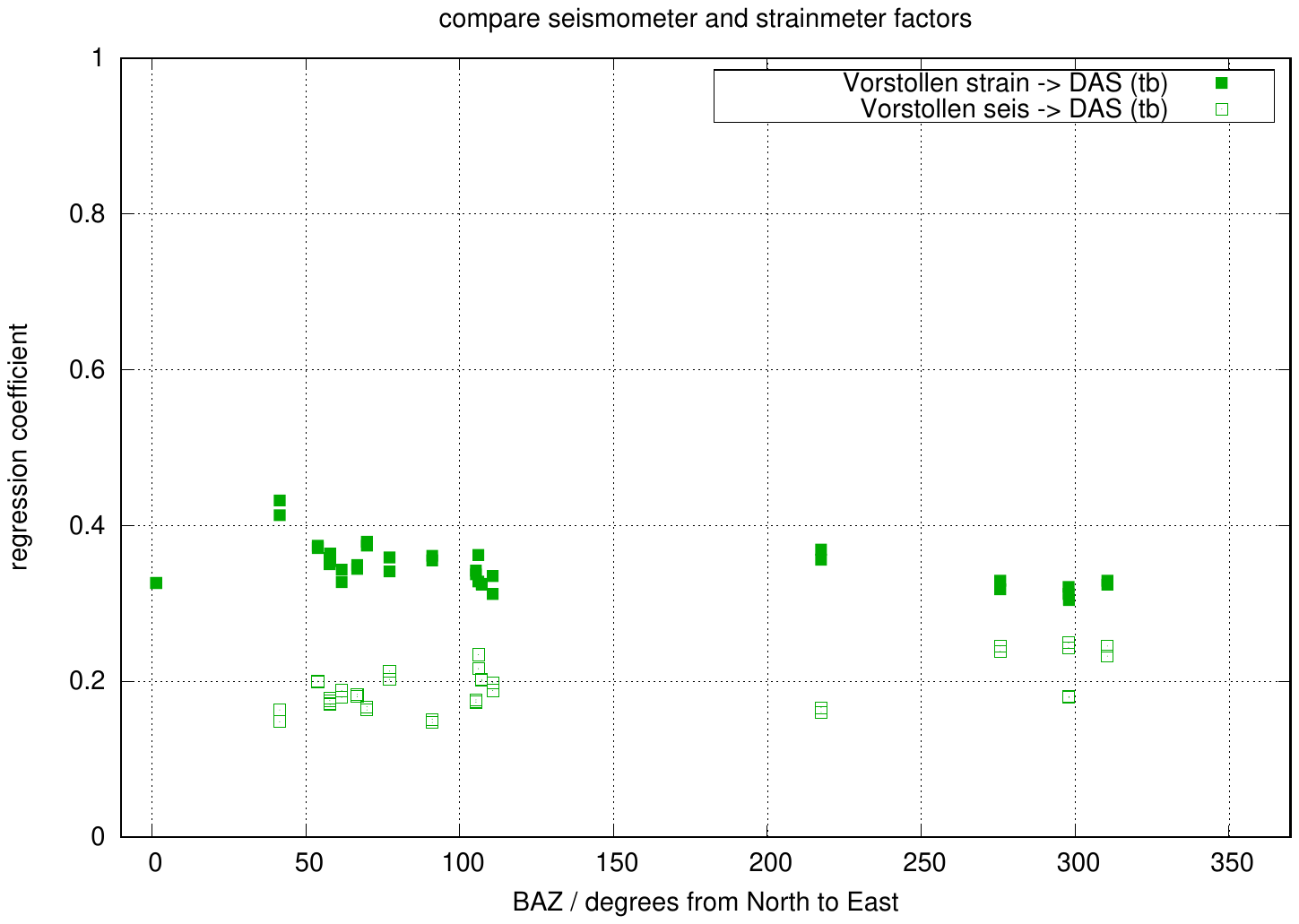}}
    &
  {\includegraphics[trim=60 60 40 90,clip,width=0.48\textwidth]{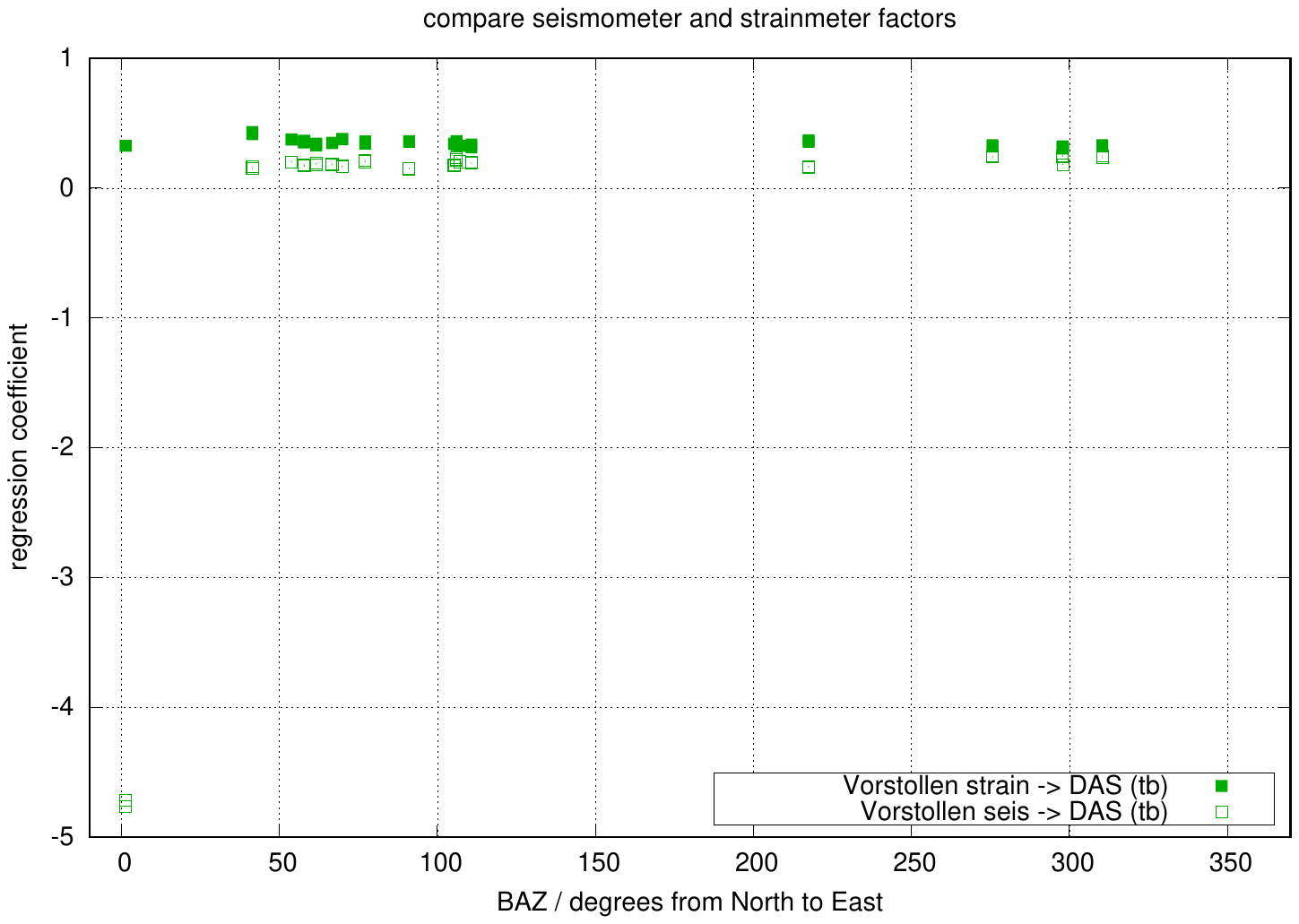}}\\
  {\includegraphics[trim=60 60 40 90,clip,width=0.48\textwidth]{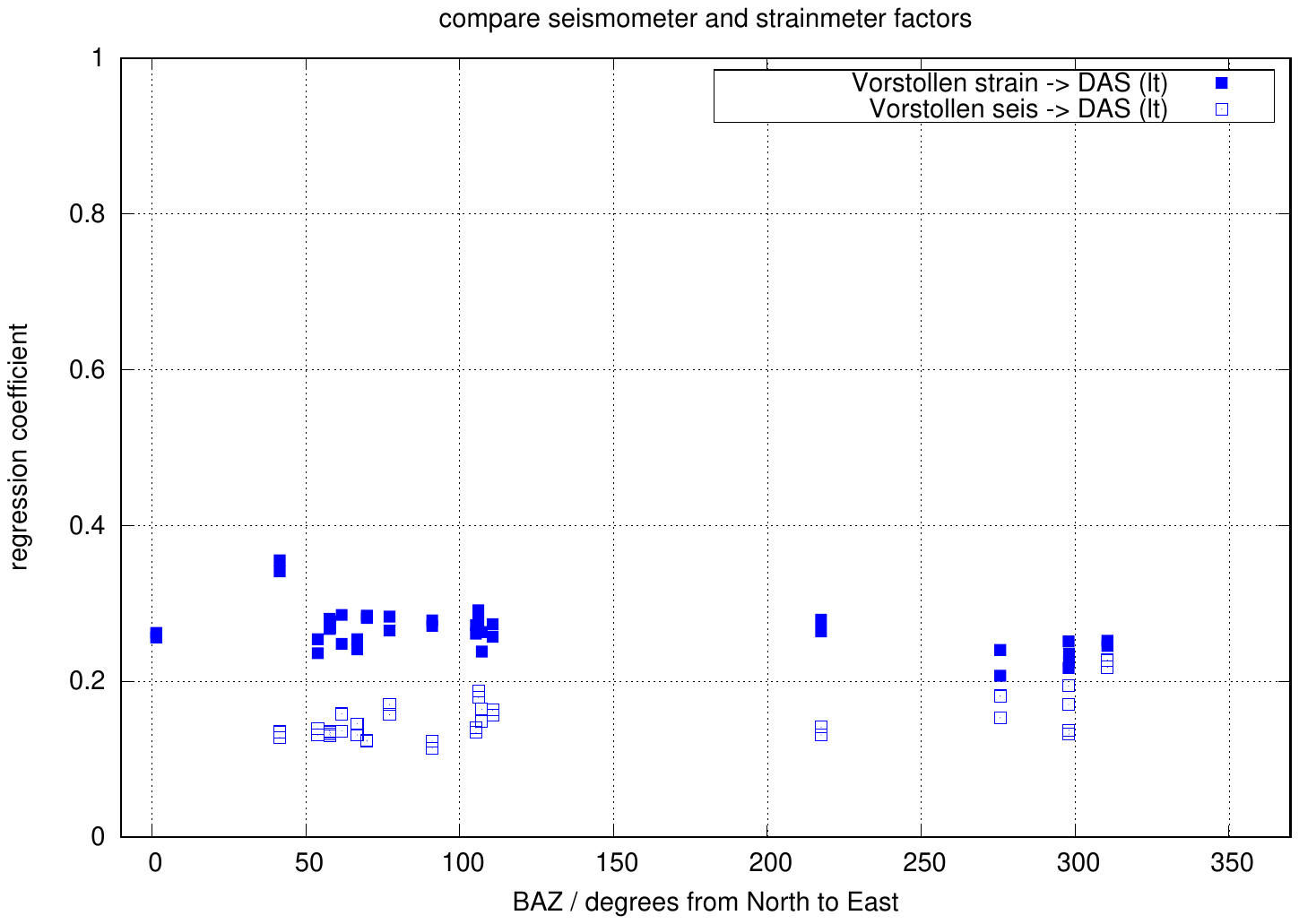}}
    &
  {\includegraphics[trim=60 60 40 90,clip,width=0.48\textwidth]{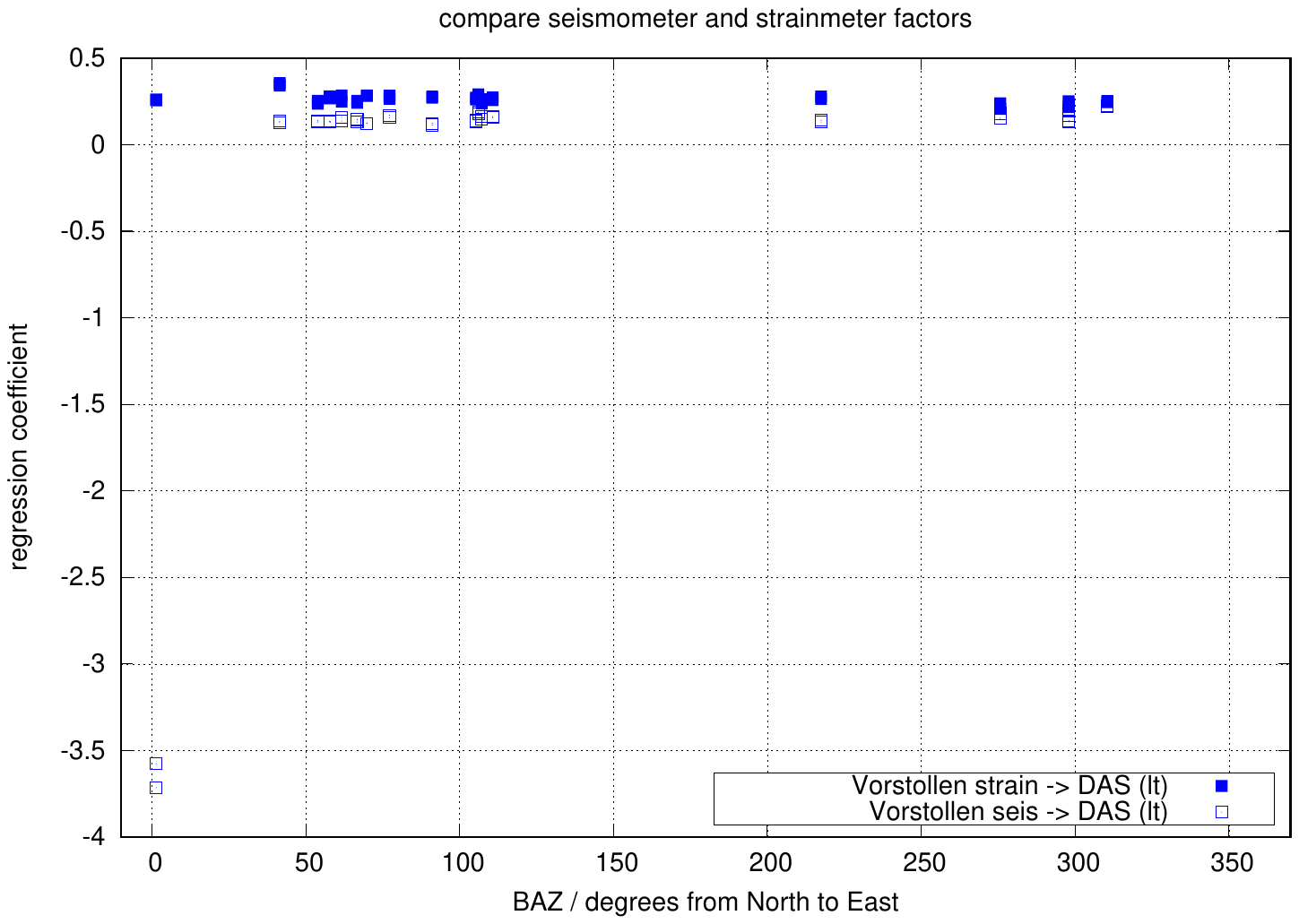}}\\
  \end{tabular}
  \end{minipage}\par
\end{center}
  \caption{Regression coefficients obtained when fitting to the DAS data and
  using the strainmeter data (filled symbols) as regressor on the one hand and
  seismometer data (open symbols) on the other hand.
  The values are plotted against the BAZ of the
  respective earthquake.
  The diagrams on the left are limited to values from 0 to 1.
  The diagrams on the right show the full scatter.
  From top to bottom: 
  tight-buffered in the \Lanton,
  loose-tube in the \Lanton,
  tight-buffered in the \Lvorstollen,
  loose-tube in the \Lvorstollen.
  The seismometer data is scaled for a slowness of 280\,ms\,km$^{-1}$,
  equivalent to a phase velocity of 3.57\,km\,s$^{-1}$.}
  \label{fig:RC:DAS:vs:seismometer:and:strainmeter}
\end{figure*}

Fig.~\ref{fig:RC:DAS:vs:seismometer:and:strainmeter} compares the regression
coefficients measured by fitting strainmeter signals to DAS signals (filled symbols), with those obtained by fitting scaled seismometer data to DAS
signals (open symbols).
The seismometer data is scaled according to \eqref{eq:supp:plane:wave:strain} to represent
linear strain in the direction of the DAS fiber in cases where the waves
propagate along the BAZ great circle with phase slowness
$s_h=280\,\text{ms}\,\text{km}^{-1}$.
This value corresponds to phase velocity of $3.57\,\text{km}\,\text{s}^{-1}$,
which is the value for Rayleigh waves at 0.05\,Hz (the lower end of the
investigated frequency band) as shown for Southern Germany by
\citet{friederich1996}.
Love-waves propagate with a higher phase velocity.
For both wave types phase velocity decreases with increasing frequency, such
that the chosen value lies within the range to be expected in the observed
wave trains.

The regression coefficients obtained by fitting the strainmeter waveform against the
DAS waveform are the STRs, as displayed in
Fig.~\ref{fig:str:vs:baz}.
They are reproduced for reference.
The regression coefficients for fitting the scaled seismometer waveform
against the DAS waveform show a considerably stronger scatter, not only for
cases where BAZ is close to perpendicular to the strain azimuth.

Cases where BAZ is close to \azimuth{60} or \azimuth{240} have a great-circle
propagation direction almost perpendicular to the azimuth of the \Lanton\
(\azimuth{330}).
For the Vorstollen (\azimuth{90}), the same applies when BAZ is close to \azimuth{0} or \azimuth{180}.
Most of these cases are illustrated in Fig.~\ref{fig:RC:DAS:vs:seismometer:and:strainmeter}. The coefficients are significantly outside reasonable limits, with some even being negative.
The scatter is more substantial for the \Lanton, where we find several cases
with BAZ almost perpendicular to the strain azimuth.
For the \Lvorstollen{} the study includes one case with BAZ near \azimuth{0},
but none with BAZ near \azimuth{180}.
In consequence the overall scatter is less strong.

For data from the \Lvorstollen{}
we also observe a systematic shift between coefficients obtained for regression
with respect to strainmeter data compared to regression with respect to
seismometer data.
The seismometer data appears to over-estimate the strain amplitude, such that
it has to be downscaled by a factor of about 0.5 with respect to the
coefficients obtained with the strainmeter data.
The P-wave signals in Figs.~\ref{fig:turkey:BWA:Vorstollen} and
\ref{fig:turkey:BWB:Vorstollen} show an amplitude mismatch of a similar size.
This might be a consequence of strain-strain coupling due to the local
topography as discussed in section~\secref{subsec:cavity:effects}.
The East-West strain measured by the strainmeters as well as by the DAS fibers
in the \Lvorstollen{} is reduced compared to what is estimated from particle
velocity (seismometer data).
From tidal analysis 
\citet[][their table~S2 in the supporting material]{zuern2015} 
estimate a factor of 0.58 for \azimuth{60} and
\citet[][their figure~5]{emter1985} derive 
a factor of 0.67 by a 2D finite element analysis.
Both are in the order of magnitude of the bias seen in 
Fig.~\ref{fig:RC:DAS:vs:seismometer:and:strainmeter} for the \Lvorstollen.

\begin{figure*}
  \begin{tabular}{rr}
  {\includegraphics[trim=60 60 40 90,clip,width=0.48\textwidth]{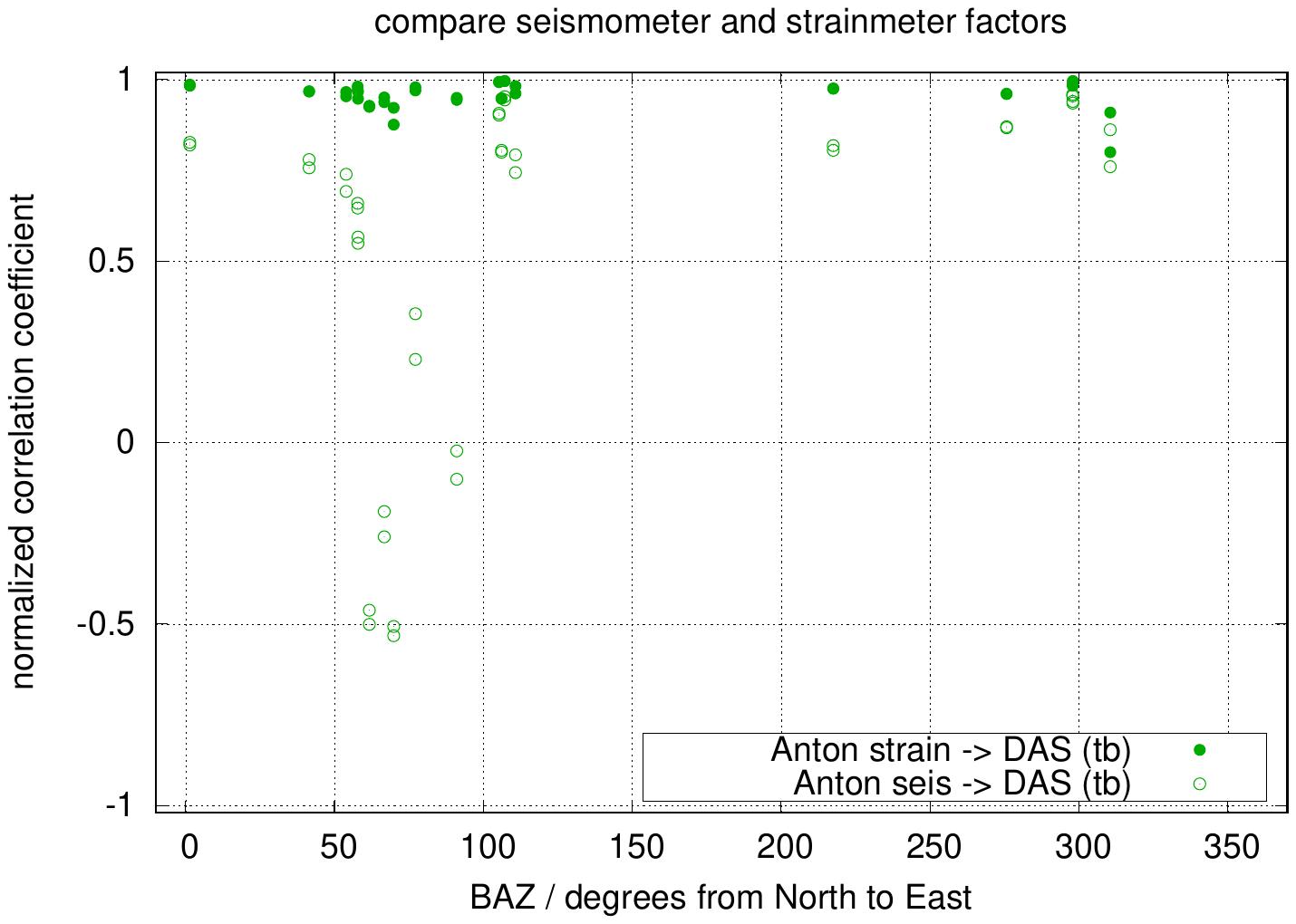}} &
  {\includegraphics[trim=60 60 40 90,clip,width=0.48\textwidth]{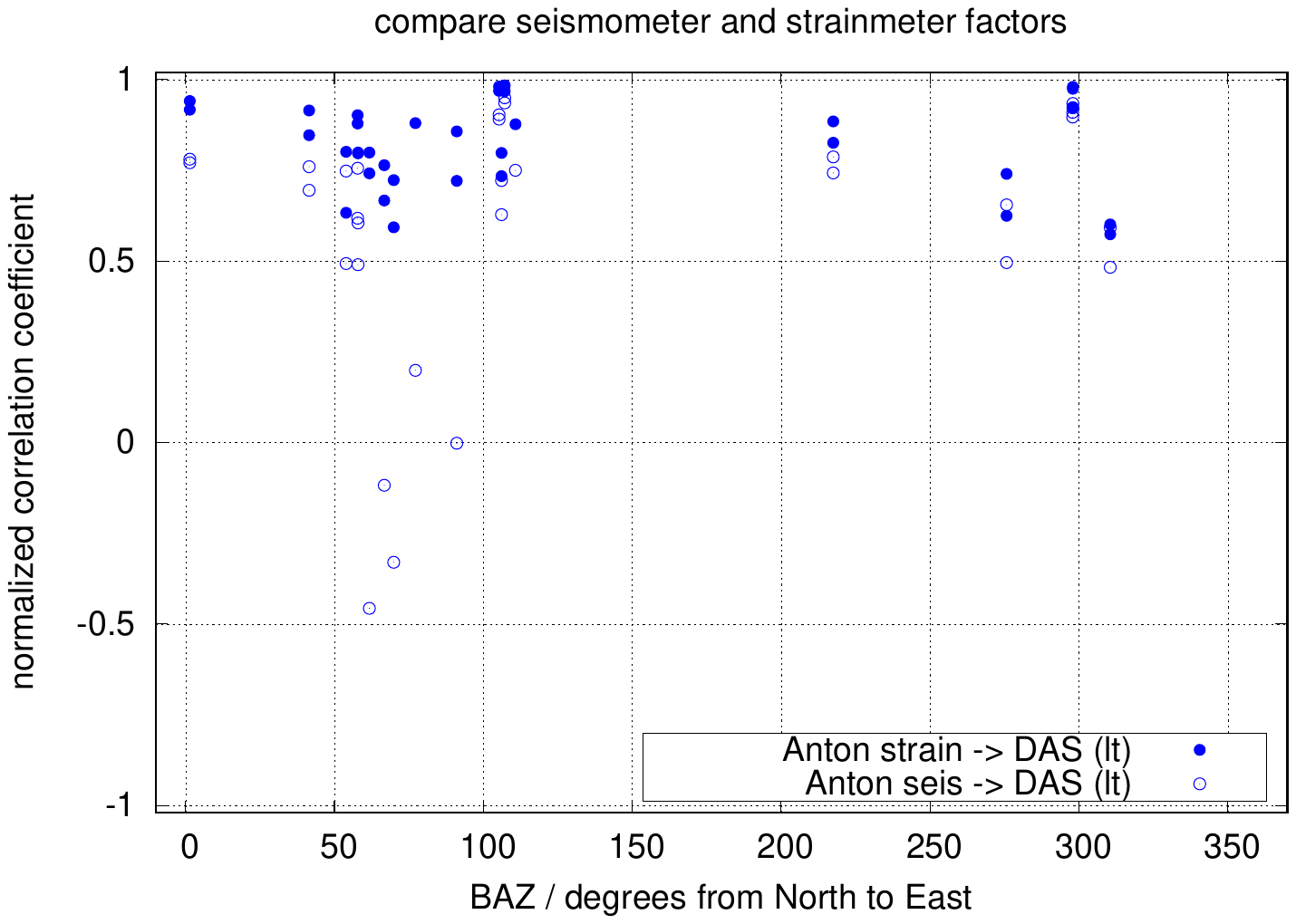}}\\
  {\includegraphics[trim=60 60 40 90,clip,width=0.48\textwidth]{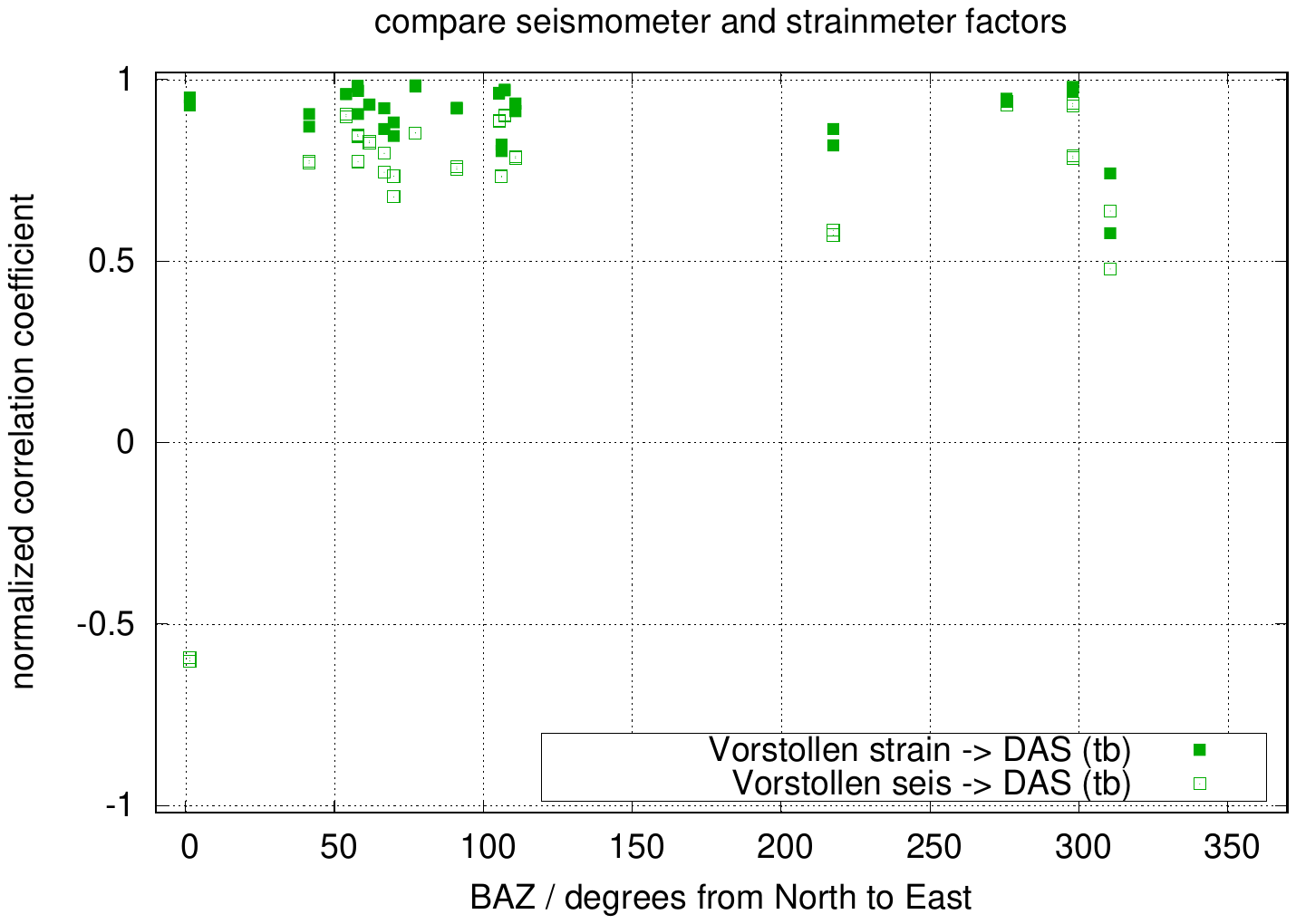}} &
  {\includegraphics[trim=60 60 40 90,clip,width=0.48\textwidth]{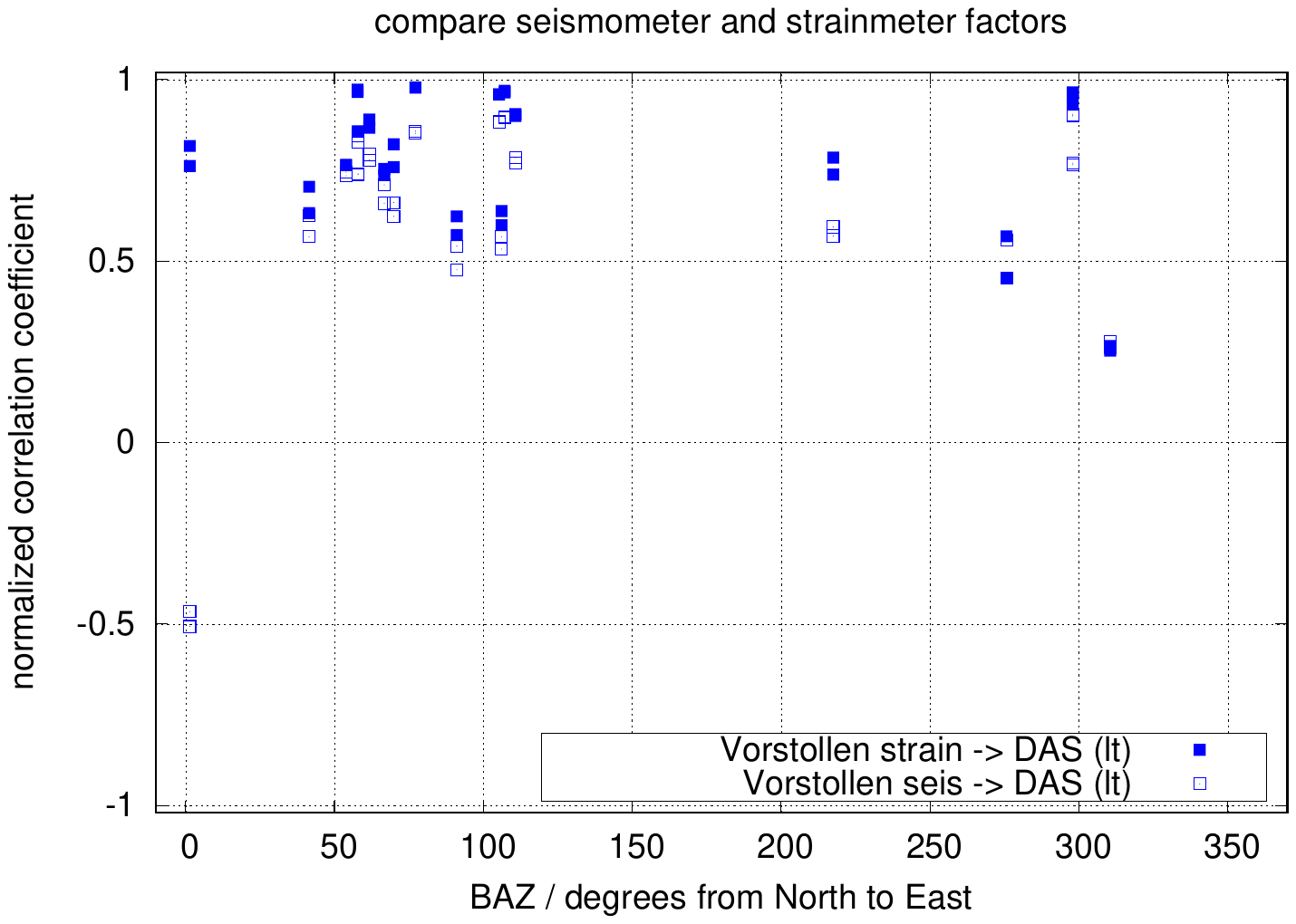}}
  \end{tabular}
  \caption{Normalized correlation coefficients for DAS data and
  strainmeter data (filled symbols) on the one hand and
  seismometer data (open symbols) on the other hand.
  Results are plotted against the backazimuth.
  Results for both cables (left: tight-buffered, right: loose-tube) and both
  locations (top: \Lanton, bottom: \Lvorstollen) are displayed.
  }
  \label{fig:NCC:DAS:seismometer:and:strainmeter}
\end{figure*}

Apart from these amplitude related issues,
the DAS waveforms also show greater similarity to the 
strainmeter waveforms than to the seismometer waveforms.
This is shown in Fig.~\ref{fig:NCC:DAS:seismometer:and:strainmeter}, where the
values of NCCs are plotted against
backazimuth.
The NCC measured by comparing DAS data with respect to seismometer data are generally smaller than
for DAS with respect to the strainmeter data.
For BAZ nearly perpendicular to the fiber optic cable azimuth, some of the
signals are even anti-correlated
(Fig.~\ref{fig:NCC:DAS:seismometer:and:strainmeter}). 
This results from the change in the sign of the cosine in
\eqref{eq:supp:plane:wave:strain}.

The dissimilarity is attributed to the non-plane components of the waves, rather than to instrumental noise or signal-to-noise ratio (SNR) issues. Both the strainmeter and seismometer data exhibit significantly better SNR than the DAS data, with the seismometer having the highest SNR among the instruments compared.
This can be seen in Figs.~\ref{fig:turkey:BWA:Anton},
\ref{fig:turkey:BWB:Anton}, \ref{fig:turkey:BWA:Vorstollen}, and
\ref{fig:turkey:BWB:Vorstollen}, where the signal level prior to the P-wave
onset is least for the seismometer data, when compared to the P-wave
amplitude.

\subsection{Comparison of seismometer and strainmeter data}
\label{subsec:regression:seismometer:and:strainmeter:data}
\begin{figure*}
  \begin{tabular}{rr}
  {\includegraphics[trim=60 60 40 90,clip,width=0.48\textwidth]{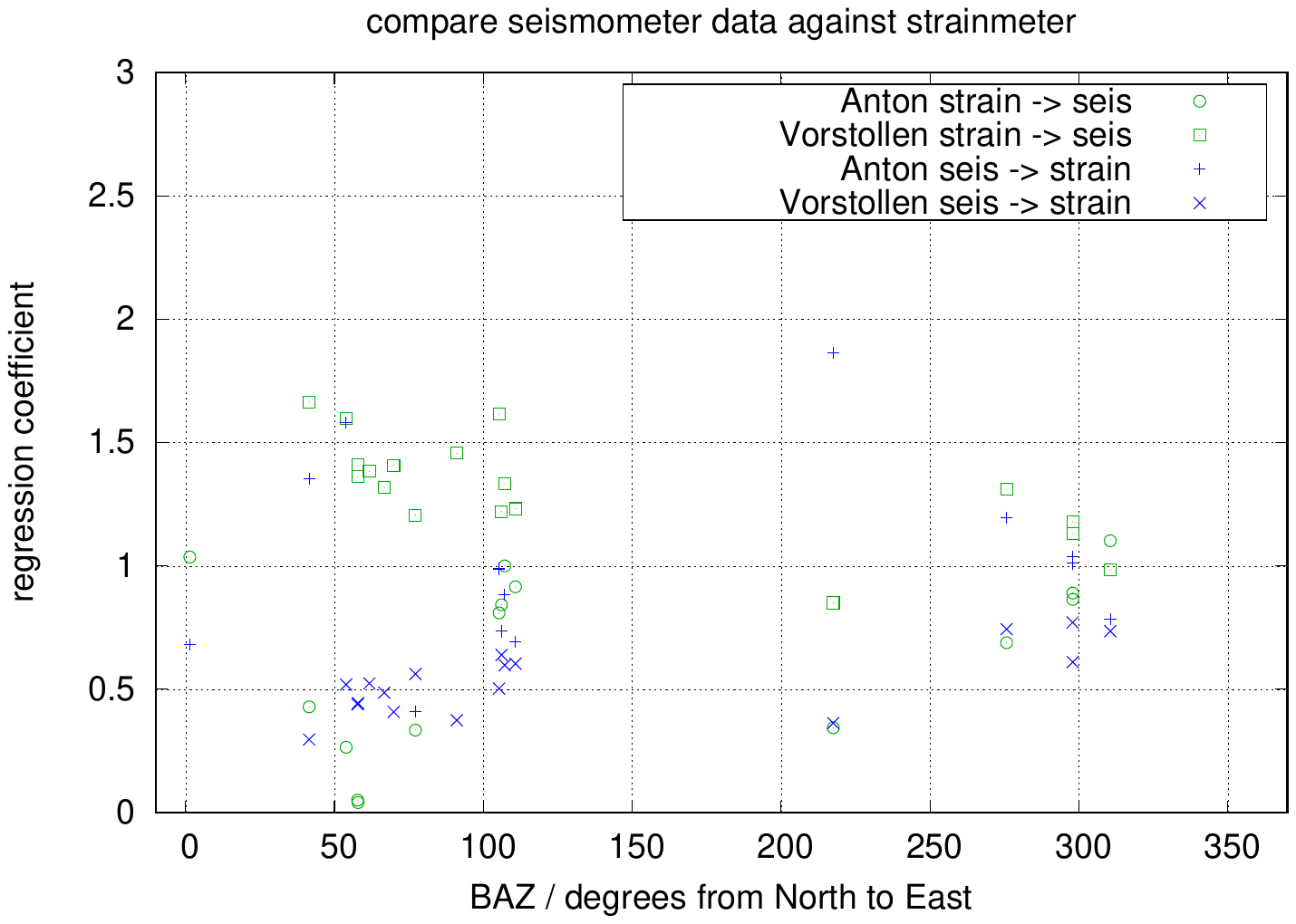}}
    &
  {\includegraphics[trim=60 60 40 90,clip,width=0.48\textwidth]{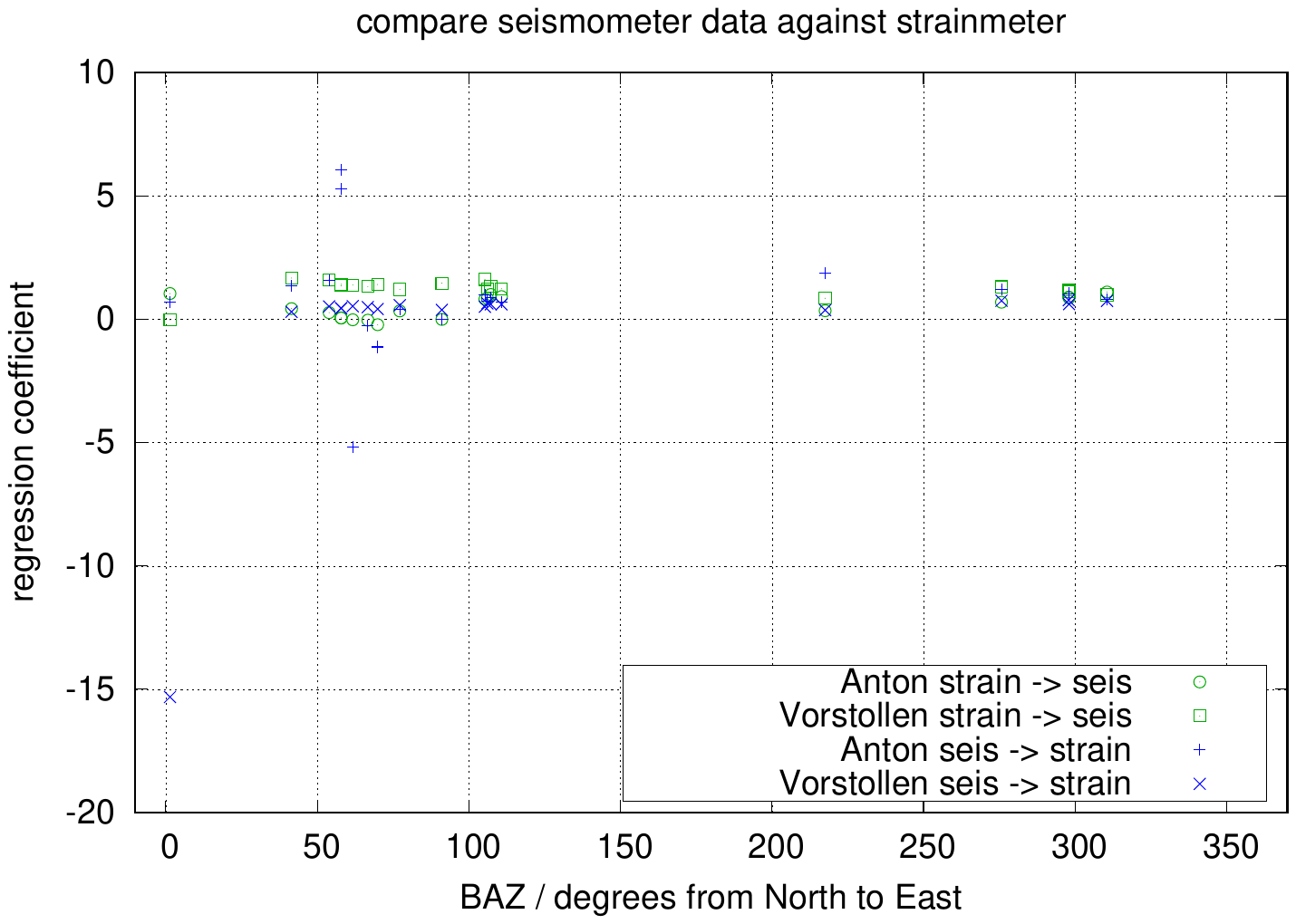}}
  \end{tabular}
  \caption{Regression coefficients when fitting surface wave strainmeter data
  to seismometer data (green) and fitting seismometer data to strainmeter data
  (blue).
  Results are displayed for the azimuths of the \Lanton\ (circles and pluses)
  and the \Lvorstollen{} (squares and crosses).
  The values are plotted against the backazimuth (BAZ) of the respective
  earthquake.
  Left: limited value range.
  Right: full range.
  In cases, where the seismometer derived strain signal represents rock
  strain, the values equal~1.
  These cases are rare.
  }
  \label{fig:RC:seis:to:strain:vs:baz}
\end{figure*}

To evaluate how well strain from particle velocity recorded by the seismometer can represent \Trockstrain{}, we perform a direct comparison between the strain waveforms from the strainmeter array and those derived from the particle velocity recorded by the seismometer. Fig.~\ref{fig:RC:seis:to:strain:vs:baz} shows the results of the comparison.
If the seismometer signal scaled by \eqref{eq:supp:plane:wave:strain} would be a
good representation of \Trockstrain, all these coefficients would equal~1.
This takes place in only very few cases.
If the mismatch would be due to an amplitude factor only, the coefficients for
fitting the strainmeter data to the seismometer data would be the reciprocals
of the coefficients computed when fitting seismometer data to strainmeter data.
This is not the case, due to a considerable waveform mismatch caused by
non-plane components of the surface waves.

\end{multicols}
\clearpage


\end{document}